\documentclass[english]{book}
\usepackage[latin1]{inputenc}
\usepackage{a4wide}
\usepackage{fancyhdr}
\usepackage{graphicx}
\usepackage{setspace}
\usepackage{amssymb}

\makeatletter

\usepackage{babel}
\makeatother
\begin{document}
\begin{titlepage} \begin{center}  \LARGE Field Theory Analysis of Laplacian Growth Models\newline \vspace*{5cm}\newline Thesis submitted for the degree of\newline  "Doctor of Philosophy"\newline by\newline Eldad Bettelheim\newline \vspace{2cm} \newline Submitted to the Senate of the Hebrew University\newline August 2004\end{center}  \newpage \LARGE This work was carried out under the supervision of \newline Prof. Oded Agam\end{titlepage} \thispagestyle{empty}  \thispagestyle{empty} \cleardoublepage  

\tableofcontents{}

\chapter*{Abstract}

We consider Laplacian growth problems using a field theory approach.
In particular we consider the Saffman-Taylor (ST) problem, in which
a non-viscous fluid is pumped into the center of a Hele-Shaw cell
filled with a highly viscous fluid generating a bubble of a distinctive
fingering fractal pattern. The idealized settings of the problem,
with vanishing surface tension between the bubble and the surrounding
fluid, is singular due to the formation of cusps after a finite time
(for generic initial conditions). A natural regularization of the
cusp, is the addition of surface tension, but this complicates the
mathematical description of the problem a great deal. 

Our first goal is to reveal the equivalence of the idealized Saffman-Taylor
problem and the semiclassical limit of the quantum Hall system in
a strong inhomogeneous magnetic field. This quantum system regularizes
the cusp by introducing a short-scale cut-off associated with a Planck
constant. Next we discuss the relation between the quantum Hall system
and an integrable system of nonlinear equations known as the two-dimensional
Toda lattice (2DTL). This relation allows us to employ methods from
soliton theory for the description of the ST problem, and in particular
it will allow us to to find a novel method for the regularization
of the cusp. 

In the language of soliton theory, the semiclassical limit, is termed
{}``the dispersionless limit'', since it may be associated with
neglecting dispersion in the nonlinear equations (the 2DTL in our
case). In the limit of small dispersion, one obtains a system of equations
known as the Whitham equations. Dropping the dispersion from the outset,
results in singular solutions for the Whitham equations. On the other
hand, one may take the limit of small dispersion correctly and obtain
well-behaved solutions. By analogy, the semiclassical limit must be
taken correctly in the quantum Hall system in order to obtain solution
which do not exhibit the singularities of the cusps. 

We study the implications of this regularization method in this thesis.
We show that it amounts to allowing new small bubbles to form in the
ST problem. These bubbles grow and eventually merge with the large
bubble. The appearance of new bubbles is a mathematical extension
of the original ST evolutions, in which fluid can {}``tunnel'' from
the large bubble into a new location, where it may form a new bubble.
During the evolution of this multi-bubble solution tunneling persists
as fluid is constantly exchanged between the two bubbles (all the
bubbles share the same pressure). 

In order to study the physical implications of this regularization
procedure we will study the evolution obtained by this method. In
particular we will study the shape of the bubble after the merging
of two bubbles, and show that for some initial conditions one obtains
tip-splitting solutions. Tip splitting is believed to be a process
important for the formation of the fractal. We discuss the scaling
form of these solutions and the possible implications for the fractal
dimensions of the ST bubble.

\chapter*{Acknowledgments}

I would like to thank my supervisor, Prof. Oded Agam, for his great
help in advising me in this work. I would also like to thank Prof.
Paul Wiegmann and Prof. Anton Zabrodin for very helpful discussions
and for cooperating with me in our research. I would also like to
thank Dr. Razvan Teodorescu for stimulating discussions. This research
was supported by the Israel Science Foundation (ISF) grant No.198/02,
and by the German Israel Foundation (GIF) grant No. I-709-58.14/2001.

\chapter{Introduction}

The goal of physics is to understand the laws of nature and how these
laws bring about the phenomena we see in the natural world. One of
the great achievements of physics is that it provides much insight
into systems in equilibrium. However the law of entropy increase,
suggests strongly that a system in equilibrium could not support very
complex phenomena, such as life. Indeed, many of the complex phenomena
which we observe in the world around us are dependent on an external
energy source, and subsequently these systems cannot be considered
to be in equilibrium. Thus to begin to understand these phenomena,
we have to gain a better understanding of systems which are out of
equilibrium. We may approach this complicated problem by considering
{}``toy models''. Namely, systems which may not display the complexity
of non-equilibrium systems in its full generality, but do capture
the prominent features of these systems. The study of these {}``toy
models'' is still a challenging task, and exact solutions of non-equilibrium
systems are few and far between. This thesis is an attempt to apply
the tool-box of theory, and in particular the application of methods
from classical integrable systems to the study of one such class of
systems, namely Laplacian growth problems. 

We will be interested in particular in Saffman-Taylor (ST) flows,
which is one example of a Laplacian growth model. The ST problem is
a two-dimensional hydrodynamical system, where two fluids occupy the
thin gap between two parallel plates (an experimental setup which
is called a Hele-Shaw cell, see Fig.~\ref{ST setup}). One fluid,
which is assumed to be of low viscosity, forms a bubble in the other
fluid which is assumed to be of high viscosity. As more fluid of low
viscosity is pumped into the bubble, this bubble develops interesting
fractal shapes (see Fig.~\ref{ST aggregate}). We will study the
case where the surface tension at the boundary between the two fluids
is zero. The zero surface-tension limit is singular in the sense that
cusp-like singularities form in the droplet for generic initial conditions,
beyond which the evolution of the droplet cannot be continued. 

In this thesis we will consider a field theory approach to Laplacian
growth. First we will make the connection between the ST problem to
the quantum Hall effect in an inhomogeneous field. We will show that
an electron droplet in the ground state, of a clean system without
interactions, evolves according to the ST dynamics in the zero surface-tension
limit. The cusp-like singularities which form in the zero-surface
tension problem, correspond to points where the semiclassical approximation
breaks down. We thus conclude that the semiclassical limit should
be taken more carefully, in order not to run into these singularities.
In order to take the semiclassical limit correctly, we shall reveal
the relation between Laplacian growth and the method of dispersive
regularization which is employed in classical integrable systems.
Dispersive regularization is an approach to finding approximate solutions
to integrable systems using the Whitham averaging method, in the limit
of low dispersion. Taking the low dispersion limit by setting the
dispersion to zero from the outset, results in singular solutions
(for example overturning waves in the KdV equation), while the dispersive
regularization technique takes a correct low dispersion limit, by
taking into account phenomena, which may occur even at vanishingly
small dispersion (e..g oscillatory KdV waves).

We will show that in order apply the dispersive regularization method
one must consider multi-bubble solutions in the ST flows. The evolution
of the multiply connected domain will be shown to be the ST growth
with equal pressure in all the bubbles. The exchange of liquid between
the bubbles, which is necessary to have equal pressure in the bubbles,
is associated with tunneling in the quantum Hall system. The naive
$\hbar\to0$ limit, neglects this tunneling, as it is a nonperturbative
effect in $\hbar,$ while the more careful limit, which is embodied
in the {}``dispersive regularization'' approach, takes into account
this effect.

The dispersive regularization technique that we will employ in this
thesis may then be viewed as an extension of the ST dynamics which
is suggested by the quantization. In order to see whether this regularization
method, of the zero surface tension problem is, physically sensible,
we will study the evolution of droplets which exhibit tip-splitting,
as this process is believed to be important to the formation of the
fractal structure of the bubble.

\chapter{Laplacian growth}

In this chapter we will give a brief review of Laplacian growth models.
In particular we will be interested in Saffman-Taylor flows in a Hele-Shaw
cell. Laplacian growth (LG) has been a widely studied system for almost
half a century \cite{Saffman:Taylor}, and the study of LG has produced
many interesting advances also in recent years\cite{43:Zabr:Wiegm,Hastings:Levitov,Feigenbaum:Procaccia,MineevWeinstein,Dvdvtch:Proccia:LG,18:Davido:Procach}.
Laplacian growth appears in many different physical settings, for
example in Hele-shaw flows\cite{ST:review}, electric deposition\cite{Sawada:Deposition},
colloidal aggregation\cite{Weitz:Colloidal}, dielectric breakdown\cite{Niemeyer:Dielectric},
dendritic crystal growth\cite{Langer:Dendritic} and diffusion limited
aggregation \cite{Witten:Sander}. LG serves as a paradigm for the
study of two-dimensional non-equilibrium systems, because of its elegant
and simple mathematical formulation, and because of the appealing
fractal geometric patterns it displays, which resemble patterns which
are ubiquitous in nature. To be more concrete, let us introduce first
Laplacian flows in Hele-Shaw flows, and then discuss briefly some
of its other Laplacian growth cousins.

\begin{figure}
\begin{center}\includegraphics[%
  width=0.30\textwidth]{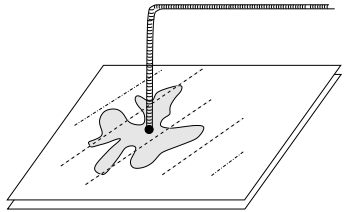}\end{center}

\caption{Schematics of a Hele-Shaw cell. Two parallel glass plates are drawn,
with a hole drilled through the upper one, which serves to inject
non-viscous fluid. The latter displaces the highly viscous fluid which
fills the gap between the plates.\label{ST setup}}
\end{figure}

As mentioned above a Hele-Shaw cell is an experimental setup where
the small distance between two solid plates is occupied by flowing
liquids or gases. In Hele-Shaw flows that we consider, known as Saffman-Taylor
(ST) evolution\cite{Saffman:Taylor}, this space is occupied by two
liquids one highly viscous and the other of low viscosity (actually
a liquid can be replaced by a flowing gas, usually assumed to be in
the incompressible regime). The liquid of lower viscosity forms a
small bubble inside the highly viscous fluid. By pumping more of the
low viscosity fluid into the bubble, the bubble increases in size
and the high viscosity fluid recedes (see Figure \ref{ST setup}).

The pressure jump across the interface between the two fluids is related
to the surface tension. A simple assumption is that the pressure jump
takes the form \cite{ST:review}:\[
\delta p=\gamma k,\]
 where $\gamma$ is the surface tension and $\kappa$ is the local
curvature of the interface. Actually this simple expression must be
modified in order to describe the physical situation in the Hele-Shaw
cell\cite{TrueSurfaceTension} , but it will not matter for us, since
we shall be interested in the limit of zero surface tension. The velocity
of the interface is given by D'Arcy's law:\begin{equation}
v=-\frac{b^{2}}{12\mu}\partial_{n}P,\label{Darcys}\end{equation}
 where $\mu$ is the fluid viscosity, $b$ is the distance between
the plates, $\partial_{n}$ is the derivative normal to the interface
and $P$ is the pressure. Since the fluid is assumed to be incompressible
one also has the condition:\begin{equation}
\Delta P=0.\label{eq:Laplace}\end{equation}
 If we neglect the pressure jump across the interface (due to surface
tension) and fix the pressure inside the less viscous fluid to be
$0$ (since the fluid is extremely non-viscous, the pressure gradients
inside the bubble are negligible) we obtain the boundary condition
for the pressure of the fluid, namely $P=0$ for the interface. Far
away from the bubble the pressure diverges logarithmically, as a drain
for the viscous fluid is present at infinity:\[
\begin{array}{ccc}
P(\vec{x}) & \longrightarrow & \log(|\vec{x}|)\\
 & ^{|\vec{x}|\to\infty}\end{array}\]
The common feature of all Laplacian growth phenomena is that the evolution
is determined by some Laplacian field (in this case the pressure $P$).
Different Laplacian growth models may have variants of D'Arcy's law
as the growth law, for example Dielectric breakdown models have the
normal velocity of the aggregate proportional to the normal derivative
of the pressure raised to some power, $\eta,$ namely $v\propto\left(\partial_{n}P\right)^{\eta}$. 

Another Laplacian growth model \cite{Witten:Sander} is Diffusion
Limited Aggregation (DLA), which may be considered as the discrete
analogue of the ST evolution \cite{Dvdvtch:Proccia:LG}. In this process,
usually realized in computer simulations, one starts with a seed that
is placed at the origin of a 2-dimensional domain. A particle is then
released far away from the origin and is allowed to perform a random
walk until it hits a cell adjacent to the origin. At this point the
particle is frozen in place - as if glued to the seed at the origin.
The next step is to release another particle which is allowed to perform
a random walk until it hits one of the frozen particles. This process
is iterated ad nauseam. The frozen particles form an aggregate (see
Figure \ref{DLA}). The relation to Laplacian growth is that the probability
for a random walker to hit the aggregate at some point may be approximated
to be the normal derivative of the Laplacian field, $\frac{\partial P}{\partial n}.$
It was shown that this approximation does not change the main geometrical
properties of the fractal in \cite{Hastings:Levitov}. 

\begin{figure}
\begin{center}\includegraphics[%
  width=0.30\textwidth]{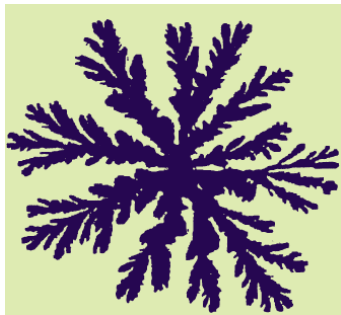}\end{center}

\caption{A typical ST aggregate.\label{ST aggregate}}
\end{figure}

The Laplacian growth models exhibit fingered fractal patterns as seen
for example in figure (\ref{DLA}) for DLA, or figure (\ref{ST setup})
for Hele-Shaw flows. The parts of the interface further away from
the origin tend to move faster than the parts closer to the center.
This instability causes the fingers (see Figure \ref{ST aggregate})
to become more and more sharp until finally they form a cusp. Of-course
in real experiments the cusp is cut-off by surface tension%
\begin{figure}
\begin{center}\includegraphics[%
  width=0.30\textwidth]{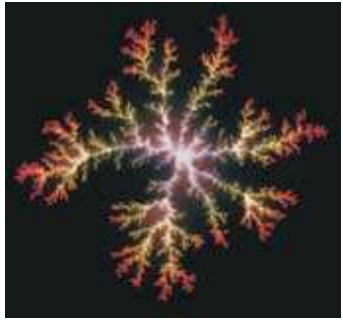}\end{center}

\caption{A typical DLA aggregate.\label{DLA}}
\end{figure}
. Our approach is to study how quantization of the ST problem, may
provide another method of regularization which keeps the integrability
of the problem intact, and thus avoids the cumbersome mathematical
complications presented by introducing surface tension \cite{144:Tanveer}
. This issue will be studied in later chapters, in the following sections
we will discuss the mathematical description of the ST bubble.

\section{Mathematical descriptions\label{sec:MathDescrp}}

In oder to analyze the evolution of, for example, the ST bubble, one
needs first a mathematical tool to describe the shape of a this two-dimensional
object. In this section we shall present two such mathematical methods:
The first one is based on conformal mappings, while the second one
employs the Schwarz function formalism.

\begin{figure}
\includegraphics[%
  width=10cm]{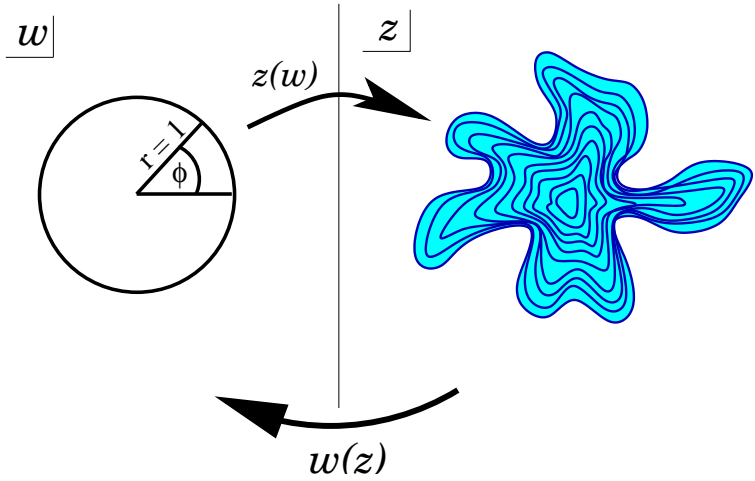}

\caption{Mapping from the exterior of the unit circle, where the complex coordinate
is denoted by $w$ to the exterior of the aggregate (where the coordinate
is $z$).\label{cap:MappingZw}}
\end{figure}

To begin with, consider the univalent conformal mapping, $z(w;t)$,
from the exterior of the unit circle, $|w|>1$, to the exterior of
the bubble at time $t$, such that the unit circle $|w|=1$ is mapped
to the interface contour, see Fig.~\ref{cap:MappingZw}. This mapping
is assumed to have the following Laurent expansion: \begin{equation}
z(w;t)=u_{1}(t)w+u_{0}(t)+u_{-1}(t)w^{-1}+u_{-2}(t)w^{-2}+...,\label{eq:conformal-mapping}\end{equation}
 where the time dependence of the expansion coefficients $u_{i}(t)$
determine the evolution of the bubble in time. The form above is derived
by demanding that the poles of the mapping reside inside the unit
circle of $w$ plane, and that it is univalent in the exterior domain%
\footnote{This is because the highest order term of the expansion is $w$ which
implies that the mapping is univalent around infinity, indeed in the
whole exterior domain.%
}.

Using the conformal mapping description it is easy to express the
solution for the Laplace equation (\ref{eq:Laplace}) as \[
P=\Re\left(\log\left(w\left(z\right)\right)\right),\]
 where $w(z)$ is the inverse mapping. Indeed, $\log(w(z))$'s real
part is a harmonic function satisfying Laplace equation. Moreover,
it also satisfies the correct boundary conditions, $P=0$ on the interface.
This is because the bubble contour is mapped to $|w(z)|=1$, consequently
$\log w(z)$ is purely imaginary, and therefore its real part trivially
vanish.

One may study the dynamics of the ST bubble by considering the evolution
of the conformal map in time, $z(w;t)$. In fact, using the conformal
mapping approach, D'Arcy's law (\ref{Darcys}) takes a very appealing
form - it can be written as the Poisson brackets of the conformal
map, $z(w,t)$ and its complex conjugate%
\footnote{The complex conjugate mapping is obtained from (\ref{eq:conformal-mapping})
by replacing the expansion coefficients $u_{i}(t)$ with their complex
conjugates $\bar{u}_{i}(t)$. Note that $\bar{w}=w^{-1}$ for $w$
on the unit circle, so that $\bar{z}(w^{-1})$ is the analytical continuation
of $\bar{z}$ away from the contour. %
} $\bar{z}(w,t)$: \begin{equation}
\{ z(w;t),\bar{z}(w^{-1};t)\}=1,\label{eq:poisson}\end{equation}
 where the canonical variables are $\log w$ and the time $t$ (which
is also area of the bubble). To see that this is indeed a manifestation
of D'Arcy's law, (\ref{Darcys}), let us choose units where $b^{2}/12\mu$
is unity, and write it in the form \begin{equation}
v_{n}=\left(\frac{\partial l}{\partial\theta}\right)^{-1}\left|\frac{\partial\vec{r}}{\partial t}\times\frac{\partial\vec{r}}{\partial\theta}\right|=-\frac{\partial P}{\partial n}\end{equation}
 where $\vec{r}=(x,y)$ denote a point along the bubble contour, $\theta$
is the angle parameterizing the unit circle $\Im\log w=\theta$ (see
Fig. ~\ref{cap:MappingZw}), and $l$ is the arc length along the
perimeter of the bubble. \texttt{$\frac{\partial\vec{r}}{\partial t}$}
is the velocity of a point on the contour. To obtain the normal velocity
we take the cross product with the tangent $\frac{\partial\vec{r}}{\partial\theta}$,
and then divide by the length of the tangent $\frac{\partial l}{\partial\theta}.$
Now using the Cauchy Riemann equations we get $\frac{\partial\theta}{\partial l}=\frac{\partial P}{\partial n}$,
which implies:\[
\left|\frac{\partial\vec{r}}{\partial t}\times\frac{\partial\vec{r}}{\partial\theta}\right|=-1.\]
Writing the cross product in complex coordinates, and analytically
continuing the equation away from the contour, we get (\ref{eq:poisson})
(where we also use the fact that the analytic continuation of $\bar{z}$
outside the perimeter of the bubble is $\bar{z}(w^{-1})$)

The second useful mathematical device we shall use in this work is
the Schwarz function. To define the Schwarz function consider, first,
an equation for the bubble interface (in the x-y plane) given by:\[
f(x,y)=0.\]
 In terms of the complex coordinates, $z=x+iy$, $\bar{z}=x-iy$,
we may rewrite this equation in the form \[
g(z,\bar{z})=f(\frac{z+\bar{z}}{2},\frac{z-\bar{z}}{2i})=0.\]
 Then solving the equation $g(z,\bar{z})=0$ for $\bar{z}$ we obtain
\[
\bar{z}=S(z).\]
 $S(z)$ when analytically continued away from the perimeter of the
bubble is known as the Schwarz function \cite{Schwartz,Vasconcelos,EJAM:Howison2,186:Howison}.
Thus $S(z)$ is an analytic function which equals $\bar{z}$ on the
bubble contour.

An important feature of the Schwarz function is its unitarity \begin{equation}
z=\bar{S}(S(z)),\label{unitarity}\end{equation}
 which follows from the fact that on the contour this condition is
trivially satisfied%
\footnote{$\bar{S}(S(z))=\bar{S}(\bar{z})=\overline{S(z)}=\bar{\bar{z}}=z$%
}, and since both sides of the equation represent analytic functions,
the equality must hold all over the complex plane.

The relation between the Schwarz function and the conformal mapping
description of the contour is \[
S(z)=\bar{z}\left(\frac{1}{w(z)}\right)\]
 where $\bar{z}(w)$ is the complex conjugate mapping to $z(w)$ and
$w(z)$ is the inverse mapping. This relation is derived by noticing
that on the contour $w=e^{i\theta},$ $1/w=\bar{w}$, and $S(z)=\bar{z}(\bar{w})=\overline{z(w)}$.

\section{\noindent Constants of motion}

The ST dynamics in the absence of surface tension possesses an infinite
number of conserved quantities known as the harmonic moments of the
system\cite{Richardson1972}. These harmonic moments are defined as
\begin{equation}
t_{k}=-\frac{1}{\pi k}\int_{\mbox{visc.fluid}}\!\!\!\!\!\!\!\!\!\!\!\! d^{2}zz^{-k},k=1,2,\cdots,\label{harmonic}\end{equation}
 where integration is over the viscous fluid (exterior) domain. 

We give the proof that $\frac{dt_{k}}{dt}=0$, where $t$ is the time,
which according to our convention will be equal to the area of the
bubble. To show this, let us introduce the complex potential, $\Psi$,
which is an analytic function such that $\Re(\Psi)=P$ is the pressure.
Its imaginary part, $\theta=\Im(\Psi)$, known as the stream function
goes to the polar angle as the distance from the origin goes to infinity
since $P\sim\log(|z|)$. Thus by the Cauchy-Riemann equations $\frac{\partial P}{\partial n}=\frac{\partial\theta}{\partial l}$
where $n$ is the direction normal to the contour and $l$ is the
direction along the contour. Therefore\[
\frac{d}{dt}t_{k}=-\frac{1}{\pi k}\frac{d}{dt}\int z^{-k}d^{2}z=-\frac{1}{\pi k}\oint z^{-k}\frac{\partial P}{\partial n}dl=\]
 \[
=-\frac{1}{\pi k}\oint z^{-k}\frac{\partial\theta}{\partial l}dl=\frac{i}{\pi k}\oint z^{-k}d\Psi,\]
 where in the last equality we have used the fact that $\Psi$ is
purely imaginary on the contour and thus $d\theta=-id\Psi$. Now by
deforming the contour of integration such that it is a large contour
around infinity, we see that the integrand vanishes quickly enough
to render the integral $0.$ Finally, we remark that the above proof
holds also for the case where there are more than one bubble of water.

\chapter{The Quantum Hall Effect}

In the idealized setting, where no surface tension is present, the
Saffman-Taylor problem confronts an obstacle. As a result of the scale
invariance, some fingers develop cusp-like singularities within a
finite time \cite{Shraiman:Bensimon}. A modification of the growth
law which introduces a mechanism curbing the curvature of the interface
at a micro scale is necessary. The form (\ref{eq:poisson}) of the
evolution equations, as Poisson brackets suggests to use a quantization
procedure, whereby a Planck scale may be introduced which will serve
as the small-scale cut-off for the theory. In this chapter we show
\cite{100:Bettelheim:Agam:Wieg} how the quantum Hall system in an
inhomogeneous magnetic field serves as a quantization of the ST problem.
In later chapters we will discuss the integrability of the quantum
system, and the way it may be employed to find solutions of the ST
problem, regularized by quantization. 

We will study a shape of a large electronic droplet on the fully occupied
lowest Landau level of a quantizing magnetic field. The magnetic field
is assumed to be nonuniform in the area away from the droplet. We
show that Aharonov-Bohm forces, associated with the nonuniform part
of the magnetic field shape the edge of the droplet in a manner similar
to a fingering interface driven by a Laplacian field.

In order to present our argument we shall neglect the interactions
among the electrons and assume that the external electrostatic potential
is zero. Under these conditions we will show that the semiclassical
dynamics of the QH droplet is governed by the same equations of viscous
fingering scaled to a nanometer scale. By the semiclassical limit
we mean a large number of electrons $N\to\infty$, small magnetic
length $\ell\equiv\sqrt{\hbar c/eB_{0}}\rightarrow0$ but a finite
area of the droplet ($\ell^{2}N\to const$). The droplets' area is
$\ell^{2}N$.

Let us first recall the physics of QH-droplets (see e.g., \cite{76:Cappelli:Wsymmetry}\cite{Laughlin}).
Consider spin polarized electrons on a plane in the lowest level of
a quantizing nonuniform magnetic field, directed perpendicular to
the plane, $B(x,y)>0$: \[
H=\frac{1}{2m}\left((-i\hbar\vec{\nabla}-\vec{A})^{2}-\hbar B\right).\]
 The lowest level of the Pauli Hamiltonian is degenerate even for
a nonuniform field. Aharonov and Casher\cite{Aharonov:Casher} showed
that the degeneracy equals the integer part of the total magnetic
flux $\Phi=\int dxdyB$ in units of flux quanta, $\Phi_{0}=2\pi\hbar$
(we set $e=c=1$)\cite{Aharonov:Casher,Avron}. To see this define
$A=A_{x}+iA_{y}$ and $\Pi=-2\hbar\partial_{\bar{z}}-i\bar{A}$. Then
$H$ can be written as $H=\frac{1}{2m}\Pi^{\dagger}\Pi,$ assuming
$\vec{\nabla}\cdot\vec{A}=0$. A zero energy solution $|\Psi\rangle$
would have \[
\left\langle \Psi\right|H\left|\Psi\right\rangle =\left\langle \Psi\right|\Pi^{\dagger}\Pi\left|\Psi\right\rangle =0\Rightarrow\Pi\left|\Psi\right\rangle =0\]
Thus, we may find the zero energy solutions, by solving the equation
$\Pi\Psi=0.$ Since $H$ is a positive operator, these solutions would
then belong to the ground state.

The equation $\Pi\Psi=0$ is first order and may be easily solved.
A solution, $\tilde{\psi}_{m}(z),$ is given by:\begin{equation}
\tilde{\psi}_{m}(z)=e^{-\frac{1}{2\hbar}V(\bar{z},z)}\tilde{P}_{m}(z),\label{eq:BA Hall}\end{equation}
 where $\tilde{P}_{n}(z)$ is an arbitrary polynomials of a degree
$n$. $V(\bar{z},z)$ is defined by the solution to the equation $B=\frac{1}{2}\nabla^{2}V$,
given by\[
V(z,\bar{z})=\frac{1}{2\pi}\int\log(|z-z'|^{2})B(z')d^{2}z'.\]
 We may now choose a gauge by taking $A=i\partial_{z}V$.

In order for this formal solution to be an eigenvalue, it must be
normalizable. As $|z|\longrightarrow\infty$, $P_{m}(z)\sim z^{m}$
and $V(z,\bar{z})\sim\Phi log(|z|),$ which gives $|\tilde{\psi}_{m}(z)|\sim|z|^{\left(m-\frac{\Phi}{\Phi_{0}}\right)},$
thus for normalizability we must have $m\leq\frac{\Phi}{\Phi_{0}}-1$.
Thus the degeneracy of the level is $\frac{\Phi}{\Phi_{0}},$ a result
obtained in \cite{Aharonov:Casher,Avron}. For later use let us define
the orthogonal polynomials with respect to the measure $e^{-\frac{1}{\hbar}V(z,\bar{z})}$:\[
\int d^{2}xP_{n}(z)P_{m}(\bar{z})e^{-\frac{1}{\hbar}V(z,\bar{z})}=e^{\phi_{n}}\delta_{nm},\]
the orthogonal polynomials are assumed to be normalized such that
their leading coefficient is 1, $P_{m}(z)=z^{m}+...$

We will consider the following arrangement: A strong uniform magnetic
field $B_{0}>0$ is situated in a large disk of radius $R_{0}$; The
disk is surrounded by a large annulus $R_{0}<|z|<R_{1}$ with a magnetic
field $B_{1}<0$ directed opposite to $B_{0}$, such that the total
magnetic flux $\Phi$ of the disk, $|z|<R_{1}$, is $N\Phi_{0}$.
The magnetic field outside the disk $|z|<R_{1}$ vanishes. The disk
is connected through a tunneling barrier to a large capacitor that
maintains a small uniform positive chemical potential slightly above
the zero energy of the lowest Landau level.

In this arrangement a circular droplet of $N$ electrons is trapped
at the center of the disk $|z|<R_{0}$. We choose the magnetic field
$B_{1}$ such that the droplet's size, $\ell\sqrt{2N}$, is much smaller
than the radius of the disk $R_{0}$.

Next we assume that a weakly nonuniform magnetic field $\delta B$
is placed inside the disk $|z|<R_{0}$ but well away from the droplet.
The nonuniform magnetic field does not change the total flux $\int\delta Bdxdy=0$.
The droplet grows when $B_{1}$ is adiabatically increased, keeping
$B_{0}$, $\delta B$ and the chemical potential fixed. Then the degeneracy
of the Landau level and, consequently, the size of the droplet increase.

For later reference it will be useful to write the potential $V(z,\bar{z})$
as:\begin{equation}
V(z,\bar{z})=\frac{|z|^{2}}{2\ell^{2}}-W(z)-\overline{W(z)}.\label{eq:V(z,zb)prequel}\end{equation}
The first term is associated with the uniform magnetic field of magnetic
length $\ell,$ the second term is the associated with the non-uniform
part of the total field, namely $\delta B$. Near the origin the field
is uniform and thus $W$ is harmonic, and is given by:

\[
W(z)=\sum_{k\geq1}t_{k}z^{k},\;\;\;\;\;\;\;\, t_{k}=\frac{1}{2\pi k}\int\delta B(z)z^{-k}d^{2}z.\]
 The parameters $t_{k}$ are, now, the harmonic moments of the deformed
part of the magnetic field. Summing up, we have, around the origin,
\begin{equation}
V(z,\bar{z})=\frac{|z|^{2}}{2\ell^{2}}-\sum_{k\geq1}\left(t_{k}z^{k}+\bar{t}_{k}\bar{z}^{k}\right).\label{eq:V(z,zb)}\end{equation}
From now on we will choose units in which the magnetic length is $\ell=\frac{1}{\sqrt{2}}.$

\section{Coulomb gas\label{sec:2DTL}}

We will now show that the electrons occupy a region in space where
the density is uniform - the electron droplet. Furthermore we will
show that the droplet evolves according to the ST dynamics. The multi-particle
wave function is given by a Slater determinant of wave functions of
the form (\ref{eq:BA Hall}). The Slater determinant gives just a
Vandermonde determinant, so that the normalization of the multi-particle
wave function may be written as:\[
\tau_{N}=\frac{1}{N!}\int\prod_{i=1}^{N}d^{2}z_{i}\left|\det_{i,j}\left(P_{i}(z_{j})\right)\right|^{2}e^{-\frac{1}{\hbar}\sum_{i=1}^{n}V(z_{i},\bar{z}_{i})}=\]
\begin{equation}
=\frac{1}{N!}\int\prod_{i=1}^{N}d^{2}z_{i}\prod_{i>j}|z_{i}-z_{j}|^{2}e^{-\frac{1}{\hbar}\sum_{i=1}^{n}V(z_{i},\bar{z}_{i})},\label{Dyson-gas}\end{equation}
where $\frac{1}{N!}$ is added for later convenience. We shall call
this normalization factor {}``the $\tau$- function''. Let us now
define the function \begin{equation}
\psi_{m}(z)=P_{m}(z)e^{\frac{1}{\hbar}\left(-\frac{|z|^{2}}{2}+t_{k}z^{k}\right)}.\label{eq:BA exact}\end{equation}
Then by the orthogonality of the polynomials, the $\tau-$function
factorizes to: \[
\tau_{N}=\Pi_{m=0}^{N}\int|\psi_{m}(z)|^{2}d^{2}z=e^{\sum_{m=0}^{N}\phi_{m}}.\]
The factorization of the $\tau-$function is related to the fact that
we are dealing with free fermions, i.e. the multi-particle wave function
can be written as a Slater determinant of single-particle wave functions. 

In understanding the relation between the quantum Hall setup described
above and the ST problem it is instructive to adopt the view of Dyson's
gas \cite{Mehta} for the $\tau$-function. According to this picture,
$z_{i}=x_{i}+iy_{i}$ denote the complex coordinate of the $i$-th
particle, the Vandermonde determinant $\prod_{i>j}|z_{i}-z_{j}|^{2}$,
when exponentiated, accounts for the logarithmic interaction between
the particles, and $V(z,\bar{z})$ is the potential energy.

To find the density of Dyson's gas, we write down the action of the
Coulomb gas in terms of the density of the particles $\rho(z)=\sum_{i}\delta(z-z_{i})$:\[
S\left[\rho(z)\right]=-\int d^{2}z\left(\frac{1}{\hbar}V(z,\bar{z})\rho(z)-\int d^{2}z'\rho(z')\log(|z-z'|)\rho(z)\right),\]
 such that by varying this action with respect to $\rho(z)$, \begin{equation}
\frac{1}{\hbar}V(z,\bar{z})-\int\rho(z')\log(|z-z'|^{2})=0,\label{eq:screening}\end{equation}
 and operating with the Laplacian on this equation we obtain:\[
\rho(z)=\frac{1}{\pi\hbar}\equiv\rho_{0}.\]
 The last equation implies that the density of the eigenvalues is
constant. However, the functional derivative makes sense only when
$\rho>0$ since the density cannot be negative. Therefore, we expect
that $\rho(z)=\rho_{0}$ only in the interior of some domain, $D_{+}$,
and that it vanishes at the exterior of this domain, $D_{-}$.

Our purpose, now, is to show that $D_{+}$ and $D_{-}$ are indeed
the interior and exterior domains of the ST bubble. For this purpose
we shall show that the harmonic moments of $D_{-}$ are the $t_{k}'s$
which enter in the potential. To show this, we will first show that
$V(z,\bar{z})$ is the potential generated by a positive charge distributed
uniformly in a region of harmonic moments $\{ t_{k},\bar{t}_{k}\}$.
Screening of this potential by Dyson's gas (see equation (\ref{eq:screening})),
which is assumed to be negatively charged, implies that Dyson's gas
is also distributed uniformly in a region of harmonic moments $\{ t_{k},\bar{t}_{k}\}$.

Consider the potential generated by a positive charge, of density
$1/\pi\hbar$, distributed uniformly in $D_{+}$: \[
\tilde{V}(z,\bar{z})=\frac{1}{\pi\hbar}\int_{D_{+}}d^{2}z'\log\left|z-z'\right|^{2}\]
 Viewing this potential as the sum of a positive charge distributed
uniformly in the whole plain and negative charge, of the same density,
in the exterior domain, we may rewrite this potential as \[
\tilde{V}(z,\bar{z})=\frac{1}{\hbar}|z|^{2}-\frac{1}{\pi\hbar}\int_{D_{-}}d^{2}z'\log\left|z-z'\right|^{2},\]
 where the first term account for the uniform charge density in the
whole plain, while the second term accounts for the opposite charge
density in the exterior domain. Notice that $z$ is located in the
interior of the domain.

Now we expand the logarithm in powers of $\frac{z}{z'}$ ( $z$ is
in the interior domain while $z'$ is in the exterior domain, and
we expand in the small parameter $\frac{z}{z'}$). Then up to an additive
constant, %
\footnote{Formally, this constant is infinity. However, we may consider the
negative charge in the exterior domain to extend up to a large finite
radius. This will not change our argument but will keep this constant
finite.%
} \[
\tilde{V}(z,\bar{z})=\frac{1}{\hbar}|z|^{2}+2\Re\frac{1}{\hbar}\int_{D_{-}}d^{2}z'\sum_{k=1}^{\infty}\frac{1}{\pi k}\left(\frac{z}{z'}\right)^{k},\]
 and from the definition of the harmonic moments (\ref{harmonic})
we arrive at (\ref{eq:V(z,zb)}). Thus Dyson's gas occupy a domain
of area $\pi\hbar N$ in the complex plane and whose harmonic moments
are the set $\{ t_{k},\bar{t}_{k}\}$. Thus the evolution of this
domain, while increasing its area is precisely the evolution of ST
bubble in the absence of surface tension.

It is interesting to compare dimensionless viscosity of liquids used
in viscous fingering experiments \cite{moore}\cite{zhao}, and a
\char`\"{}viscous\char`\"{} effect of quantum interference of at the
first Landau level. The parameter of the dimension of length controlling
viscous fingers in fluids is $\frac{2\pi}{q}\frac{b^{2}}{12\eta}\sigma$,
where $q$ is the flow rate, $b$ is the thickness of the cell, $\eta$
is the viscosity and $\sigma$ is the surface tension. In recent experiments\cite{moore}\cite{zhao}
, this length stays in high hundred of nanometers, but can be decreased
by increasing the flow rate. This length is to be compared with the
magnetic length $\ell$. At magnetic field about 2T it is about 50
nm. Semiconductor devices imitating a channel geometry of the original
Saffman-Taylor experiment \cite{Saffman:Taylor} may facilitate fingering
instability.

\section{Relation to the 2d Toda lattice equations.}

An interesting feature of the quantum Hall problem described above
is that the $\tau-$function , $\tau_{n}=\prod_{m=0}^{n}e^{\phi_{m}},$
satisfies the integrable nonlinear equation \cite{43:Zabr:Wiegm,42:Zabr:Wiegm,44:Wiegmann:Zabrodin,45:Mineev:Zabrodin,MineevWeinstein,173:Kazakov:Marshakov},
associated with the 2d Toda lattice (2DTL).:\begin{equation}
\frac{\partial^{2}}{\partial t_{1}\partial\bar{t}_{1}}\log\left(\frac{\tau_{n}}{\tau_{n-1}}\right)=\frac{\tau_{n-1}^{2}}{\tau_{n-2}\tau_{n}}-\frac{\tau_{n}^{2}}{\tau_{n-1}\tau_{n+1}},\label{eq:1st2DTL}\end{equation}
 where $t_{1}$ is the time (here assumed to be a complex number)
and $n$ is an integer which can be regarded as a discretized coordinate
of space (see appendix \ref{cha:Proof-of-2DTL} for a proof). This
equation is a natural extension of the one dimensional Toda lattice
where $t_{1}$ is real. The full hierarchy of nonlinear equations
can be derived from the Lax equations which are derived in appendix
\ref{apndx:Sting2DTL}. In the next chapter we will discuss how the
quantized system may be treated as a regularization of the ST system.
We will use methods from the study of classical integrable systems
to describe this regularization.

\chapter{Dispersive regularization\label{sec:Dispersive-Regularization}}

In the previous chapter we established the relation between the 2DTL
and the ST problem. The purpose of this chapter is to study the implementation
of a method known as dispersive regularization \cite{Gurevich:Pitavsk,Whitham:1966:NDW,89:Kodama:Bloch}.
We shall borrow methods from soliton theory in order to find solutions
to the 2DTL and find their classical analogue which describe the ST
bubble. As was explained, the need for regularization in the ST problem
arises from the appearance of cusps in the shape of the bubble, in
finite time, for generic initial conditions. Surface tension, in this
respect, is the most natural candidate, since no matter how small
it is, as long as it is nonzero, it hinders the formation of a cusp
. Indeed, the real system avoids cusps by repeated processes of tip
splittings. Yet, the introduction of surface tension destroys the
integrable structure of the idealized problem, and any analytic treatment
becomes a complicated task\cite{144:Tanveer}.

\begin{figure}
\includegraphics[%
  width=12cm]{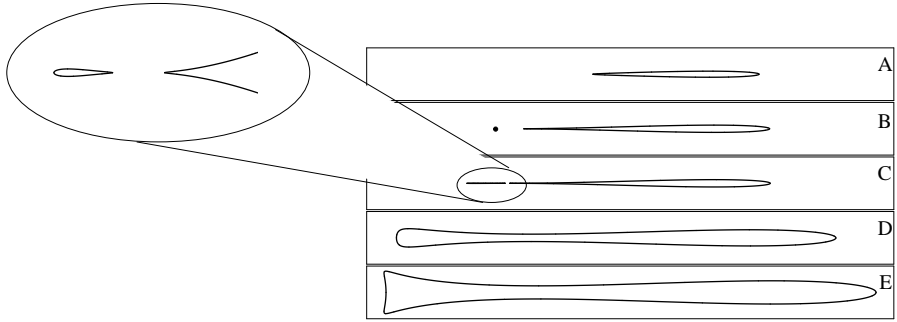}

\caption{Evolution of the droplet regularized by the addition of a droplet.
Panel A shows a single bubble. Panels B and C show an additional bubble
forming near the tip of the first droplet. In panel D the two bubbles
have merged to form a single bubble again.\label{cap:Evolution-of-the}}
\end{figure}

We suggest using dispersive regularization as an alternative method
by which the dynamics of the idealized problem can be continued beyond
the cusp, for some large set of initial conditions. This is the set
which can be associated with the cases where the cusp forms due to
the merging of the initial bubble with another small bubble, as illustrated
in Fig.~\ref{cap:Evolution-of-the}. In other words, dispersive regularization
is an extension of the ST dynamics in which more than one bubble may
exist, for small time intervals, and the evolution passes through
repeated events of formation of new bubbles which merge with the original
one.

\begin{figure}
\includegraphics[%
  width=15cm]{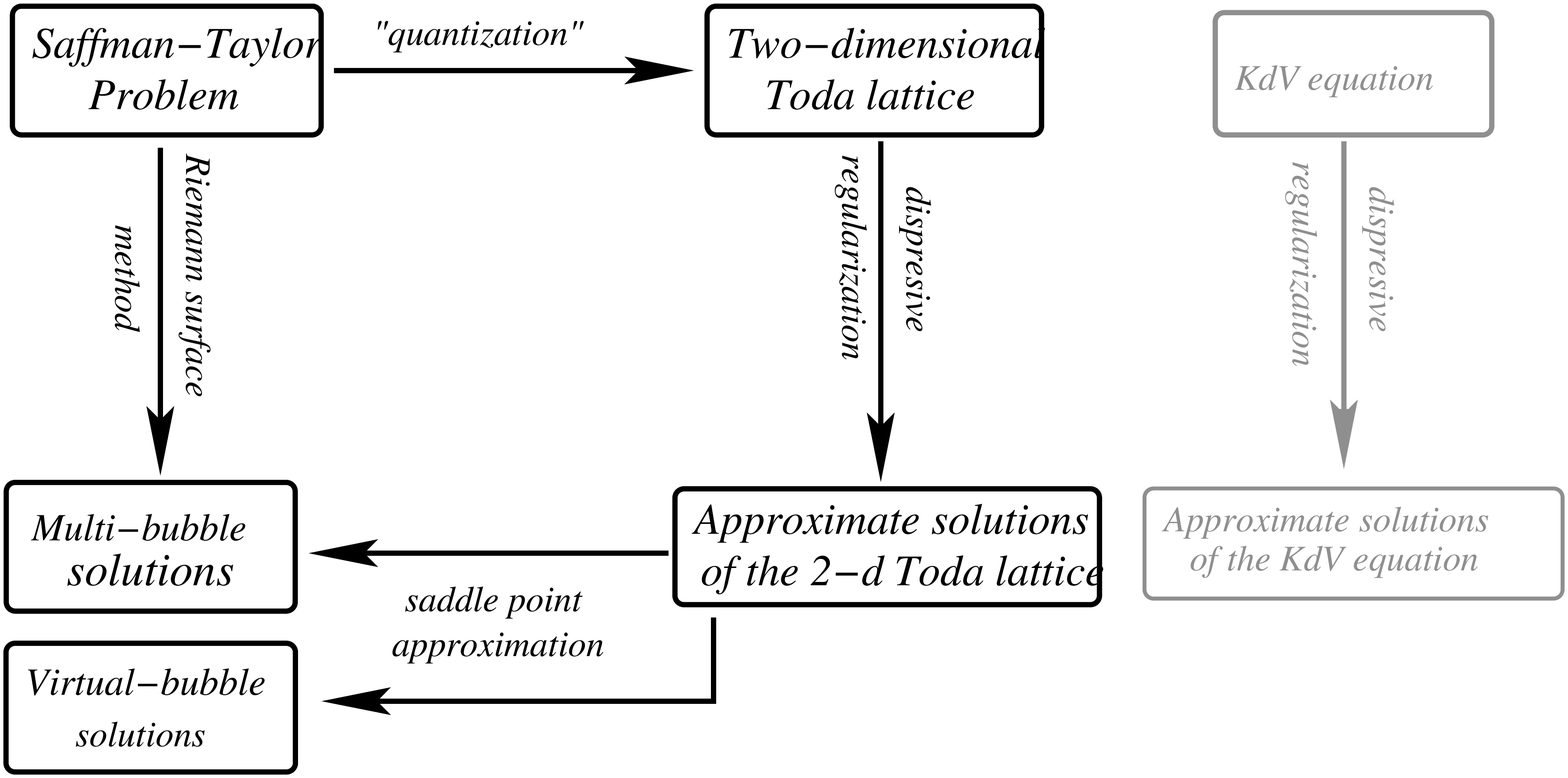}

\caption{Flow chart for organization of chapters \ref{sec:Dispersive-Regularization}-\ref{sec:Multi-Droplet}.
We find solutions of the 2DTL by the dispersive regularization approach
which leads to the introduction of multi-bubbles solutions for the
ST dynamics. We also discuss the role of virtual bubbles in the solutions
thus found. We introduce some of the methods which are used in this
article in the settings of the KdV equations and use the analogy to
the 2DTL to gain insight to the ST problem. \label{cap:Flow-chart}}
\end{figure}

In the general context dispersive regularization is a method for finding
approximate solutions for nonlinear wave equations, such as the KdV
equation (where it is also known as the Gurevich-Pitaevskii method
\cite{Soliton:Book,Gurevich:Pitavsk}). The application of dispersive
regularization to the ST problem follows from the observation that
the equations governing the latter problem are the saddle point equations
of the two dimensional Toda lattice (2DTL). Putting it differently,
''quantization'' of the ST problem leads to the 2DTL. Thus, using
dispersive regularization one can construct approximate solutions
for the 2DTL, and taking the saddle point approximation (equivalent
to the quasi-classical limit) yields general solutions for the ST
problem, see the flow diagram in Fig. \ref{cap:Flow-chart}. As we
will show some of the solutions obtained via dispersive regularization
describe situations where the system may consists of more than one
bubble, as illustrated in Fig.~(\ref{cap:Evolution-of-the}). These
turn out to be exact solutions of the multi-bubble generalization
of the ST problem.

In this chapter we will first discuss the simpler case of the KdV
equation. Then we show how to generalized the method for the 2DTL.
In Chapter \ref{sec:Richardson}, following Richardson, we develop
an alternative, equivalent, method for describing the evolution of
ST bubbles. As we will show the solutions of the equations may not
comply with the ST dynamics - a situation which refer to as {}``virtual
bubble'' solutions, see Fig.~\ref{cap:Flow-chart}. We will discuss
these virtual bubbles and explain their behavior by employing a physical
viewpoint based on the description of the ST problem as a noninteracting
fermion system. In chapter \ref{sec:Multi-Droplet} we show how to
construct general two-bubble solutions for the ST problem, and describe
their evolution. We hall also explain how these solutions resolve
the problem of cusp-like singularities.%

\section{Dispersive regularization for the KdV Equation \label{sub:KdV}}

Dispersive regularization is a general method for calculating approximate
solutions for nonlinear wave equations, for a large set of initial
conditions. Conceptually, it consists of two stages: In the first
stage a set of exact solutions for the nonlinear wave equation is
constructed. These solutions depends on parameters, $\{\lambda_{i}\}$,
which are, in fact, constants of motion of the problem. In the second
stage, this set of solutions is used as an approximate local description
of more general solutions, by allowing the parameters $\{\lambda_{i}\}$
to have slow time and space dependences. Thus the aim of this section
is to construct solutions for the 2DTL using the dispersive regularization
approach. It will be instructive, however, to overview first the milestones
of this approach for the simpler case of the KdV equation\cite{88:Flaschka:KdV:Avging}.
We shall begin this section by describing the analogy between the
KdV system and the ST problem. then we shall consider the inverse
scattering approach, which will allow us to construct modulated oscillatory
solutions for the KdV equation, the equations for the modulated waves,
known as the Whitham equations will be discussed next.

The KdV (Korteweg de-Vries) equation, \begin{equation}
u_{t}-6uu_{x}-\beta u_{xxx}=0,\label{eq:KdV}\end{equation}
 is a nonlinear one-dimensional wave equation devised in order to
model water waves is a shallow canal. Although very different from
the ST problem considered in this thesis, the dynamics of the KdV
equation is, in some sense, analogous to our problem. To clarify this
analogy consider the evolution of $u(x,t)$ in the absence of dispersion
$(\beta=0)$, \[
u_{t}-6uu_{x}=0.\]
 with general initial conditions: \[
u(x,t=0)=f(x).\]
 The Riemann solution for this problem is given by the implicit form:
\begin{equation}
u(x,t)=f(x+6u(x,t)\cdot t).\label{eq:RiemannSolution}\end{equation}

\begin{figure}
\includegraphics[%
  width=10cm]{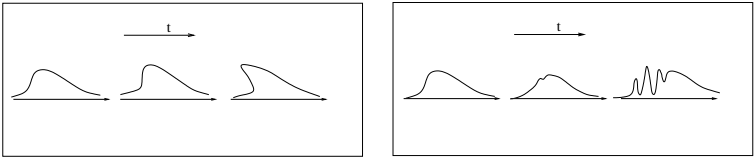}

\caption{left panel shows the evolution of the dispersionless KdV equations,
while the right panel shows the effect of dispersion which replaces
the overhang with an oscillatory region.\label{cap:Riemann solution}}
\end{figure}

However, this solution for a general function $f(x)$ becomes non-physical
beyond some finite time: It develops overhangs as illustrated in Fig.~(\ref{cap:Riemann solution}).
The dispersive term, however small it is, hinders such situations
since as the system approaches a overhang the dispersive term $u_{xxx}$
diverges. Indeed, instead of overhangs the system develops oscillations
as demonstrated in Fig.~(\ref{cap:Riemann solution}).

\begin{figure}
\includegraphics[%
  width=6cm]{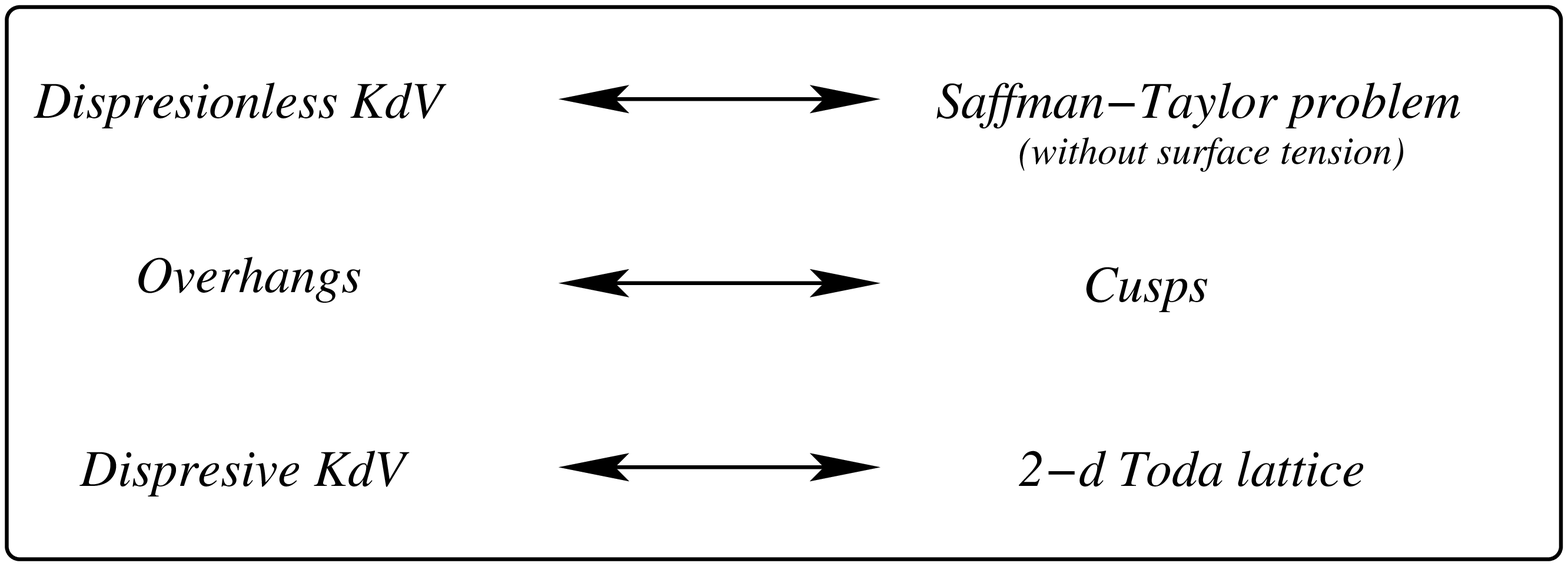}

\caption{Table showing analogy between the KdV equations and the ST dynamics.
The dispersionless KdV which is analogous to the ST problem develops
overhangs in much the same way as the ST dynamics develops cusps.
The resolution of this problem may be achieved in the KdV equations
by the introduction of dispersion, which is analogous to finding dispersively
regularized solutions of the 2DTL. \label{cap:analogy-table}}
\end{figure}

The analogy between the KdV equation and ST problem is summarized
in Fig.~(\ref{cap:analogy-table}). The dispersionless KdV equation
is analogous to the ST problem in the absence of surface tension;
Overhangs are analogous to cusps; And the dispersive KdV equation
is analogous to the 2DTL.

We turn now to describe how dispersive regularization resolves the
problem of overhangs. The first stage is the construction of a set
of exact solutions for the problem. This construction, is based on
the remarkable fact that if a solution $u(x,t)$ of the KdV equation
is considered as a time-dependent potential in a Schr\"{o}edinger
operator,\[
{\mathcal{H}}=-\frac{d^{2}}{dx^{2}}+u(x,t),\]
 then the resulting spectrum of the equation \[
{\mathcal{H}}\psi_{\lambda}=\lambda\psi_{\lambda}\]
 is independent of time, $(d\lambda/dt=0)$.

Thus one may assume a given form of the eigenvalue-spectrum, as well
as some analytic properties of the wave function (the {}``scattering
data''), and try to construct the corresponding potential $u(x,t)$
(i.e. the {}``scattering potential''). This reconstruction of the
potential from the scattering data is known as the inverse scattering
approach.

The idea is first to construct eigenfunction $\psi_{\lambda}$, as
a the Baker-Akhiezer (BA) function, from its analytic properties as
function of $\lambda$. Then the potential (which is the solution
of KdV equation) is straightforwardly extracted by substitution into
the Schr\"{o}edinger equation: \begin{equation}
u(x,t)=\frac{1}{\psi_{\lambda}}\frac{d^{2}\psi_{\lambda}}{dx^{2}}+\lambda\label{eq:uFromPsi}\end{equation}

\subsection*{Inverse Scattering }

To demonstrate this inverse scattering approach, let us consider,
first, the simplest situation where $u(x,t)=u_{0}$ is a constant
depending neither on time nor a space (this is clearly a trivial solution
of the KdV equation). The spectrum in this case is continuous for
$\lambda>u_{0}$, and the corresponding eigenfunction takes the form
$\psi_{\lambda}=e^{\pm i\sqrt{\lambda-u_{0}}x}$. Thus for general
complex parameter, $\lambda$, there are two solutions $\psi_{\lambda}^{(+)}$
and $\psi_{\lambda}^{(-)}$ corresponding to the two values of the
square root function. These solutions have branch cuts which may be
taken to coincide precisely with the spectrum.

\begin{figure}
\includegraphics[%
  height=3cm]{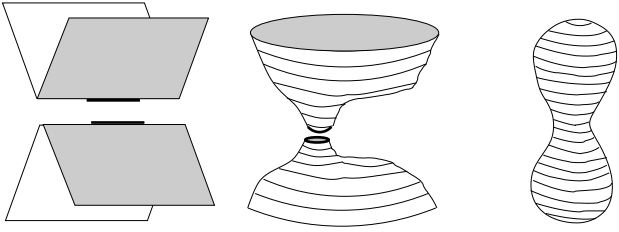}

\caption{Construction of a genus-0 Riemann surface: On the left the two sheets
of the Riemann surface are presented with branch cut in heavy line.
In the middle figure, the two sheets are deformed and the branch cuts
are opened up. The drawing on the right shows the surface which is
obtained by pasting the two surfaces along the branch cuts and after
compactification have been performed (by incorporating the infinities
in the Riemann surface. \label{cap:genus0construct}}
\end{figure}

Instead of considering the eigenfunctions, $\psi_{\lambda}$ to be
a multi-valued function of $\lambda$, we may define a Riemann surface
over which the BA function is single valued: We may take two copies
of the complex plane with branch cuts along the ray $\lambda>u_{0}$
and assign the value $e^{i\sqrt{\lambda-u_{0}}x}$ , on one copy and
$e^{-i\sqrt{\lambda-u_{0}}x}$ on the other. We can now cut the two
copies of the complex plane along the branch cut and glue the upper
part of the branch cut of one sheet to the lower part of the branch
cut on the other sheet (and vice versa) so that now $\psi_{\lambda}$
is a smooth function on the Riemann surface obtained by this cut and
paste procedure (see Fig. \ref{cap:genus0construct}). One can also
assign proper coordinate systems around each point so that the surface
has the structure of a Riemann surface. This is done by choosing a
coordinate system $\lambda^{-\frac{1}{2}}$ around the infinities,
$\sqrt{\lambda-u_{0}}$ around the branch cuts, and $\lambda-\lambda_{0}$
near a general point $\lambda_{0}$. The Riemann surface obtained
by this method is the algebraic Riemann surface associated with the
equation $y^{2}=\lambda-u_{0}$. We shall generally refer to the Riemann
surface associated with the Hamiltonian ${\mathcal{H}}$ as the spectral
surface.

\begin{figure}
\includegraphics[%
  height=3cm]{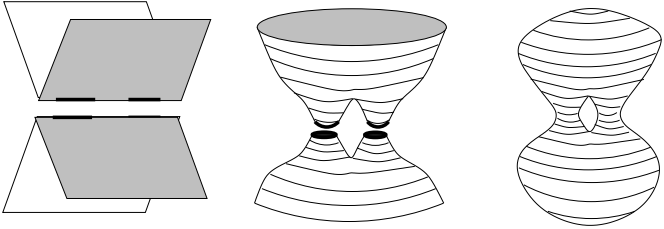}

\caption{Schematic drawing of construction of genus-1 Riemann surface, from
the curve $y^{2}=\Pi_{i=1}^{4}(z-\lambda_{i})$.The construction is
analogous to that demonstrated in Fig~\ref{cap:genus0construct}.
\label{cap:EllipticConstruct}}
\end{figure}

Consider now a more general situation where the eigenvalue spectrum
of ${\mathcal{H}}$ consists of disconnected pieces, or in other words,
the spectrum has a finite number of gaps. Our goal, now, is to calculate
the corresponding BA function from which we can deduce the potential
$u(x,t)$, according to equation (\ref{eq:uFromPsi}). Our strategy
is to construct, first, the corresponding Riemann surface over which
the BA function is single valued, and then to deduce its form from
the analytic properties on this surface, together with a choice of
normalization.

In understanding the structure of the spectral surface for finite
gap solutions, it is important to notice that it is always composed
of two Riemann sheets. This follows from the property that the Schr\"{o}edinger
operator ${\mathcal{H}}$ is of second order. If we denote by $\lambda_{i}$
the endpoints of the pieces of the spectrum, then the spectral surface
has the form: $y^{2}=\Pi_{i=1}^{2m+1}(\lambda-\lambda_{i})$, where
$m$ is the number of gaps. Such a Riemann surface is called hyper-elliptic
see Fig.~\ref{cap:EllipticConstruct}.

The BA function may now be constructed by demanding proper analytic
behavior on this hyper-elliptic Riemann surface, and that in the large
$\lambda$ limit it takes the asymptotic form \[
e^{\pm i\sqrt{\lambda}x},\]
 a behavior which reflects an essential singularity at $\lambda\rightarrow\infty$,
on both Riemann sheets. We defer these details to Appendix \ref{cha:APNDXThe-KdV-equation}.
We just mention the result that the finite gap solutions of the KdV
equation obtained by this method have the form of quasi-periodic traveling
waves \begin{equation}
u(x,t)=u(kx+\omega t;\{\lambda_{i}\}),\label{eq:TravelingSol}\end{equation}
 where the wavenumber $k(\{\lambda_{i}\})$ and the frequency $\omega(\{\lambda_{i}\})$
are functions of the branch points, $\{\lambda_{i}\}$.

\subsection*{Whitham equations}

We showed above how to construct a set of finite gap solutions for
the KdV equation, by the inverse scattering method. Our next step
is to sew together these solutions to approximate more general cases.
Namely, we would like to use exact solutions found in the previous
stage as a local (space-time) description of a general function, as
illustrated in Fig.~ (\ref{cap:Riemann solution}).

Since the solutions of the KdV equation (\ref{eq:TravelingSol}) are
uniquely determined by the values of branch points, $\{\lambda_{i}\}$
one may generalize these solutions by endowing these branch points
a slow time and space dependence. A more general solution having this
property is \[
u(x,t)=u(S(x,t);\{\lambda_{i}\})\]
 where the set $\{\lambda_{i}\}$ which determines the form of the
solution, is slowly changing in space and time, and the function $S(x,t)$
replaces the argument $k\cdot x+\omega\cdot t$ of the exact solutions.
To ensure that this solution is locally described by one of the exact
solution we demand that the first order expansion of $S(x,t)$ reproduce
the argument of the finite gap solutions, thus \begin{equation}
\frac{\partial S}{\partial x}=k(\{\lambda_{i}\})~~~~~~~\frac{\partial S}{\partial t}=\omega(\{\lambda_{i}\}).\label{whitham-KdV}\end{equation}
 where $k(\{\lambda_{i}\})$, and $\omega(\{\lambda_{i}\})$ are the
wave number and frequency of the KdV solution characterized by the
set $\{\lambda_{i}\}$. For constant $\{\lambda_{i}\}$, these equations
trivially reduce to the finite gap solutions (\ref{eq:TravelingSol}).
However, when $\{\lambda_{i}\}$ have slow time and space dependences,
these equations, known as the Whitham equations, yield a more general
behavior which is described by slow evolution of the finite gap solutions.
The treatment above of the Whitham equations may only be considered
as a motivation, while the rigorous proof of these equations usually
proves more technically involved\cite{Whitham:1966:NDW,Whitham:Lagrangian}.

The evolution of the branch points $\{\lambda_{i}\}$ as function
of the time can be deduced from the compatability condition for Whitham
equations, \begin{equation}
\frac{\partial k(\{\lambda_{i}\}}{\partial t}=\frac{\partial\omega(\{\lambda_{i}\}}{\partial x}.\label{competability}\end{equation}
 Thus, given initial conditions of the KdV equation, $u(x,t=0)$,
one can calculate the set $\{\lambda_{i}(x,t=0)\}$ and hence the
evolution in time $\{\lambda_{i}(x,t)\}$. Then solution of the Whitham
equations yield the approximate solution (\ref{eq:TravelingSol})
for any time $t$.

By this approach the complicated problem of describing general solutions
for the initial value problem of the KdV equation (which in general
depends on a very large set of constant $\lambda_{i}'$s) has been
approximated by equations for a small number of parameters, which
are the branch points of the Riemann surface. These equations which
describe the evolution of spectral surface, allow for appearances
and disappearances of gaps in the spectrum, i.e. changing the genus
of the spectral surface.

Thus by dispersive regularization, one may obtain solutions for the
KdV equation which are relevant also for long times, and which do
not exhibit the singular behavior of the Riemann solution. These solutions
display a breakup of the wave into oscillations (described by finite
gap solutions) near points where the dispersion term, $u_{xxx}$ becomes
large. A gap in the spectrum is opened up when the system approaches
an overhang. Near this space-time point the behavior is described
by a genus one surface, associated with the oscillatory behavior.
Far from this point, where the Riemann solution (\ref{eq:RiemannSolution})
gives a good approximation, the corresponding spectral surface is
of genus zero.

\section{The 2d Toda Lattice equations\label{sub:2DTL}}

\noindent Our purpose now is to use dispersive regularization approach
to find solutions of the 2DTL. This is the central part of this chapter
and it will be organized as follows: We first present the spectral
equation associated with the 2DTL, then discuss the meaning of the
BA function and show that this function defined in the complex plane
is peaked along the contour of the ST bubble. Next we construct the
spectral surface, and present the general procedure for constructing
the BA function based on the Krichever approach \cite{80:Krichever:Spect:Theory}.
Finally we will show that this procedure is equivalent to the Whitham
equations.

\subsection*{\noindent The spectral equation}

\noindent As explained above, the dispersive regularization approach
is based on the existence of a related spectral equation. Indeed,
there is such an equation for the 2DTL, which is similar to that of
the KdV, \begin{equation}
{\mathcal{L}}\vec{\psi}=z\vec{\psi},\label{SE2DTL}\end{equation}
 but in contrast with the KdV case, where the spectral operator (the
Schr\"{o}edinger operator) is Hermitian, here the operator ${\mathcal{L}}$
is non-Hermitian, and the analogue of the eigenvalue spectrum assumes
general complex values, $z$.

The noninteracting fermion picture of the 2DTL provides a simple way
for understanding the meaning of the spectral equation - it reflects
the recursion relations of the corresponding orthogonal polynomials,
$P_{m}(z)$\cite{Teodorescu}. To explain this, let us choose the
single particle wave functions, $\psi_{n}(z)$ to be the elements
of the BA function, which in our case should be viewed as a vector:
\[
\left(\vec{\psi}\right)_{n}=\psi_{n}(z)=P_{n}(z)e^{\frac{1}{\hbar}\left(-\frac{|z|^{2}}{2}+t_{k}z^{k}\right)}.\]
 Multiplying $\psi_{n}(z)$ by $z$ results in the replacement of
$P_{n}(z)$ by another polynomial $zP_{n}(z)$. Since $zP_{n}(z)$
can be written as a sum of the orthogonal polynomials of degree $m\leq n+1,$
we have:\[
z\psi_{n}(z)=\psi_{n+1}+a_{n,n}\psi_{n}+a_{n,n-1}\psi_{n-1}...+a_{n,0}\psi_{0}.\]
 which can be recast in vector form as the spectral equation (\ref{SE2DTL}).%
\footnote{Notice that the coefficient in front of $\psi_{n+1}$, on the right
hand side of the equation, is unity because of the choice of normalization
of of the orthogonal polynomials $P_{m}(z)=z^{m}+$ lower order terms.%
} where the matrix elements of $\mathcal{L}_{mn}=a_{nm}$ form a triangular
matrix, since ${\mathcal{L}}_{nm}=0$ for $m>n+1$. In appendix \ref{apndx:Sting2DTL}
we prove that this construction of ${\mathcal{L}}$ indeed satisfies
the conditions required in order to be considered as the Lax operator
of the 2DTL.

A different view of the spectral operator employs the quantization
relation between the ST problem and the 2DTL. Namely the spectral
operator ${\mathcal{L}}$ may be considered as the operator associated
with quantization of the classical conformal mapping (\ref{eq:conformal-mapping}).
This quantization follows form the identification of $\log(w)$ and
the time $t$ as classical conjugate variables, as discuss in Sec.~\ref{sec:MathDescrp}.
Thus demanding non-vanishing commutation relations between them, $[\log(w),t]=i\hbar$
allows representing $\log(w)$ as $i\hbar\partial_{t}$, and from
expression (\ref{eq:conformal-mapping}) for the conformal mapping
we have, \begin{equation}
{\mathcal{L}}=z\left(e^{i\hbar\frac{\partial}{\partial t}}\right)=u_{1}e^{i\hbar\frac{\partial}{\partial t}}+u_{0}+u_{-1}e^{-i\hbar\frac{\partial}{\partial t}}+u_{-2}e^{-i2\hbar\frac{\partial}{\partial t}}+...\label{L}\end{equation}
 Notice that this expression for the spectral operator agrees with
the triangular form of the matrix ${\mathcal{L}}_{nm}$ discussed
above, since $e^{i\hbar\frac{\partial}{\partial t}}$ is a shift operator
\[
\left(e^{i\hbar\frac{\partial}{\partial t}}\vec{v}\right)_{t}=\vec{v}_{t+\hbar}.\]
 and $t=\pi\hbar n$ is the area of the droplet containing $n$ fermions.

\noindent Thus in the dispersionless limit, $\hbar\rightarrow0$,
${\mathcal{L}}$ reduces to the conformal mapping $z(w;t)$. Moreover
the relation $\{ z(w;t),\bar{z}(w;t)\}$ suggests that the spectral
operator satisfies analogous quantum relation, known as the string
equation\cite{48:DiFrancesco:2dgravity}: \begin{equation}
[{\mathcal{L}},{\mathcal{L}}^{\dagger}]=\hbar.\label{eq:string-equation}\end{equation}
 Notice that the string equation, which will be proved in Appendix
\ref{apndx:Sting2DTL}, imposes an additional constraint on the form
of the spectral operator of the 2DTL. This constraint selects a subset
of solutions out of the general solutions of the 2DTL, as we expect
the corresponding eigenfunction, $\psi_{n}(z)$, to be associated
with the shape of the classical bubble.

\subsection*{\noindent The Baker Akhiezer function of the 2DTL}

\noindent As we saw above, the spectral equation can be thought as
a quantization of the conformal map formalism. Below we take a closer
look at the wave functions $\psi_{n}(z)$ themselves and see that
they already encode the shape of the bubble in the ST evolution in
a quite natural way. The consequence of this will be that the spectral
surface ($z$ plane) can be identified as the physical plane (the
2D plane on which the physical evolution takes place), in contrast
with the KdV case where the spectral surface served only as an auxiliary
mathematical construction.

We begin by arguing%
\footnote{we provided a more rigorous proof in \cite{100:Bettelheim:Agam:Wieg}%
} that the $N$-th component of the BA function, i.e. $\psi_{N}(z)$,
is peeked along the contour of ST bubble of area $\pi\hbar N$. Consider
the particle density associated with the noninteracting fermionic
systems, which is a sum over the single particle densities: \[
\rho(z)=\sum_{n=1}^{N}|\tilde{\psi}_{n}(z)|^{2},\]
 where $\tilde{\psi}_{n}(z)$ denote the normalized $\psi_{n}(z)$
(namely $\int d^{2}z|\tilde{\psi}_{n}(z)|^{2}=1$). As we showed in
Section \ref{sec:2DTL}, using Dyson's gas picture, this density,
in the limit $\hbar\to0$, $N\rightarrow\infty$ keeping $\hbar N$
constant, vanishes outside the ST bubble and is constant within the
bubble. Moreover, we have shown that by increasing the number of fermions
$N$, the droplet area increases in accordance with the ST dynamics,
i.e. keeping the harmonic moments, $\{ t_{k},\bar{t}_{k}\}$ constant.
Therefore the last particle added to the droplet must be distributed
along the droplet perimeter, i.e. $\tilde{\psi}_{N}(z)$ is peeked
along the contour of ST bubble of area $\pi\hbar N$.

Thus, in the dispersionless limit we expect the BA function to have
the form \begin{equation}
\psi(z)\sim e^{-\frac{1}{\hbar}\left(\frac{1}{2}|z|^{2}-\int dzS(z)\right)},\label{eq:BA approx}\end{equation}
 where $S(z)$ is the Schwarz function of the ST bubble. This form
emerges because the saddle manifold, where the modulus of this function
assumes its maximal value, is the droplet contour $\bar{z}=S(z)$.
Thus by constructing the BA function of the 2DTL on various types
of spectral surfaces, and looking at its saddle manifolds as function
of $\pi\hbar N$ (which is the bubble area $t$) we expect to be able
to extract the regularized dynamics of the ST problem.

To further clarify the analytic structure of the BA function, consider
its form given by the formula, \[
\psi(z)=\psi_{N}(z)=e^{-\frac{1}{\hbar}\left(\frac{|z|^{2}}{2}-\sum_{k}t_{k}z^{k}\right)}P_{N}(z),\]
 where $P_{N}(z)$ denotes the orthogonal polynomial of problem as
discussed in Sec.~\ref{sec:2DTL}. Since this polynomial is of order
$N$, it has $N$ roots associated with $N$ zeros located in the
complex plane. Thus lifting the polynomial $P_{N}(z)$ to the exponent\[
\psi(z)=e^{-\frac{1}{\hbar}\left(\frac{|z|^{2}}{2}-\sum_{k}t_{k}z^{k}\right)+\log(P_{N}(z))},\]
 and comparing to the expected form of the BA function (\ref{eq:BA approx}),
we conclude that $S(z)$ has a pole singularity at each zero of the
polynomials.

When $N\rightarrow\infty$ the number of zeros of $P_{N}(z)$ becomes
infinite, and a well defined limit for their distribution is reached
when $\hbar\rightarrow0$ keeping $\hbar N$ constant. In this limit
the zeros of the polynomial may be described by a line density on
a set of contours in the complex plane. These lines may form the branch
cuts of $S(z)$, which are analogous to the spectrum in the KdV problem.

\subsection*{The spectral surface}

Since the BA function determine the shape of the ST bubble, our goal
now is to compute this function and describe its evolution in time.
For this purpose we shall employ the dispersive regularization approach
as was done for solving the KdV equation. Namely, we will reduce the
problem to the evolution of the spectral surface (i.e. to the evolution
of its branch points). This procedure amounts to the assumption that
the Schwarz function, $S(z)$, is an algebraic function, namely it
is defined on an algebraic Riemann surface, defined by an equation
of the form $Q(z,y)=0,$ where $Q$ is a polynomial in its both arguments,
e.g. $Q(z,y)=y^{2}-\prod_{i=1}^{4}(z-\lambda_{i})$.

In order to construct the Riemann surface from this polynomial, one
first solves the equation $Q(z,y)=0$ for $y$. The resulting function,
$y(z)$, is multi-valued due to the presence of roots in the solution
(e.g. $y=\sqrt{\Pi_{i=1}^{4}(z-\lambda_{i})}$ in the example above).
In general there will be $n$ solutions of $y$ for each $z$. Thus
one may introduce $n$ copies of the complex plane, where each copy
is associated with a well defined value $y(z)$. Clearly, on each
copy of the complex plane $y(z)$ is discontinuous along the branch
cuts. To make it continuous we may paste together the various sheets
of the complex plane along the branch cuts. Then, together with a
choice of local coordinate systems around each point one obtains a
Riemann surface composed of $n-$sheets glued together along the branch
cuts, similar to the construction demonstrated in Fig.~(\ref{cap:EllipticConstruct}).

To each point on the Riemann surface, we have constructed, we would
like to assign a value of the Schwarz function, $S$. In order to
specify a point on the Riemann surface we must indicate an index $i$
(where $1\leq i\leq n$), and a complex number $z$. The index, $i$,
will specify on which copy of the complex plane the point lies, and
a complex number, $z$, will specify the coordinate on that copy.
There will be one copy of the complex plane, which will be termed
as {}``the physical sheet'', on which the bubble lies. On this Riemann
sheet $S(z)$ will be equal to $\bar{z}$ on the perimeter of the
bubble. On the other copies of the complex plane, $S(z)$ should be
understood as an analytic continuation of $\bar{z}$.

\subsection*{Construction of $S(z)$}

The next step of the dispersive regularization approach is the construction
of the BA function which in our case is equivalent to the construction
of the Schwarz function, $S(z)$. Two conceptual steps constitute
this construction. First, the identification of the singular behavior
of $S(z)$ outside the bubble (on the physical sheet). This is done
using the relation \begin{equation}
S(z)\sim\sum_{k=1}^{\infty}kt_{k}z^{k-1}+\frac{t_{0}}{z}\label{analytic-out}\end{equation}
 which holds for large $z$ outside the bubble, where $t_{k}$ and
$t$ are the harmonic moments and the area of the bubble respectively.
The above relation follows from the definition of harmonic moments
(\ref{harmonic}) which by using Green's theorem implies that \begin{equation}
t_{k}=-\frac{1}{\pi k}\int_{D_{-}}\frac{d^{2}z}{z^{k}}=\frac{1}{2i\pi k}\oint dzS(z)z^{-k},\label{eq:HarmonicIntegral}\end{equation}
 where the contour integral is taken along the bubble perimeter. By
deforming the contour on the physical sheet one can verify that $S(z)$
must satisfy (\ref{analytic-out}). Notice that by definition of the
physical sheet, the deformation of contour does not cross any branch
cuts. If there are no singularities of $S(z)$ except at infinity,
a similar argument shows that $t_{0}$ is equal to the area of the
bubble $t$: $t_{0}=t=\int_{D_{+}}d^{2}z$.

The second step is to identify the singularities of $S(z)$ on the
other parts of the Riemann surface (including all non-physical sheets).
This is achieved using the unitarity condition $z=\bar{S}(S(z))$.
Recall that the Schwarz function maps the region outside the bubble
onto the region inside the bubble which includes all non-physical
copies of the complex plain. Thus knowing the analytic properties
of $S(z)$ and the structure of the Riemann surface one may construct
$S(z)$ and in particular its dependence on the parameters $t_{k}$
as well as in their anti-holomorphic counterparts, $\bar{t}_{k}$,
and the area $t$.

We turn now to describe how the above procedure is implemented in
practice. For this purpose it is more convenient to construct the
differential $\hat{S}=S(z)dz$ and then deduce the Schwarz function
from the relation $S(z)=\hat{S}/dz$. 

We begin by identifying all singular points of the Riemann surface.
To this end we use formula (\ref{analytic-out}) on the physical sheet,
outside the bubble, and the unitarity condition, $z=\bar{S}(S(z))$,
to reveal the singularities inside the bubble including the non-physical
sheets. The next step is to associate with the singular points meromorphic
differential, on the Riemann surface, $\hat{\omega}_{i}$ which comply
with a given singular structure at these points. The differential
$\hat{S}$ is then obtained as the sum of these meromorphic differentials.

The meromorphic differentials on a Riemann surface, $\hat{\omega}_{i}$,
are defined uniquely, given their singular structure (analytic behavior
around the poles of the Riemann surface) and their normalization.
This normalization is determined by the integral of $\hat{\omega}_{k}$
around cycles of the Riemann surface. Below, we shall elaborate on
this point. For the time being we shall only state that different
choices of this normalization amounts to different choices of the
types of evolution of the ST dynamics.

For an arbitrary Riemann surface the procedure above will produce
a function which satisfies the unitarity condition around the singular
points, but in general will not satisfy the unitarity condition globally.
But if the Schwarz function exists on a given Riemann surface the
procedure above must produce this function. Thus to obtain a Schwarz
function for a given set of harmonic moments, we must first find the
Riemann surface on which it is defined, in other words we must find
the location of the branch points (which determine the Riemann surface
uniquely if we also know how the different sheets are interlaced).

To find the location of the branch points we may use the fact that
unitarity must be satisfied around the branch points. This leads on
fairly general conditions to the conclusion that $S(z)$ may not diverge
at the branch points. This property alone will be enough to find the
location of the branch points. To show that the Schwarz function does
not diverge at the branch points, we assume that $\overline{S(z)}$
maps the exterior of the bubbles to the interior, where the interior
also includes the unphysical sheets. We also assume that all branch
cuts are in the interior. If we assume that $S(z)$ diverges on a
branch points, $\lambda_{i}$, then $\overline{S(z)}$ will also diverge
on the branch point. This in turn implies, by unitarity, that $S(z)$
has the following form around infinity:\[
S(z)\sim\bar{\lambda}_{i}+O(z^{-1}).\]
Which gives that $t_{1}=\bar{\lambda}_{i}$. Thus we have that $S(z)$
diverges on a branch point only for special choices of the harmonic
moments.

Let us look at the implication of the non-divergence property of $S(z)$
near the branch points. Consider the Schwarz function, $S(z)=\hat{S}/dz$.
Near one of its branch points, the local coordinate is $\delta=\sqrt{z-\lambda_{i}}$
and therefore $dz=2\delta d\delta$. Since $\hat{S}$ is meromorphic,
near the branch point it can be expanded as $\hat{S}\simeq a_{1}d\delta+a_{2}(d\delta)^{2}+...$,
and therefore $S(z)=\hat{S}/dz\sim a_{1}/\delta=a_{1}/\sqrt{z-\lambda_{i}}$.
The requirement that $S(z)$ is not singular near the branch points
implies that $a_{1}=0$. This condition may be recast as follows:\begin{equation}
\oint\frac{S(z)dz}{\sqrt{z-\lambda_{i}}}=0,\label{eq:Krichever}\end{equation}
 where the integral is over small circles around each one of the branch
points of the Riemann surface. The number of conditions which these
integrals gives is the same as the number of branch points, which
is enough to fix the form of the Riemann surface. Eq. (\ref{eq:Krichever}),
which is the central result of this section, has been constructed
in different context by Krichever \cite{83:Krichever:Averaging}.
It gives the evolution of the spectral surface as a function of the
times. From this information one can deduce the behavior of the Schwarz
differential, $\hat{S}\equiv S(z)dz$, and obtain the time dependence
of Schwarz function which describes the bubble dynamics.

\subsection*{Example: One Miwa variable on genus-0 surface}

Let us consider an example in order to demonstrate the procedure described
above. We focus on the case where the set of harmonic moments is given
by:\[
t_{k}=\frac{\mu}{k}q^{-k}\]
 for $k>1$, and for $k=1$ \[
t_{1}=t'_{1}+\frac{\mu}{q}\]
 where $\mu$ is real, while $t'_{1}$ and $q$ are arbitrary complex
numbers. From formula (\ref{analytic-out}) which holds outside the
bubble we have \[
S(z)=\frac{t_{0}}{z}+t'_{1}+\mu\sum_{k=1}^{\infty}\frac{z^{k-1}}{q^{k}}\]
 and summing over $k$ yields \[
S(z)\sim\frac{t_{0}}{z}+t'_{1}-\frac{\mu}{z-q}~~~~~~\mbox{in the exterior domain}\]
 Thus Schwarz function has a simple pole at $q$ with residue $\mu$.
Since now $S(z)$ has an additional singularity at the point $q$,
the equality $t_{0}=t$ (where $t$ is the area of the bubble). In
fact one may derive $t_{0}=t-\mu$. The pole of $S(z)$ in the the
exterior domain is called {}``Miwa variable''\cite{28:Jimbo:Miwa,33:Morozov:Matrix:Intgrbl,65:Miwa:Hirota,29:Kostov:bilinear,21:Kostov:Confrml:Mtrx,61:Hirota:Difference,65:B:Miwa:Hirota,65:C:Miwa:Hirota,65:D:Miwa:Hirota}.
It is characterized by its location, $q$ and weight $\mu$. The case
of one Miwa variable is one of the simplest nontrivial cases to consider.

Having assumed the singular structure at the exterior of the bubble
we may now use the unitarity condition, $\bar{S}(S(z))=z$, to find
the singular structure in the interior of the bubble %
\footnote{by interior of the droplet we mean the interior on the physical sheet
as well as all the non-physical sheets%
}. Consider the limit $z\rightarrow\infty_{+}$, where by $\infty_{+}$
we refer to the infinity on the physical sheet. From the unitarity
condition we have $\bar{S}(t'_{1})=\infty_{+}$, therefore $S(z)$
must have a pole at $\bar{t}'_{1}$. From a local analysis near $z=\infty_{+}$
it is also easy to show that the residue of the pole is $\mu-t$,
Thus \[
S(z)\sim-\frac{t-\mu}{z-\bar{t}'_{1}}~~~~\mbox{ near $\bar{t}'_{1}$}\]
 which is inside the bubble either on the physical sheet or on the
non-physical one.

Consider now the limit $z\rightarrow q$ then we have $S(z)\rightarrow\infty_{-}$,
where $\infty_{-}$ denotes the infinity on the non physical sheet
(recall that $S(z)$ maps points form the exterior domain to the interior
domain which includes all non physical sheets). Thus $\bar{S}(\infty_{-})=q$
and from a local analysis near $z=\infty_{-}$ we get that \[
S(z)\sim\bar{q}-\frac{\mu}{z}~~~\mbox{ near $\infty_{-}$}.\]

Thus the poles of $\hat{S}$ on the exterior domain (at $z=\infty_{+}$
and $z=q$) have been mapped onto other poles in the interior domain
(at $z=\bar{t}_{1}$ and $z=\infty_{-}$), and the singular behavior
can be summarized by the relations \begin{eqnarray}
S\sim t'_{1}+\frac{t-\mu}{z}~~\mbox{near $\infty_{+}$};~~~~~~S\sim\bar{q}-\frac{\mu}{z}~~\mbox{near $\infty_{-}$};\label{eq:Ssing1}\\
S\sim-\frac{t-\mu}{z-\bar{t}'_{1}}~~\mbox{near $\bar{t}'_{1}$};~~~~~~~~S\sim\frac{-\mu}{z-q}~~\mbox{near $q$}.\label{eq:Ssing2}\end{eqnarray}
 From here one concludes that the Schwarz differential $\hat{S}=S(z)dz$
has six singularities: One simple pole at $q$ with residue $-\mu$;
A second simple pole at $\bar{t}'_{1}$ with residue $\mu-t$; Two
simple poles at $\infty_{+}$ and $\infty_{-}$ with residues $\mu-t$
and $\mu$ respectively (because the differential $dz/z$ in the local
coordinate system $s=z^{-1}$ is $dz/z=-ds/s$); Finally two double
poles at $\infty_{+}$ and $\infty_{-}$ (since in the local coordinate
system $s=z^{-1}$, $dz=-ds/s^{2}$).

\begin{figure}
\includegraphics[%
  width=8cm]{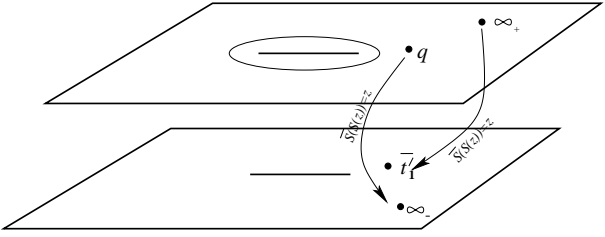}

\caption{Riemann surface for the 1-Miwa case. The Riemann surface is composed
of two sheets (two copies of the complex plane). The physical sheet
(the upper plane in the drawing) contains a bubble (drawn schematically
as an oval) inside which lies a branch cut ( which is also drawn on
the unphysical sheet). The physical sheet also contains the Miwa point
$q$, an a point at infinity marked as $\infty_{+}$. $S(z)$ maps
$\infty_{+}$ to $\bar{t}_{1}'$, while the infinity on the lower
(unphysical) sheet $\infty_{-}$ is mapped to $q$. \label{1MiwaSheeting}}
\end{figure}

Specifying also the location of the branch cuts defines the full analytic
structure of the spectral surface. In this example we consider the
simple case where the spectral surface is of genus zero, i.e. there
is one branch cut between the points $\lambda_{1}$ and $\lambda_{2}$.
The process of unraveling this analytic structure is illustrated in
Fig.~(\ref{1MiwaSheeting})

Our purpose now is to construct the meromorphic differentials associated
with the singular points which we have identified. Consider first
the differentials associated with the double poles near $\infty_{\pm}$.
The meromorphic differential which satisfies this behavior and does
not have any other singularities takes the form \[
\hat{\omega}_{\infty_{\pm}}^{(2)}=-\frac{1}{2}\left(1\pm\frac{z-\frac{1}{2}(\lambda_{1}+\lambda_{2})}{\sqrt{z-\lambda_{1}}\sqrt{z-\lambda_{2}}}\right)dz.\]
 Apart from the branch points and infinity, where we may need to take
a closer look, it is obvious that the differential is holomorphic.
At the branch points the local parameter is $\delta\equiv\sqrt{z-\lambda_{i}}$,
and by writing $dz=2\delta d\delta$ we see that indeed the differential
is holomorphic around the branch points too. Turning our attention
to the points at infinity we first observe that $\sqrt{z-\lambda_{1}}\sqrt{z-\lambda_{2}}\sim\pm z$
where the sign depends on whether we are on the first or second sheet.
The local parameter is $s\equiv\frac{1}{z}$ , which gives, e.g. $\hat{\omega}_{\infty_{+}}^{(2)}\sim\frac{ds}{s^{2}}$
on the infinity on the upper sheet and no singularity on the lower
sheet.

Now we would like to construct two more meromorphic differentials.
The first one, which we denote by $\omega_{\bar{t}'_{1},\infty_{+}}$,
has two simple poles: One at $\bar{t}'_{1}$ and the other at $\infty_{+}$.
The second meromorphic differential, $\omega_{q,\infty_{-}}$ has
one pole at $\bar{q}$ and second pole at $\infty_{-}$, and as before,
both poles have the same residue, $\mu$. These meromorphic differentials
have the form: \[
\hat{\omega}_{\zeta,\infty_{\pm}}=\pm\frac{1}{2}\frac{(z-\zeta)\pm\sigma_{\zeta}\sqrt{\zeta-\lambda_{1}}\sqrt{\zeta-\lambda_{2}}\pm\sqrt{z-\lambda_{1}}\sqrt{z-\lambda_{2}}}{(z-\zeta)\sqrt{z-\lambda_{1}}\sqrt{z-\lambda_{2}}}dz.\]
 where \[
\sigma_{\zeta}=\left\{ \begin{array}{rl}
1 & \mbox{ if $\zeta$ on the physical sheet}\\
-1 & \mbox{ if $\zeta$ on the nonphysical sheet}\end{array}\right.\]
 Note that the behavior of $\omega_{\zeta,\infty_{\pm}}$ near $\infty_{\pm}$
is given by $\omega_{p,\infty_{\pm}}\sim dz/z=-ds/s$ while in the
vicinity of $\zeta$, it is $\omega_{\zeta,\infty_{\pm}}\sim dz/(z-\zeta)$.

Having the meromorphic differential we can now construct the Schwarz
differential by summing over these differentials with the appropriate
weights dictated by (\ref{eq:Ssing1},\ref{eq:Ssing2}). \[
\hat{S}=-t'_{1}\hat{\omega}_{\infty_{+}}^{(2)}-\bar{q}\hat{\omega}_{\infty_{-}}^{(2)}+t\hat{\omega}_{\bar{t}'_{1},\infty_{+}}-\mu\hat{\omega}_{q,\infty_{-}}.\]

Finally using local coordinates of the Riemann surface we see that
$S(z)=\frac{\hat{S}(z)}{dz}$ in general singular near $\lambda_{i}$,
i.e. $S(z)\sim\frac{C_{i}}{\sqrt{z-\lambda_{i}}}$ where $C_{i}$
is a function of the parameters of the system and the branch points
$\lambda_{i}$. The Krichever equations (\ref{eq:Krichever}) imply
that $C_{i}$ must vanish. Thus the equations $C_{i}=0$ determine
the evolution of th branch points as function of the times. In particular
the Krichever equation associated with $C_{1}=0$ is \[
\frac{1}{2}(t'_{1}-\bar{q})(\lambda_{1}-\lambda_{2})+t\left(1-\sqrt{\frac{\bar{t}'_{1}-\lambda_{2}}{\bar{t}'_{1}-\lambda_{1}}}\right)+\mu\left(1-\sqrt{\frac{q-\lambda_{2}}{q-\lambda_{1}}}\right)=0\]
 where, for simplicity, we assumed that $\bar{t}'_{1}$ is located
on the non-physical sheet. The second Krichever equation, associated
with $C_{2}=0$ is obtained by interchanging the branch points $\lambda_{1}\leftrightarrow\lambda_{2}$,
since $\hat{S}$ is symmetric in these variables.

The solution of the above nonlinear equations for $\lambda_{i}$ as
function of the time $t$ (keeping all other parameters fixed) allows
us to construct the Schwarz function, which describes the evolution
of the bubble in time. We shall describe this evolution in detail
in section \ref{sec:Richardson}.

\subsection*{The normalization of the meromorphic differentials}

As mentioned before in the general case of spectral surface of nonzero
genus the meromorphic differentials should be normalized. The mathematical
reason for this necessity is that Riemann surfaces of genus $g$ has
additional degree of freedom associated with the existence of holomorphic
differentials.%
\footnote{In the simple example of a Riemann surface of genus $g$, defined
by $y^{2}=\prod_{i=1}^{2g+2}(z-\lambda_{i})$, the holomorphic differentials
are $\hat{\Omega}_{k}=z^{k-1}dz/y$ where $1\leq k\leq g$. The check
of holomorphicity is done using the local coordinate on the Riemann
surface: For $z\neq\lambda_{i}$ and $z\neq\infty$, clearly $z^{k-1}/y$
is holomorphic for $k\geq1.$ Around infinity we use local coordinates
$z=s^{-1}$, therefore $dz=-s^{-2}ds$ and obtain $\hat{\Omega}_{k}\rightarrow-s^{g-k}ds$
which is nonsingular for $k\leq g$. Near the branch point $\lambda_{i}$,
the local is coordinate $\delta=\sqrt{z-\lambda_{i}}$, thus $dz=2\delta d\delta$,
and as $\delta\rightarrow0$, one obtains $\hat{\Omega}_{k}\rightarrow\prod_{j\neq i}(\lambda_{i}-\lambda_{j})^{-1}\lambda_{i}d\delta$
which is again nonsingular.%
} These holomorphic differentials are nonsingular and may be added
to the meromorphic differentials without changing their analytic properties.
Therefore an additional information (normalization) is required in
order to specify the Schwarz differential uniquely.

The physical origin for the appearance of additional degrees of freedom,
is that in the nonzero genus case the Schwarz function describes $g+1$
bubbles. Therefore to specify the multi-bubble dynamics uniquely,
additional information is required about the relations among the bubbles.
For instance, one may specify the rate at which the area of each bubble
grows, or set the internal pressure to be the same for all bubbles,
etc. As we will show the normalization of the meromorphic differentials
determine the inter-relations among the bubbles.

\begin{figure}
\includegraphics[%
  width=6cm]{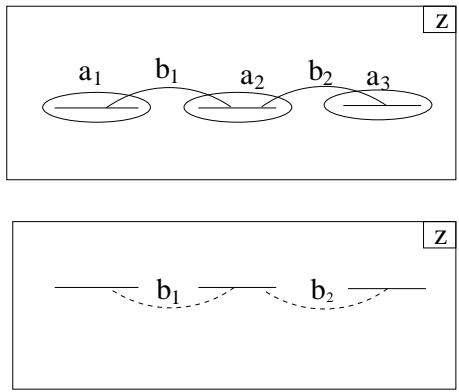}

\caption{$a$ and $b$ cycles for a genus-2 surface. The choice of cycles
is not the canonical one. Figure show two copies of the complex plane,
a portion of the $b$ cycles lies on the unphysical sheet, while the
$a$ cycles are drawn on the physical sheets only.\label{cap:abcycles}}
\end{figure}

We begin by specifying the types of cycles on the spectral surface.
let us denote by $a_{i}$ the cycle which is equivalent (that can
be deformed without crossing one of the points where $S$ has singularities)
to the boundary of the $i$-th bubble, see Fig.~(\ref{cap:abcycles})
for an example. These cycles stay on one Riemann sheet only. The second
type of cycles, the $b$-cycles, has parts on both Riemann sheets,
it crosses the bubbles $a_{i}$ and $a_{i+1}$ as shown schematically
in Fig.~(\ref{cap:abcycles}).

The integral of $S(z)$ over the $a_{i}$ cycle is proportional to
the area, ${\mathcal{A}}_{i}$, of the $i$-th bubble. This is an
immediate consequence of Green's theorem: \[
\oint_{a_{i}}S(z)dz=\oint_{a_{i}}\bar{z}dz=2i{\mathcal{A}}_{i}.\]
 Thus we may choose a normalization by specifying the integrals along
the $a$ cycles, and this would be equivalent to choosing the bubbles
area. However, this normalization, which does not allow for a transfer
of the non-viscous liquid between different bubbles, is problematic.
It implies, for instance, that the internal pressure of each bubble
is generally different, and therefore two bubbles cannot merge smoothly.
Thus if one would like to view the dispersive regularization approach
as regularization of the idealized ST dynamics which allows temporary
formation of new bubbles, we must seek another normalization allowing
for merging of bubbles. The most obvious candidate is the normalization
which ensures that all bubbles share the same pressure. From here
on, for simplicity we consider the genus one case consisting of two
bubbles. The generalization to multi-bubble situation (i.e. higher
genus) is straightforward.

A normalization of meromorphic differentials which sets a vanishing
pressure difference between the two bubble, is likely to be associated
with the $b$ cycle which connects the two bubbles. Indeed, we will
now show that the equal pressure condition is equivalent to the requirement
that the integral of the Schwarz differential along the $b$ cycle
vanish: \begin{equation}
\oint_{b}S(z)dz=0\label{eq:normalization}\end{equation}

\begin{figure}
\includegraphics[%
  width=6cm]{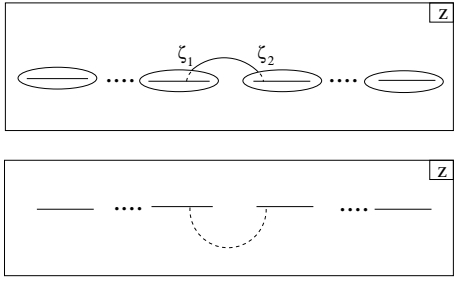}

\caption{$Q(\zeta_{1})-Q(\zeta_{2})$, where $\zeta_{1}$ and $\zeta_{2}$
lie on contours of different droplets, is shown in this article to
be equal to $\frac{1}{2}\Re\oint S(z)dz$, where the contour integral
is taken over the cycle which connects the two points in the drawing.
The part of the contour on the exterior of the droplets is drawn in
a solid line, while the part lying on the interior (and the unphysical
sheet) is drawn in a dashed line. The dashed part of the contour can
be thought of as the image under $\overline{S(z)}$ of the solid line.
\label{cap:ProofFig}}
\end{figure}

To understand the meaning of this normalization, let us separate the
contour integral along the $b$ cycle into two contributions: One
is a line integral from $\zeta_{1}$ to $\zeta_{2}$, and the other
$\zeta_{2}$ to $\zeta_{1}$ along the second branch of the cycle.
We set $\zeta_{1}$ and $\zeta_{2}$ to be on the contours of the
first and the second bubble, respectively, see Fig.~(\ref{cap:ProofFig}).
Then a straightforward algebraic manipulation%
\footnote{The second integral in $\oint_{b}S(z)dz=\int_{\zeta_{1}}^{\zeta_{2}}S(z)dz+\int_{\zeta_{2}}^{\zeta_{1}}S(z)dz$
is evaluated by parts, and using the fact that on the droplets contours
$S(z)=\bar{z}$, one obtains \[
\oint_{b}S(z)dz=\int_{\zeta_{1}}^{\zeta_{2}}S(z)dz-\int_{\zeta_{2}}^{\zeta_{1}}zdS(z)+|\zeta_{1}|^{2}-|\zeta_{2}|^{2}\]
 Then changing the integration variable in the second integral to
$y=\bar{S}(\bar{z})$ gives \[
\oint_{b}S(z)dz=\int_{\zeta_{1}}^{\zeta_{2}}S(z)dz+\overline{\int_{\zeta_{1}}^{\zeta_{2}}S(y)dy}+|\zeta_{1}|^{2}-|\zeta_{2}|^{2}\]
 which is equivalent to (\ref{eq-argument}).%
} yields the relation \[
\oint_{b}S(z)dz=2\Re\left.\left(\int_{\zeta_{1}}^{\zeta_{2}}S(z)dz-\frac{|z|^{2}}{2}\right|_{\zeta_{1}}^{\zeta_{2}}\right).\]
 Thus the normalization condition (\ref{eq:normalization}) implies
the equality \begin{equation}
Q(\zeta_{1})=Q(\zeta_{2})\label{eq-argument}\end{equation}
 where \begin{equation}
Q(z)=\Re\left(-\frac{|z|^{2}}{2}+\int_{\zeta_{0}}^{z}S(z')dz'\right),\label{eq:DefinitionQz}\end{equation}
 and $\zeta_{0}$ is some arbitrary point. Notice that $Q(z)/\hbar$
is the real part of the exponent of the BA function, see (\ref{eq:BA approx}).
Thus the normalization (\ref{eq:normalization}) is equivalent to
the condition that the amplitude of the BA function on the contours
of both droplets is the same. In other words, both droplets have the
same fermion density. If we had, say, $Q(\zeta_{1})>Q(\zeta_{2})$
then the height of the wave function at contour of the second droplet
would be exponentially small, as compared to the height on the contour
of the first droplet, and in fact it will be vanishingly small in
the $\hbar\rightarrow0$ limit. Thus equation (\ref{eq-argument})
ensures that our solution in the dispersionless limit, $\hbar\rightarrow0$,
indeed describes two classical droplets.

Finally, by differentiating equation (\ref{eq-argument}) with respect
to time we show in Appendix \ref{sec:AppendixPressure} that $dQ(\zeta_{i})/dt=P_{i}$
where $P_{i}$ denote the pressure inside the $i$-th bubble and therefore
$P_{1}=P_{2}$, implying that the pressure difference between the
bubbles is zero.

\noindent The normalization condition associated with equal pressure
on all bubbles may be written in a more general form as \begin{equation}
\Re\oint S(z)dz=0.\label{eq:normailzation}\end{equation}
 where the contour integrals is over any cycle of the Riemann surface.
For the $b-$cycle integrals this is the normalization discussed above.
For the $a-$cycle integrals we saw that the result is purely imaginary
(recall that an $a-$cycle integral equals $2i{\mathcal{A}}$, where
${\mathcal{A}}$ is the area of the corresponding bubble) and therefore
the above normalization trivially holds. The normalization, which
has the property that it does not depend on the choice of cycle on
the Riemann surface, was discovered by Krichever in the general context
of dispersionless limit of integrable systems\cite{83:Krichever:Averaging}.

\subsection*{\noindent Example: One Miwa variable on genus-1 surface}

\noindent Consider again the one Miwa case, this time on surface of
genus one, $y^{2}=\Pi_{i=1}^{4}(z-\lambda_{i})$, characterized by
four branch points, $\lambda_{i}$, where $i=1,..4$. The meromorphic
differentials in this case, appear with free constants associated
with the existence of the holomorphic differential: \[
\hat{\Omega}_{1}=\frac{1}{\sqrt{\Pi_{i=1}^{4}(z-\lambda_{i})}}dz\]
 which is free of singular behavior on the spectral surface.

Thus the meromorphic differentials on this Riemann surface, having
the same singular dependence described in the previous example (of
1-Miwa on genus-0 surface) are: \[
\hat{\omega}_{\pm\infty}^{(2)}=\frac{1}{2}\left(1\pm\frac{z^{2}-z\frac{1}{2}\sum_{i=1}^{4}\lambda_{i}+\alpha}{\sqrt{\prod_{i=1}^{4}(z-\lambda_{i})}}\right)dz\]
 associated with the double poles at the infinities of the physical
and non-physical sheets, and \[
\hat{\omega}_{\zeta,\infty_{\pm}}=\frac{1}{2}\left(\frac{1}{z-\zeta}+\frac{\sigma_{\zeta}\sqrt{\prod_{i=1}^{4}(\zeta-\lambda_{i})}}{(z-\zeta)\sqrt{\prod_{i=1}^{4}(z-\lambda_{i})}}\pm\frac{z-\gamma}{\sqrt{\prod_{i=1}^{4}(z-\lambda_{i})}}\right)dz\]
 associated with the two simple poles at $\zeta$ and at infinity.
The constant $\alpha$ and $\gamma$, which are the coefficients of
the holomorphic differential $\hat{\Omega}_{1}$ are fixed by the
normalization conditions. \[
\Re\oint\hat{\omega}_{\pm\infty}^{(2)}=\Re\oint\hat{\omega}_{\zeta,\infty_{\pm}}=0\]
 where the integration is along any cycle of the Riemann surface.
Notice that the normalization of each meromorphic differential ensures
the normalization of the Schwarz differential: \[
\hat{S}=-t'_{1}\hat{\omega}_{\infty_{+}}^{(2)}-\bar{q}\hat{\omega}_{\infty_{-}}^{(2)}+t\hat{\omega}_{\bar{t}'_{1},\infty_{+}}-\mu\hat{\omega}_{q,\infty_{-}}.\]

However the normalization constants, in general, are rather complicated
functions of the branch points, and therefore the construction of
Krichever equations (\ref{eq:Krichever}) becomes cumbersome. We will
show how to go around this difficulty in the next section by the alternative
approach of Richardson\cite{175:Richardson:Multiply}\cite{ST:why:becomes:o4}.

As an illustration for the behavior of the normalization constants,
consider the simplest example of the differential, $\hat{\omega}$
having two simple poles at $\infty_{+}$ and $\infty_{-}$ with residue
$1$ and $-1$, respectively. This differential takes the form:\[
\hat{\omega}=\frac{z-c}{\sqrt{\Pi_{i=1}^{4}(z-\lambda_{i})}}dz\]
 where $c$ is the constant which should be determined by normalization.
The Krichever normalization yields:\[
c=(\lambda_{2}-\lambda_{1})\frac{\Pi\left(\frac{\lambda_{3}-\lambda_{2}}{\lambda_{3}-\lambda_{1}},k\right)}{K(k)}+\lambda_{1},\]
 where \[
k=\sqrt{\frac{\lambda_{3}-\lambda_{2}}{\lambda_{4}-\lambda_{2}}\frac{\lambda_{4}-\lambda_{1}}{\lambda_{3}-\lambda_{1}}}.\]
 Here $K$ and $\Pi$ are the complete elliptic integrals of the first
and third kind respectively.

\subsection*{Whitham equations for the 2DTL}

We conclude this section by introducing the Whitham equations for
the 2DTL, and clarifying their relation to the Krichever equations
(\ref{eq:Krichever})

To present the Whitham equation, let us introduce the set of meromorphic
differentials $\hat{\omega}_{k}$ having the following singular behavior
near the infinity on the physical sheet, $\infty_{+}$: \begin{equation}
\hat{\omega}_{k}\sim dz\cdot z^{k},k\geq0,\label{behavior-near-infinity}\end{equation}
 and properly normalized, i.e. \begin{equation}
\Re\oint\hat{\omega}_{k}=0\label{normalization-Whitham}\end{equation}
 along any cycle of the spectral surface. Then the Whitham equations
of the 2DTL takes the form \begin{equation}
\frac{\partial\hat{S}}{\partial t_{k}}=k\hat{\omega}_{k-1},\,\,\,\,\,\,\frac{\partial\hat{S}}{\partial t}=\hat{\omega}_{\infty_{+},\bar{t}_{1}}\label{eq:whitham}\end{equation}
 These equations determine the evolution of the Schwarz differential
as function of the times $t_{k}$. They are set such that near $\infty_{+}$
the behavior coincide with that dictated by Eq.~(\ref{analytic-out}).

Equations (\ref{eq:whitham}) are analogous to the Whitham equations
(\ref{whitham-KdV}) in the KdV case, and our purpose now is to show
that the Krichever equations (\ref{eq:Krichever}) play a role similar
to the compatibility condition (\ref{competability}) of the Whitham
equations for the KdV case.

To derive (\ref{eq:whitham}), first note that $\frac{\partial\hat{S}(z)}{\partial t_{k}}$
is a differential which has the same singular behavior at infinity
as $\hat{\omega}_{k}$, and has the same normalization dictated by
(\ref{normalization-Whitham}). Thus if $\frac{\partial\hat{S}(z)}{\partial t_{k}}$
does not have any other singularities apart from the singularity at
infinity it must be equal to $\hat{\omega_{k}}$ by the uniqueness
of the meromorphic differential.

The only possible additional location, where $\frac{\partial\hat{S}(z)}{\partial t_{k}}$
may have singularities are the branch points. To exclude this possibility
have to show that, for all $j\geq0$, \[
\oint\partial_{t_{k}}\hat{S}(z)\sqrt{z-\lambda_{i}}^{j}=0.\]
 where the integration is over a small contour around the branch points
$\lambda_{i}$. We use the chain rule to obtain \[
\oint\partial_{t_{k}}\hat{S}(z)\sqrt{z-\lambda_{i}}^{j}=\frac{\partial}{\partial t_{k}}\left(\oint S(z)\sqrt{z-\lambda_{i}}^{j}dz\right)-\frac{\partial\lambda_{i}}{\partial t_{k}}\oint S(z)\frac{\partial}{\partial\lambda_{i}}\sqrt{z-\lambda_{i}}^{j}dz.\]
 The first integral on the RHS vanishes for all $j$ because of the
holomorphicity of $S(z)$ near the branch points. The second integral
vanishes trivially for $j=0$, while for $j>1$ the integral vanishes
because the $\hat{S}$ is holomorphic around $\lambda_{i}$. Finally
for $j=1$, the integral can be shown to vanish using the Krichever
equations (\ref{eq:Krichever}).

\chapter{The Richardson approach\label{sec:Richardson}}

As we have shown in the previous chapter the Whitham and Krichever
equations provide an unambiguous procedure for constructing the Schwarz
function. However, this approach becomes cumbersome when dealing with
a general algebraic Riemann surfaces, because the construction of
the normalized meromorphic differentials for such surfaces is a complicated
task. In this chapter we give an alternative approach for constructing
the Schwarz function $S(z)$ which avoids these difficulties. This
approach has been introduced by Richardson for the study of multi-bubble
dynamics\cite{142:Richardson,ST:why:becomes:o4}. We begin this section
by presenting the general formalism of Richardson approach, then we
shall present the solution of the one Miwa system (both genus-0 and
genus-1 cases), and finally, we will introduce the concept of {}``virtual
bubbles'' (bubbles of negative area) and interpret them using the
non-interacting fermion system.

\section{General formalism}

Richardson's approach for calculating the Schwarz function, $S(z)$,
avoids the need for construction of meromorphic differentials by mapping
the problem into another Riemann space where the meromorphic functions
(with a given singular behavior) are well known. For instance, an
$n$ sheet spectral surface of genus zero can be mapped to a cylinder
$-\infty<r<\infty$ and $0\leq\theta<2\pi$ where function should
be periodic in $\theta$. If, on the other hand, this surface has
a genus one topology, it may be mapped to a torus (rectangle whose
opposite sides are identified), where the meromorphic functions may
be combined from Weierstrass zeta functions. In the general case,
the spectral surface will be mapped onto a $g$-torus surface where
the meromorphic functions can be expressed in terms of the Riemann
$\Theta$-functions.

The mapping from the physical space ($z-$plane) to the fixed Riemann
space, which we shall call $u-$space, can be viewed as a conformal
mapping, $u(z)$, and the ST evolution amounts to finding the time
dependence of the inverse map $z(u)$. Since the action of the Schwarz
function, $S(z)$, in $z$ space is to reflect points outside the
bubble to ones inside, a similar reflection must be imposed in $u-$space,
and one may choose it to be $u\rightarrow-u$. This implies that the
Schwarz function is given by \begin{equation}
S(z)=\overline{z}\left(-u\left(z\right)\right).\label{z(-u(z))}\end{equation}
 Thus the image of the ST contour, $\bar{z}=S(z)$, in $u$-space
satisfies $\bar{u}=-u$.

\begin{figure}
\includegraphics[%
  width=8cm]{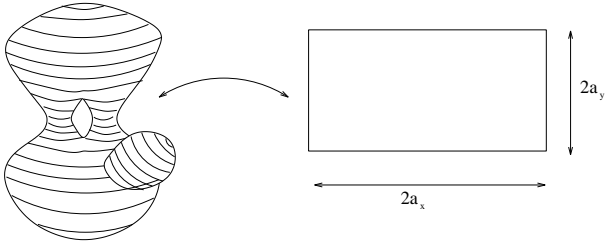}

\caption{A torus made up of three sheets, may be mapped to a standard torus
which is obtained by identifying opposite points on a rectangle.\label{cap:torusMapRec}}
\end{figure}

Consider the case where the Riemann surface, defined by the algebraic
equation $P(z,y)=0$ and comprised of $n$ copies of the complex plain
have the topology of a genus-1 surface. Let $u(z)$ maps this surface
to a torus, as illustrated in Fig.~\ref{cap:torusMapRec}. The inverse
mapping, $z(u)$ maps $u-$plane into the multiple values of the complex
plane that comprise the $z-$plane Riemann surface. Since on each
sheet of the Riemann surface one may identify a point corresponding
to infinity, $z(u)$ must have simple poles at several points, $u_{i}$.
We shall take $u_{0}$ to be the pre-image of the infinity of the
physical sheet, while $u_{i}$ for $i>1$ will be the pre-images of
the infinities on the unphysical sheets. The residues of these poles
will be denoted by $\alpha_{i}$. Thus the mapping $z(u)$ will be
a function of $2n$ parameters: $n$ associated with the locations
of the poles, $q_{i}$; $n-1$ associated with the residues;%
\footnote{Recall that the sum of residues of a meromorphic function on a Riemann
surface vanish, $\sum_{i}\alpha_{i}=0$, as can be seen by taking
a small contour around nonsingular point and deforming it to encircle
all the poles. Thus if there are $n$ poles, only $n-1$ residues
can be considered as independent variables.%
} and an additional constant term which we denote by $\beta$.

Given the mapping, $z(u)$, one may now compute $S(z)$ using Eq.~(\ref{z(-u(z))}).
$S(z)$ will have $n-1$ simple poles, outside the bubble, at points
which we denote by $q_{i}$. Sufficiently close to these poles $S(z)$
takes the form \begin{equation}
S(z)\sim-\frac{\mu_{i}}{z-q_{i}},\,\,\,\,\,\, i=1,2\cdots n-1,\label{S-poles}\end{equation}
 Thus $q_{i}$ and $\mu_{i}$, may be viewed as the Miwa variables'
locations and weights, respectively. In addition, the singular behavior
near infinity is \begin{equation}
S(z)\sim t'_{1}+\frac{t-\sum_{i=1}^{n-1}\mu_{i}}{z}.\label{S-infinity}\end{equation}

By identifying the behavior near the singular points of Schwarz function
from (\ref{z(-u(z))}), and comparing this behavior with (\ref{S-poles})
and (\ref{S-infinity}), one may construct a set of nonlinear equations
which express the variables $t$, $t'_{1}$ ,$\mu_{i}$ and $q_{i}$
in terms of parameters, $u_{i}$, $\alpha_{i}$, and $\beta$, of
the map $z(u)$. The solution of these equations enables one to express
the the conformal mapping parameters, $u_{i}$, $\alpha_{i}$, and
$\beta$, as function of the time $t$, $t'_{1}$ and the Miwa parameters.

Now, the Miwa variables, and $t'_{1}$ are constants of motion.%
\footnote{As follows from deformation of the contour in the integral representation
of the harmonic moments, (\ref{eq:HarmonicIntegral}), which imply
that the the harmonic moments, $t_{k}$, can be expressed in terms
of $\mu_{i}$ and $q_{i}$. Therefore the conservation of the harmonic
moments, $\frac{dt_{k}}{dt}=0$, induces conservation of the Miwa
variables: $\mu_{i}$ and $q_{i}$.%
} Thus keeping $q_{i}$, $\mu_{i}$ and $t'_{1}$ fixed, and letting
$t$ change, we obtain the time evolution of conformal mapping $z(u)$
which describes, in turn, the shape of the ST bubble as a function
of the time.

To be more concrete, let us construct the nonlinear equations relating
the conformal mapping parameters to the Miwa variables, and the times
$t'_{1}$ and $t$. Consider first the locations of the Miwa variables,
$q_{i}$. Since $z(u)$ has a simple pole at $u_{i}$, where $1\leq i\leq n-1$,
the Schwarz function $S(z)=\bar{z}(-u)$ would have poles at the images
of $-\bar{u}_{i}$. These points are on the exterior of the bubble
(due to the reflection property of $S(z)$ and since $u_{i}$, for
$1\leq i\leq n-1$ have been chosen to be the pre-images of the {}``non-physical''
infinities which can be viewed as located in the interior of the bubble)
and should be identified with the location of the Miwa variables,
Thus: \begin{equation}
q_{i}=z(-\bar{u}_{i})~~~\mbox{for}~~~i=1,...n-1.\label{eq:qi}\end{equation}
 The weight of the Miwa variables, $\mu_{i}$, can be extracted by
a local analysis of the behavior in the vicinity of the singularities.
The result is \begin{equation}
\mu_{i}=\left.\frac{\partial z}{\partial u}\right|_{-\bar{u}_{i}}\lim_{\delta u\rightarrow0}\left(\bar{z}(\bar{u}_{i}+\delta u)\delta u\right).\label{eq:mui}\end{equation}

Consider now the pole $u_{0}$ of $z(u)$ which is the pre-image of
the physical infinity. As follows from (\ref{S-infinity}), near the
physical infinity, $S(z)\rightarrow t'_{1}$, since also $S(z)=\bar{z}(-u)$,
we conclude that \begin{equation}
t'_{1}=\bar{z}(-u_{0}).\label{eq:t1}\end{equation}
 A local analysis near the physical infinity allows one to extract
the residue of the second term in (\ref{S-infinity}). The result
is \begin{equation}
t-\sum_{i=1}^{n-1}\mu_{i}=-\left.\frac{\partial z}{\partial u}\right|_{-\bar{u}_{0}}\lim_{\delta u\rightarrow0}\left(\bar{z}(\bar{u}_{0}+\delta u)\delta u\right).\label{eq:t}\end{equation}
 The equations above, are the general equations of the Richardson
approach.

\section{One Miwa variable}

\begin{figure}
\includegraphics[%
  width=13cm]{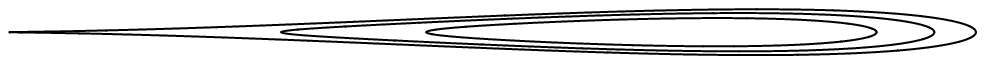}

\caption{Evolution of a 1-Miwa contour.\label{cap:Evolution1Miwa}.}
\end{figure}

We turn now to illustrate the approach outlined above for the case
of 1-Miwa variable. We consider first the simplest case of a spectral
surface of genus zero, comprised of two copies of the complex plain.
As explained above the $u$-space is chosen to be the cylinder $-\infty<\Re(u)<\infty$
and $0\leq\Im(u)<2\pi$, where a period of $2\pi i$ is implied. The
conformal mapping takes the form: \begin{equation}
z(u)=\alpha_{0}\left(\frac{1}{e^{u-u_{0}}-1}-\frac{1}{e^{u-u_{1}}-1}\right)+\beta,\label{eq:CM1MG0}\end{equation}
 where $\alpha_{0}$, $u_{0}$, $u_{1}$ and $\beta$, are the conformal
mapping parameters whose time dependence is to be determined. Notice
that the above form of $z(u)$ is indeed correctly defined on the
cylinder, namely it is periodic in $u\rightarrow u+2\pi i$. We can
use equations (\ref{eq:qi}-\ref{eq:t}) to express $q_{1},$ $t'_{1},$
$\mu_{1},$ and $t$ in terms the conformal mapping parameters. The
equations take the form: \[
q_{1}=\alpha_{0}\left(\frac{1}{e^{-\bar{u}_{1}-u_{0}}-1}-\frac{1}{e^{-\bar{u}_{1}-u_{1}}-1}\right)+\beta\]
 \[
t'_{1}=\bar{\alpha}_{0}\left(\frac{1}{e^{-u_{0}-\bar{u}_{0}}-1}-\frac{1}{e^{-u_{0}-\bar{u}_{1}}-1}\right)+\bar{\beta}\]
 \[
\mu_{1}=-|\alpha_{0}|^{2}\left(\frac{e^{-u_{1}}}{\left(e^{-\bar{u}_{1}-u_{1}}-1\right)^{2}}-\frac{e^{-u_{0}}}{\left(e^{-\bar{u}_{1}-u_{0}}-1\right)^{2}}\right)\]
 \[
t-\mu_{1}=-|\alpha_{0}|^{2}\left(\frac{e^{-u_{1}}}{\left(e^{-\bar{u}_{0}-u_{1}}-1\right)^{2}}-\frac{e^{-u_{0}}}{\left(e^{-\bar{u}_{0}-u_{0}}-1\right)^{2}}\right)\]
 Thus we have 4 unknowns and 4 equations. Solving these equations
for the parameters of the mapping in terms of $\mu_{1}$, $q_{1}$,
$t'_{1}$ (which are constants) and the time $t$, gives the evolution
of the conformal mapping (\ref{eq:CM1MG0}) as function of the time.
The contour of the bubble, ${\mathcal{C}}$ is the image of the line
$\bar{u}=-u$ which in our case is $\Re(u)=0$, i.e. \[
{\mathcal{C}}=\left\{ z\right|\left.z=z(u),\Re(u)=0\right\} \]
 The evolution of the contour $\mathcal{C}$ is demonstrated in Fig.
\ref{cap:Evolution1Miwa}. At some time $t^{*}$ the contour reaches
a cusp beyond which the equations do not possess a physical solution.%

We turn now to consider the one-Miwa problem on genus-1 surface. Here
we take $u$ space to be a torus realized as the rectangle. $-a_{x}\leq\Re(u)\leq a_{x}$
and $-a_{y}\leq\Im(u)\leq a_{y}$ with the usual identification of
opposite sides. Provided that we choose the ratio $\frac{{a_{y}}}{{a_{x}}}$
correctly we may find a isomorphism between the rectangle and the
spectral surface. In the one Miwa case such a mapping is provided
by: \begin{equation}
z(u)=\alpha_{0}\left(\zeta(u-u_{0})-\zeta(u-u_{1})\right)+\beta,\label{eq:CM1MG1}\end{equation}
 where $\alpha_{0}$, $u_{0}$, $u_{1}$ and $\beta$ are the parameters
of the mapping and $\zeta$ is the Weierstrass zeta function, which
is quasi-periodic with quasi-periods $2a_{x}$ and $2ia_{y}$:\[
\zeta(u+2a_{x})=\zeta(u)+2\eta_{x}\]
 \[
\zeta(u+2a_{y})=\zeta(u)+2\eta_{y}\]
 Thus $z(u)$ is also doubly periodic as it must be in order for it
to be defined on the rectangle with the opposite sides identified.%
\footnote{This mapping has been constructed by noticing that $z(u)$ must diverge
at two points on the rectangle, since $z$ diverges near the infinity
of both Riemann sheets comprising the spectral surface. The divergence
must be that of a simple pole for the mapping to be univalent. A periodic
function on the rectangle with two simple poles can be written as
a sum of Weierstrass $\zeta$ functions.%
}

Equations (\ref{eq:qi}-\ref{eq:t}) can be written explicitly as
in the genus zero case, and their solutions yields the time dependence
of the parameters of the conformal mapping (\ref{eq:CM1MG1}). The
evolution of the bubble contour can now be computed as the image of
$\bar{u}=-u$.

However, in the genus-1 case, $\bar{u}=-u$ defines two contours.
\[
\Re(u)=0,\,\,\mbox{ and}\,\,\,\Re(u)=a_{x}.\]
 The first solution is already familiar from the genus-0 case it is
the imaginary axis in $u$ plane which is mapped to the perimeter
of the first bubble. The second solution, $\Re(u)=a_{x}$ follows
from the torus topology where the line $\Re(u)=a_{x}$ is identified
with the line $\Re(u)=-a_{x}$. Consequently, we arrive at the conclusion
that one contour, ${\mathcal{C}}_{1}$, is the image of $\Re(u)=0,$
while the other contour, ${\mathcal{C}}_{2}$ is the image of $\Re(u)=a_{x}$:
\[
{\mathcal{C}}_{1}=\left\{ z\right|\left.z=z(u),\Re(u)=0\right\} ~~~~~~{\mathcal{C}}_{2}=\left\{ z\right|\left.z=z(u),\Re(u)=a_{x}\right\} \]
 We still have arbitrariness as for the choice whether the domain
$0<\Re(u)<a_{x}$ will be considered the exterior of the domain or
the interior. The choice must be made in such a way that the solution
becomes physical, if possible. In the 1-Miwa case the genus-1 solutions
are never physical, a point on which we will elaborate below.

\section{Virtual bubbles\label{sub:virtual}}

We turn now to explain why the genus-1 solution found for the 1-Miwa
case are non physical. As we will show this non physicality may be
understood as a situation where one of the bubble is of negative area.
We refer to such bubbles as virtual bubbles, and show that if follows
from the fact that the two closed contours described by $S(z)=\bar{z}$
are located on a different sheets of the Riemann surface.

\begin{figure}
\includegraphics[%
  width=15cm]{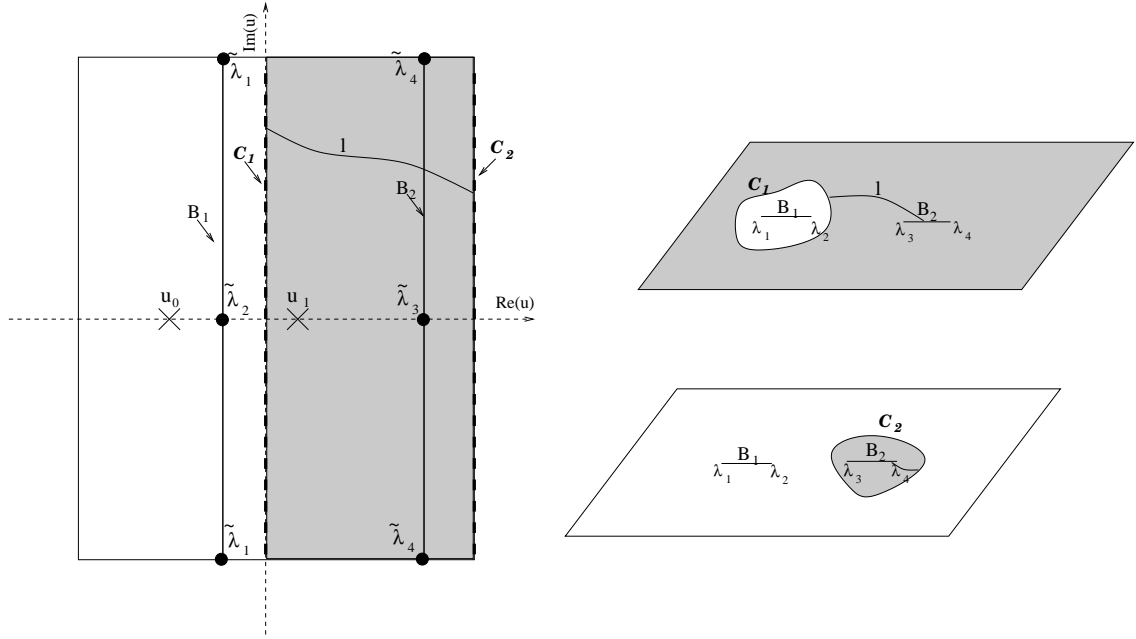}

\caption{Schematic drawing of torus to the left and the two-sheeted Riemann
surface to the right. Points $u_{0}$ and $u_{1}$ on the torus are
mapped to the infinities on the upper and lower sheets respectively,
while $\tilde{\lambda_{i}}$ are mapped to the branch points, $\lambda_{i}$.
The branch cuts are $B_{i}$ drawn on the two sheeted Riemann surface
as well as on the torus. The two contours are $C_{1}$ and $C_{2}$.
A line connecting $C_{1}$ and $C_{2}$ ($l$ for example) must cross
one of the branch cuts, which implies that the two contours lie on
different sheets of the Riemann surface.\label{cap:TorusPoints}}
\end{figure}

\begin{figure}
\includegraphics[%
  width=8cm]{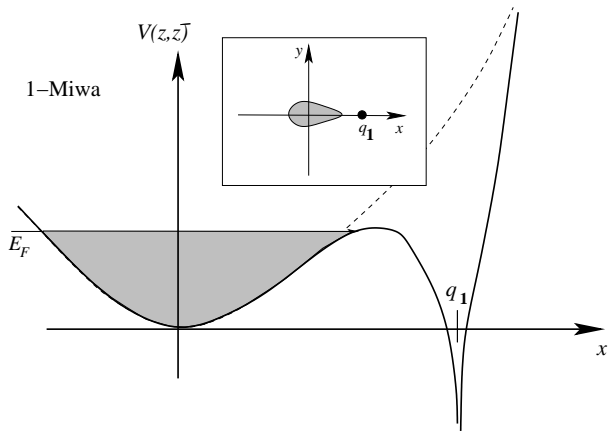}

\caption{Schematic drawing of a Fermi sea filling up the 1-Miwa potential.
The dotted line shows the potential with no Miwa variable. The Miwa
variable introduces an infinite well at point $q_{1}$ (the location
of the Miwa variable). The inset shows the shape of the droplet in
this case. The cusps forms when the Fermi sea reaches a saddle point
of the potential. \label{1MiwaPotential}}
\end{figure}

If the two contours, ${\mathcal{C}}_{1}$ and ${\mathcal{C}}_{2}$
would lie on the same sheet then we would refer to this sheet as the
physical plane, and consider the other sheets only as mathematical
surfaces on which we extend the definition of $S(z)$. This would
also determine the choice of domain on the rectangle that would represent
the exterior of the bubbles - we would choose this domain as to contain
the infinity which is on the same sheet as both the bubbles. This
domain, which is on the exterior of the bubbles, naturally does not
contain any branch points of the Schwarz function. To see that indeed
the solutions that we have obtained are non-physical, we show that
no matter which of the two possible domains we choose, we would have
branch points of the Schwarz function in it. This is seen by the symmetry
of $z(u):$\[
z\left(\frac{u_{0}+u_{1}}{2}+u\right)=z\left(\frac{u_{0}+u_{1}}{2}-u\right),\]
 which follows from the anti-symmetry properties of the Weierstrass
$\zeta$ function $\zeta(-u)=-\zeta(u)$. The four fixed points of
this symmetry on the torus are the branch cuts (where the mapping
$z(u)$ is not conformal, hence $z'(u)=0$), these are the points:\[
u_{c},u_{c}+ia_{y},u_{c}+a_{x},u_{c}+a_{x}+ia_{y},\]
 where $u_{c}=\frac{u_{0}+u_{1}}{2}$. The consequence of the fact
that these are the branch points is that it does not matter whether
we choose $0\leq\Re(u)\leq a_{x}$ as the exterior of the bubble or
its complement, there will always be a branch point on the exterior
of the bubble. This is not consistent with two physical bubbles, where
one has the branch cuts inside each one of the bubbles. In fact in
the genus-1 1-Miwa case, one must pass through the branch cut in order
to get from the physical sheet to a point on the unphysical bubble.
This is demonstrated in Fig. \ref{cap:TorusPoints}. In the fermionic
picture (see Fig. \ref{1MiwaPotential}) we may interpret the in-existence
of physical solutions for the 1-Miwa case by the fact that there are
no minima of the potential in which another bubble may form (there
is one minimum which is occupied by the original bubble).

\chapter{Multi-bubble solutions\label{sec:Multi-Droplet}}

In the previous section we have solved the 1-Miwa case and demonstrated
that the generalization to the genus-1 solutions does not solve the
problem of going beyond the cusp since it generates non-physical solutions
associated with virtual bubbles. Employing the noninteracting fermions
picture we saw that this behavior signals the breakdown of the ST
description of the problem. 

\begin{figure}
\includegraphics[%
  width=8cm]{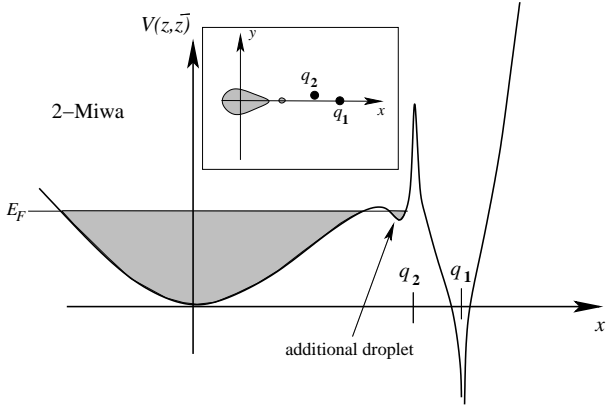}

\caption{Schematic drawing of Fermi sea filling up the two-Miwa potential.
an additional droplet now forms at the second minimum of the potential.
The cusp that would have formed near the saddle between the two minima
is regularized by this droplet. $q_{1}$ is the Miwa pole while $q_{2}$
is the Miwa zero. \label{2MiwaPotetial}}
\end{figure}

We may resolve this problem by introducing an additional weak Miwa
variable with negative weight which generate a local minimum of the
potential near the cusp, as illustrated in Fig.~(\ref{2MiwaPotetial}).%
\footnote{The potential is given by Eq. (\ref{eq:V(z,zb)}), $V(z,\bar{z)}=|z|^{2}-\sum_{k}t_{k}z^{k}+c.c.$
In the one Miwa case $t_{k}=-\mu_{1}q_{1}^{-k}/k$ yields $V(z,\bar{z})=|z|^{2}+\mu_{1}\log|1-z/q_{1}|^{2}$
This potential diverges to $-\infty$ as $z\rightarrow q_{1}$. An
additional Miwa variable with negative weight, $t_{k}=-\mu_{1}q_{1}^{-k}/k+\mu_{2}q_{2}^{-k}/k$
gives the following potential $V(z,\bar{z})=|z|^{2}+\mu_{1}\log|1-z/q_{1}|^{2}-\mu_{2}\log|1-z/q_{2}|^{2}$
which diverges to $+\infty$ as $z\rightarrow q_{2}$. The we may
arrange $q_{2}$ and $\mu_{2}$ to have a local minima as illustrated
in Fig.~\ref{2MiwaPotetial}.%
} Thus as we increase the Fermi level, at some point, fermions will
tunnel to the other minimum of the potential, and an additional droplet
will form. In this situation we expect the additional droplet to be
physical, see illustration in Fig.~\ref{2MiwaPotetial}%
\footnote{On a mathematical level, the appearance of the virtual droplet in
the genus-1 1-Miwa was a consequence of the symmetry on the torus
as explained in the previous chapter. The introduction of an additional
Miwa variable breaks this symmetry.%
}%

In this section we shall present the solution of the 2-Miwa variable
problem. Namely, we look for the solution of the problem whose Schwarz
function on the exterior of the bubble has the singular structure:
\[
S(z)\sim\frac{t-\mu_{1}+\mu_{2}}{z}+t'_{1}-\frac{\mu_{1}}{z-q_{1}}+\frac{\mu_{2}}{z-q_{2}}\]
 Following the Richardson approach one can solve this problem. In
what follows we describe this solution.

By counting the number of points for which $z'(u)=0$, one sees that
there are $3$ branch cuts in this case. The number of copies of the
complex plane of which the spectral surface is composed is $3$ (this
can be seen from the fact that $z(u)$ diverges at three different
points on the torus, which correspond to infinities on each one of
the copies of the complex plane). This is consistent with the Riemann-Hurwitz
formula which relates the number of branch points, $B$, the number
of sheets, $n$, and genus $g$ of the Riemann surface: \[
g=1+\frac{B}{2}-n.\]
The difficulty which is introduced by taking a $3$-sheeted genus-1
Riemann surface is that it is not straightforward to write down the
holomorphic differentials in this case. Thus using the Krichever equations
to solve the evolution is more difficult .

\subsection*{Early stages: genus-0 solution}

\begin{figure}
\includegraphics[%
  width=3cm]{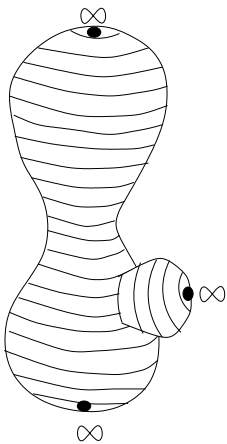}

\caption{The Riemann surface of genus-0 2-Miwa aggregate. The Riemann surface
is composed of three sheets (the points at infinity on each of the
sheets are shown). The third sheet is schematically shown on the graph
as an additional surface tacked on the Riemann sphere. \label{cap:Riemann2Miwa}}
\end{figure}

The early stages of the evolution, in the noninteracting fermion picture,
can be associated with the situation were the Fermi energy is sufficiently
low, such that the additional minimum of the potential is left unoccupied.
This means that a mapping $u(z)$ is from the cylinder $-\infty<r<\infty$,
$0\leq\theta<2\pi$ to a genus-0 surface composed of 3 copies of the
complex plain (1 plus the number of Miwa variables), as illustrated
in Fig.~(\ref{cap:Riemann2Miwa}).%

\noindent Thus the mapping, $z(u)$, from the cylinder to the spectral
surface is given by: \[
z(u)=\frac{\alpha_{0}}{e^{u-u_{0}}-1}+\frac{\alpha_{1}}{e^{u-u_{1}}-1}-\frac{\alpha_{0}+\alpha_{1}}{e^{u-u_{2}}-1}+\beta\]
 where $\alpha_{i}$, $u_{i}$ and $\beta$ are parameters whose time
evolution is to be determined. The extraction of this time dependence
using Richardson's approach, shows that the bubble develops a cusp
as in the one-Miwa case. This cusp, in the noninteracting fermion
picture, appears when the Fermi level reaches the local maximum of
the potential, see Fig.~(\ref{1MiwaPotential}).

\subsection*{\noindent Mid-stage: genus-1 solution}

\noindent The lesson from the Gurevich-Pitaevskii scenario, is that
at some point, before the cusp, we must allow for a topological transition
to take place, i.e. let the genus-1 solution take over, by forming
an additional bubble. In the noninteracting fermion picture it is
the time where the Fermi level reaches the local minimum of the potential.
A smooth transition between the two solutions can happen only if the
area of the second bubble vanishes in the genus-1 solution.

\noindent The mapping during this time is from the torus to the spectral
surface of genus-1 topology \[
z(u)=\alpha_{0}\zeta(u-u_{0})+\alpha_{1}\zeta(u-u_{1})-(\alpha_{0}+\alpha_{1})\zeta(u-u_{2})+\beta.\]
 where as before the evolution of the contour can be found by applying
the Richardson approach to extract the time dependence of the coefficients
$\alpha_{i}$, $u_{i}$ and $\beta$, keeping the normalization \[
\Re\left(\oint S(z)dz\right)=0,\]
 for every cycle on the Riemann surface, where $S(z)=\bar{z}(-u(z))$.
Details of this calculation can be found in Appendix \ref{sec:Appendix:-The-2-Miwa}

\subsection*{\noindent Late stage: genus-0 solution}

\noindent Increasing further the time causes another topological transition
to take place. This time one switches from a genus-1 solution back
to a genus-0 solution (in analogy to the GP scenario in KdV). The
reason for this transition is the merging of the two bubbles as the
Fermi surface becomes higher than the sadle points between the two
minima of the potential. The two bubbles merge through a cusp, a treatment
of the shape of the cusp is given in Appendix \ref{cha:Merging-Droplets.}.

\chapter{Tip-splitting}

\begin{figure}
\includegraphics[%
  width=5cm]{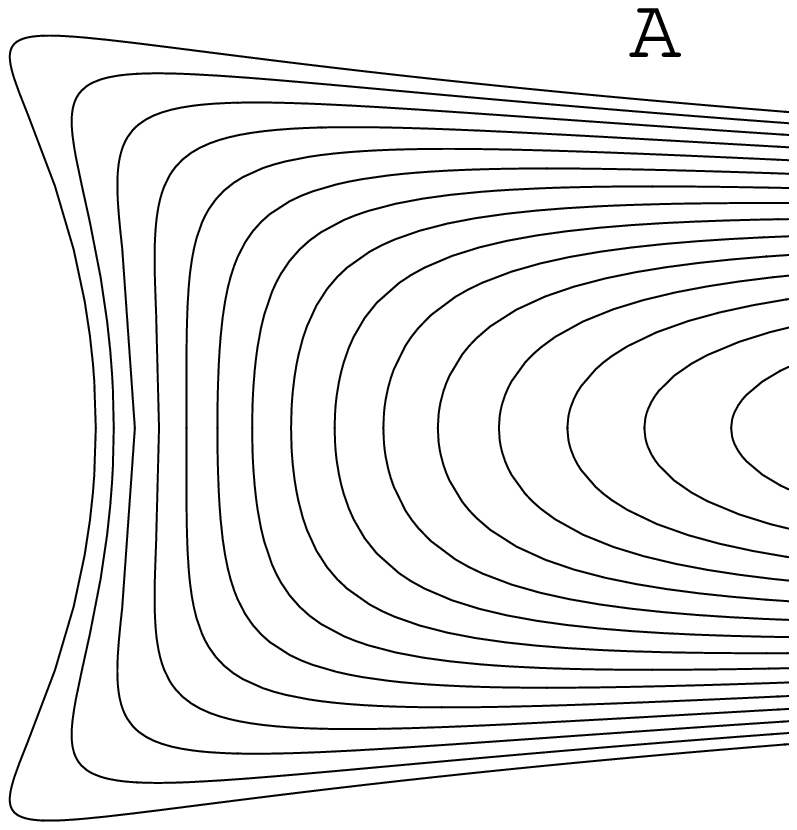}\includegraphics[%
  width=5cm]{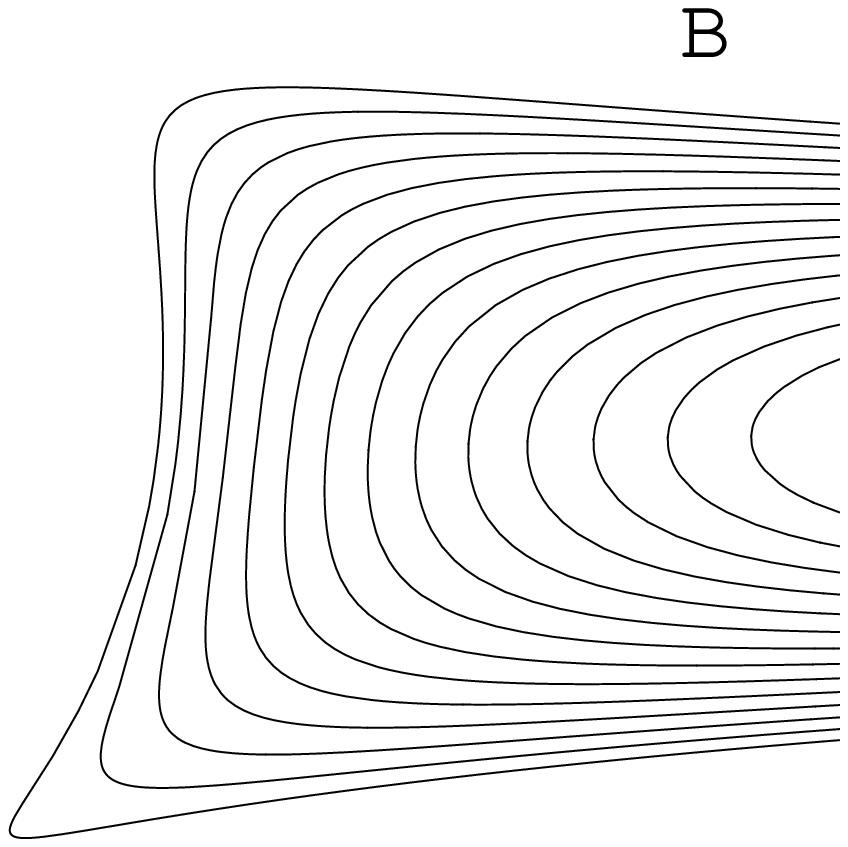}

\caption{The evolution of tip splittings in the Saffman-Taylor problem at
zero surface tension. The contours represent snapshots of the evolution
as function of the time. The evolution given in (\ref{tip-form}),
is characterized by one parameter, $\phi$, which controls the asymmetry
between the two generated fingers. Panel A shows the symmetric evolution,
while panel B represents a typical asymmetric behavior. \label{cap:tipasymtipsym}}
\end{figure}

In this chapter we focus our attention on the tip-splitting shape
which is formed after dispersive regularization of the cusp. We are
lead to study tip-splitting since one expects that an understanding
of the global structure of the ST bubble can be deduced from properties
of the basic elements of the growing process, such as tip-splitting
and side-branching. Indeed, studies of theoretical models of diffusion
limited aggregation\cite{Witten:Sander}, and their generalizations
(e.g. the dielectric breakdown model) demonstrate that the fractal
dimension of the corresponding patterns is related to characteristics
of the tip-splitting events\cite{170:Halsey} (side-branching in these
models is negligible). The central result of this chapter is a formula
which describes the evolution of tip splitting in time: \begin{eqnarray}
\frac{z(s,t)}{L}=s+u_{\phi}(t)+\frac{v_{\phi}(t)}{w_{\phi}(t)\pm\sqrt{s}}.\label{tip-form}\end{eqnarray}
 Here $z=x+iy$ is a complex coordinate, $L$ is the scale of the
tip-split, $s>0$ parametrizes the curve, and $u_{\phi}(t)$, $v_{\phi}(t)$
and $w_{\phi}(t)$ are functions of the time, $t$. These functions
which will be calculated in what follows, depend on a single parameter,
$\phi$, governing the asymmetric shape of the evolution. In Fig.~\ref{cap:tipasymtipsym}
we depict contours obtained from (\ref{tip-form}) which represent
snapshots of the tip-splitting evolution as function of the time.

To begin with, it is important first to clarify what precisely we
mean by a {}``tip''. In the ST problem with zero surface tension
there is a natural way to define a tip. The reason is that in this
case, for almost any initial bubble shape, fingers develop into cusps
(see, e.g., panels A and B in Fig.~\ref{cap:tipasymtipsym}). Here
we focus our study at the solution of the ST problem well after the
merging took place, at the period of the tip-splitting scenario illustrated
in Fig.~\ref{cap:tipasymtipsym}. We use the 2-Miwa potential introduced
in Chapter \ref{sec:Multi-Droplet}, to obtain such a tip-splitting
solution.

For actual calculation of the contour's evolution it will be convenient
to use a conformal mapping which maps the exterior of some {}``source
domain'' (in $\zeta$-plane) to the exterior of the physical bubble,
the {}``target domain'' (in $z$-space). The source domain is usually
taken to be the unit circle, however, we found it more convenient
to use a bubble with a cusp. The advantage in using this source domain
is that we can choose the mapping to be non-trivial only around the
tip, while everywhere else it would be approximately proportional
to the identity map. 

We will study the evolution near the tip in the case where the weight
of one of the Miwa variables, is small compared to the other. This
suggests choosing the source domain to be the one Miwa-variable bubble
at the point where a cusp is formed. The Riemann surface in this case
is composed of two sheets as described above. If we assume this surface
to be of genus zero, then by the Riemann-Hurwitz theorem, the number
of branch points is two. Thus the Riemann surface is given by the
polynomial equation $R^{2}=(\zeta-\lambda_{1})(\zeta-\lambda_{2})$.
A function defined on this surface (the Schwarz function in particular)
contains a branch cut which extends from point $\lambda_{1}$ to the
point $\lambda_{2}$. An example of such a function is $R(\zeta)$. 

The 1-Miwa bubble is characterized by four parameters $t_{1},t,q$
and $\mu$, since the location of the branch points depends on these
parameters, one can calculate the area of the bubble and its first
harmonic moment, $t_{1}$, from the location of the branch points
$\lambda_{1}$ and $\lambda_{2}.$ We will fix $\mu=-q=1$ and treat
$\lambda_{1}$ and $\lambda_{2}$ as the parameters which describe
the bubble. If we also demand that the bubble is at the moment where
the cusp forms, we can characterize the bubble by a single parameter,
say $\lambda_{1}$. We assume that $\lambda_{1}<\lambda_{2}$ and
that $\lambda_{1}$ is of order $\delta,$ where $\delta\ll1$. This
assumption implies that the global bubble shape is dominated by the
cusp. Then the bubble area is of order $\delta^{3}$ while the first
harmonic moment, $t_{1},$ is $1+O(\delta^{2})$. The Schwarz function,
$\sigma(\zeta)$, of such a bubble is: \[
2\sigma(\zeta)=\left(t_{1}-1\right)+\frac{t-1}{\zeta-\bar{t}_{1}}-\frac{1}{\zeta+1}+\frac{(t_{1}+1)(\zeta-\lambda_{1})R(\zeta)}{(\zeta-t_{1})(\zeta+1)},\]
 as can be ascertained by examining this function's singular structure
and that it satisfies the unitarity condition (\ref{unitarity}).
The fact that this function describes a bubble with a cusp can be
checked by considering the behavior of the solution of $\sigma(\zeta)=\bar{\zeta}$
near $\lambda_{1}$ (where the cusp is located). 

Up till now we have characterized the {}``source domain''. We would
like, now, to specify the physical bubble, or the {}``target domain''
associated with two Miwa variables. For this purpose we must give
the mapping between the source and target domains. This mapping is
taken to be: \begin{equation}
z(\zeta)=c_{1}\left(\zeta+\frac{\alpha}{2}\frac{R(\theta)-R(\zeta)}{\zeta-\theta}+\beta R(\zeta)\right)+c_{2},\label{mapping}\end{equation}
 where $c_{1}$, $c_{2}$, $\alpha$, $\beta$, and $\theta$ are
parameters of the mapping and $R(\zeta)\equiv\sqrt{\zeta-\lambda_{1}}\sqrt{\zeta-\lambda_{2}}$.
This mapping can be considered as a mapping from the 1-Miwa Riemann
surface to the 2-Miwa Riemann surface. The mapping has singularities
at the infinities on each of the sheets of the Riemann surface associated
with the source domain, which are mapped to infinities on different
sheets on the target domain, as well as at the point $\theta$ on
the unphysical sheet. This point, $\theta,$ is mapped to an infinity
on a third sheet of the target Riemann surface. Thus (\ref{mapping})
indeed maps a two-sheeted Riemann surface, which is associated with
a 1-Miwa bubble, to a three sheeted Riemann surface. 

That, indeed, the target Riemann surface is associated with a 2-Miwa
variable bubble, can be deduced from the singular structure of the
Schwarz function of the target domain. The latter is given by $S(z)=\bar{z}\left(\sigma\left(\zeta\left(z\right)\right)\right)$,
where $\zeta(z)$ is the inverse map of $z(\zeta)$, and $\sigma(\zeta)$
is the Schwarz function of the source domain. 

Having $S(z)$ one may extract all constants of motion (Miwa variables,
$q_{1}$, and $q_{2}$; Miwa weights, $\mu_{1}$ and $\mu_{2}$; and
$t_{1}$), and the area $t$, from its singular structure. These will
be expressed as functions of the parameters of the mapping (\ref{mapping})
and $\lambda_{1}$. Solving these relations one may express the parameters
of the mapping as functions of the time and thus obtain the evolution
of the contour. We take $\mu_{1}=-q_{1}=1$ by fixing $c_{1}$ and
$c_{2}$. Thus the parameters whose time evolution is to be determined
are $\alpha$, $\beta$, $\theta$ and $\lambda_{1}$. 

To obtain reasonably simple equations, we expand all quantities in
orders of $\delta$ and take the leading order. Let us assume that
for some initial moment, $t^{(0)}$, around the formation of the tip,
$\lambda_{1}$ assumes the value $\lambda$, to leading order in $\delta$.
We now make the following scaling ansatz: $\alpha\sim\delta^{4}$,
$\beta\sim\delta^{3}$, $\nu\equiv\lambda_{1}-\theta\sim\delta^{3}$
and $\delta\lambda\equiv\lambda_{1}-\lambda\sim\delta^{3}$. Then
the equations we obtain, to leading order in $\delta$, for the constants
of motion $q_{2}-\lambda$ and $\mu_{2}$ are: \begin{eqnarray}
q_{2}-\lambda=\beta+\delta\lambda-\bar{\nu}-\frac{\alpha\sqrt{-\lambda}}{\sqrt{\nu}+\sqrt{\bar{\nu}}}~~~~~~~\label{first}\\
\mu_{2}=-2\bar{\alpha}\sqrt{-\lambda}\sqrt{\nu}\left(1-\frac{\alpha\sqrt{-\lambda}}{2\sqrt{\bar{\nu}}\left(\sqrt{\nu}+\sqrt{\bar{\nu}}\right)}\right)\end{eqnarray}

Let us now define $T_{1}=t_{1}-\left(1+3\lambda^{2}-7\lambda^{3}+\frac{33}{2}\lambda^{4}\right)$
(the difference between the first harmonic moment of the target bubble
and a source bubble with $\lambda_{1}=\lambda$ to order $O(\delta^{5})$),
and similarly $T=t-(-4\lambda^{3}+18\lambda^{4}-63\lambda^{5})$.
With these definitions, analysis of the singularities of $S(z)$ yields:
\begin{eqnarray}
T=2\Re(\alpha)\lambda+4\beta\lambda^{2}-12\lambda^{2}\delta\lambda,\\
T_{1}=\Re(\alpha)-2\beta\lambda+6\lambda\delta\lambda.\label{last}\end{eqnarray}

The solution of Eqs.~(\ref{first}-\ref{last}) give the parameters
$\alpha,$ $\beta,$ $\nu$ and $\delta\lambda$ in terms of $q_{2},$
$\mu_{2}$, $T$ and $T_{1}$, which define, in turn, the conformal
mapping of the contour on the target space as function of the time.
To write down the solution of these equations it will be convenient
to define a shifted rescaled time $\delta t\equiv-\frac{T+2T_{1}\lambda}{4\mu_{2}^{3/4}\sqrt{-\lambda}}$
and introduce two functions $\xi$ and $\eta$ which satisfy the nonlinear
equations: \[
\left(\frac{\sqrt{2}\delta t}{\sqrt{\xi}}+\frac{\delta t^{2}}{2\xi^{2}}\right)\left(\eta+\xi\right)=1;~~\eta=\left(\frac{\phi}{\sqrt{\xi}-\frac{\delta t}{\sqrt{2}\xi}}\right)^{2},\]
 where $\phi=\Im\left(\frac{q_{2}}{\sqrt{\mu_{2}}}\right)$ is the
asymmetry parameter. The solution of these equations gives $\eta$
and $\xi$ as functions of the rescaled time, $\delta t$, and $\phi$.
With the help of these functions we may write the solution of Eqs.~(\ref{first}-\ref{last})
as: \begin{eqnarray*}
\alpha=\frac{\mu_{2}^{3/4}\delta t}{\sqrt{-\lambda}}\left(-1+i\sqrt{\frac{\eta}{\xi}}\right),~~~~~~~~~~~~~~~~~~~~~~~~~~~~~\\
4\beta=3\Re(q_{2})-3\lambda-\frac{3}{2}(\eta-\xi)-\frac{3\delta t}{\sqrt{2\xi}}-\frac{T_{1}}{4\lambda}+\frac{T}{8\lambda^{2}},\\
\nu=\frac{\xi-\eta}{2}+i\sqrt{\xi\eta},~~~~~~~~~~~~~~~~~~~~~~~~~~~~~~~~~~~~~~~~\\
4\delta\lambda=3\Re(q_{2})-3\lambda-\frac{3}{2}(\eta-\xi)-\frac{3\delta t}{\sqrt{2\xi}}+\frac{T_{1}}{4\lambda}-\frac{T}{8\lambda^{2}}.\\
\end{eqnarray*}

Given this time dependence of the parameters of the conformal mapping
we are in a position to describe the contour dynamics in the vicinity
of the tip-splitting. For this purpose it is sufficient to focus on
the image of the source domain around the cusp. The shape of the source
domain near the cusp is given by the universal form $y=A\delta x^{3/2}$,
were $\delta x\equiv x-\lambda_{1}$, and $A$ is some constant. Close
enough to the cusp $y\ll\delta x,$ and therefore we may assume that
$y\simeq0$ (this assumption can be proved to be consistent with the
$\delta$ expansion performed here). Thus one has to find the image
of the ray $x>\lambda_{1}$ under the mapping $z(\zeta)$ to leading
order in $\delta$. The result is given by Eq.~(\ref{tip-form}),
where $L=\sqrt{\mu_{2}}$ and the functions $u_{\phi}(t)$, $v_{\phi}(t)$,
and $w_{\phi}(t)$ are: \begin{eqnarray*}
u_{\phi}(t)=\frac{\xi-\eta}{2}-\frac{\delta t}{\sqrt{2\xi}},\\
v_{\phi}(t)=\delta t\left(1-i\sqrt{\frac{\eta}{\xi}}\right),\\
w_{\phi}(t)=\frac{\xi-\eta}{2}+i\sqrt{\eta\xi}.\end{eqnarray*}
 The above equations together with (\ref{tip-form}) describe the
evolution of a tip-splitting of the ST problem at zero surface tension
(Fig.~1). The form of the tip-splitting depends on a single parameter,
$\phi$, which controls the asymmetric shape of the evolution. Since
our derivation of the tip-splitting formula makes use only of local
properties near the tip, it is suggestive that this evolution is universal.
Namely, the tip splitting evolution is characterized by one parameter,
$\phi$, independent of the shape of the bubble on large scales. The
shortcomings of our result is that it does not include the influence
of surface tension. Therefore it breaks down after a short time due
to the formation of cusps. The procedure of dispersive regularization
must be repeated again and again in order to obtain large aggregates.

\section{Tip-Splitting Scaling and Fractal Dimension}

As noted above, tip-splitting is believed to be important in the understanding
of the fractal properties of the ST aggregate. Indeed, Halsey has
argued that the DLA may be viewed as repeated tip splitting events
which give rise the branched growth of the fractal\cite{170:Halsey}.
This prompts us to study the tip-splitting we have obtained in this
light. In particular, we want to understand what are the scaling properties
of the tip splitting solution. In order to do this, we first note
that the length scale of the tip-split, namely $L\equiv\sqrt{\mu_{2}}$
scales as $\delta^{3}$ while the shifted time, $T$, scales as $\delta^{5}$.
This suggests that the shape of the tip split would depend on time
through the parameter $\frac{\tau}{L^{5/3}}$, where $\tau\equiv T+2T_{1}\lambda$.
Indeed, the parametric form of the tip split shape (\ref{tip-form})
may be rewritten as:\begin{equation}
\left(\begin{array}{c}
x(t,s)\\
y(t,s)\end{array}\right)=L\vec{f}\left(s,\frac{\tau}{L^{5/3}}E,\phi\right),\label{eq:TipScaling}\end{equation}
where $E=\sqrt{-\frac{\mu_{2}^{1/6}}{\lambda}}$ does not scale with
$\delta$ as can be straightforwardly checked. We will assume that
parameters which do not scale with $\delta$ depend only on the large
scale structure of the bubble. $E$ may be interpreted as a parameter
which determines the fraction of the area which is added to the vicinity
of the tip split. $E$ will be effected by how much the large scale
structure of the bubble screens the tip-split area, and on how much
the tip-split protrudes from the bubble. Equation (\ref{eq:TipScaling})
suggests that the lengths scale as $\tau^{3/5}.$ In DLA clusters,
the linear size of the aggregate scales with the number of particles
as $t^{\frac{1}{D}},$ where $D=1.71,$ is the fractal dimension of
the aggregate. 

We come to the conclusion that from the naive scaling argument, the
fractal dimension of the bubble would be $5/3=1.66\ldots,$ close
to the measured fractal dimension of the DLA aggregate. However, the
linear size which appears in equation (\ref{eq:TipScaling}), $L$,
is the size of a single tip split, while in the fractal, it is the
overall size of the cluster that would scale as $t^{\frac{1}{D}}$.
Thus in order to relate the scaling of a single tip-split to that
of the fractal we have to assume that many small tip-split events
may be coarse grained into a large one. Then if we assume self-similarity,
the relation between the time $\tau$ and the size $L$ would be retained,
also for tip-splits whose size is of the order of the whole aggregate.
In particular, the radius of the bubble which is of order of the strongest
branch would scale as $t^{3/5}$, where $t$ is the area of the droplet.
Thus we would obtain $5/3$ for the fractal dimension of the aggregate.
Apart from the assumption of self-similarity, to obtain this result
we have also assumed that the parameters $E$ and $\phi$ do not change
in the renormalization process, as they do not scale with $\delta$.
A more rigorous treatment, which relaxes this {}``mean filed'' assumption
may take into account different effects neglected here, in a way that
would produce a more accurate estimate of the fractal dimension.

\chapter{Discussion}

In this thesis we made use of the relation between the quantum Hall
effect, the integrable hierarchy of the 2DTL, and the idealized ST
problem in order to describe its dynamics. For this purpose we have
used the dispersive regularization method of soliton theory. This
method, in essence, is an extension of the Whitham equations to spectral
surfaces of non-trivial topology (nonzero genus). This approach amounts
to the generalization of the idealized ST problem to the multi-bubble
case. The bubbles share the same pressure, and their evolution is
dictated by the requirement that the harmonic moments are fixed, and
the total area of the bubbles is proportional to the time.

In the picture of noninteracting fermions, this regularization has
a simple interpretation: Consider a potential with multiple minima,
and let us fill the potential with fermions injected adiabatically
at the lowest minimum of the potential. We may consider now two procedures
of taking the semi-classical limit. In the first one $\hbar\rightarrow0$
and then time is sent to infinity (notice that here time has nothing
to do with the area of the droplet which is determined by the number
of fermions). In this case the fermions cannot tunnel into other minima,
and once the filling is such that the fermion droplets reaches the
saddle point, the dynamics cannot be continued adiabatically any longer.
In the second limiting procedure, we first take time to infinity and
only then $\hbar\rightarrow0$, with this order of limits, tunneling
is allowed at any step, and fermions may form multiple droplets. These
multiple droplets are the multi-bubble solutions of the idealized
ST problem.

As we show, in the multi-bubble solutions, situations may occur where
bubbles of negative area are developed. One may try, in this case,
to focus on the dynamics of only the physical bubbles. However, this
viewpoint implies that pressure in the exterior domain is not harmonic
and that the harmonic moment of the physical bubbles are, in general,
not conserved. Therefore virtual bubble solutions do not represent
a clear regularization of the ST evolution, and their physical meaning
is still obscure.

\begin{figure}
\includegraphics[%
  width=8cm]{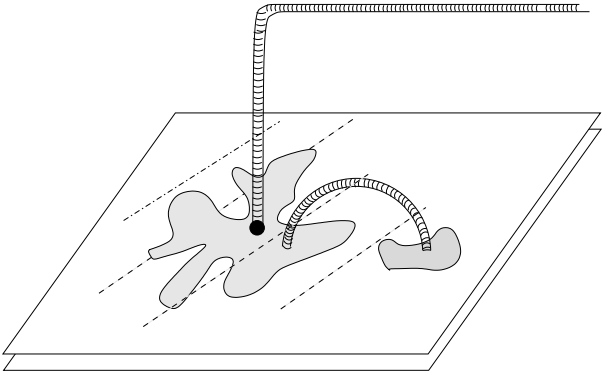}

\caption{Multi-bubble dynamics realized as an ST problem in which the non-viscous
fluid is injected at more than one point into the Hele-Shaw cell.
\label{STsetup2}}
\end{figure}

Nevertheless, one may choose a set of initial conditions in which
the integral representation of the $\tau$-function converge. In this
case, all bubbles which may appear are real (have positive area).
The dynamics of this generalized system may be viewed as a situation
in which pumping of the non-viscous fluid is done simultaneously into
more than one point of the Hele Show cell, as illustrated in Fig.~\ref{STsetup2}.

The main problem is to what extent this type of solutions may be viewed
as regularization of the idealized ST problem. An evolution where
new bubbles appear and merge with the original bubble is clearly not
equivalent to the evolution of the ST problem with finite small surface
tension, since in the latter only one bubble exist. Nevertheless,
one may hope that the global structure of both problems is statistically
the same, analogous to the arguable equivalence between some statistical
features (e.g., the fractal dimension) of diffusion limited aggregation
(DLA) and the ST with small surface tension. If this is indeed, the
situation, then one may regard the multi-bubble model as an integrable
regularization of the ST problem. This model can be viewed as a hybridization
of DLA and the idealized ST dynamics, since the new bubbles which
appear near the tips of the fingers may be regarded as DLA particles
sticking to the aggregate.

Yet, one can easily imagine simple examples where the multi-bubble
solutions clearly cannot be considered as a regularization. For instance,
the merging of two big bubbles, or situations where bubbles engulf
large droplets of viscous fluid. To avoid these situations one must
find a method which eliminates these cases, but nevertheless, allows
one to choose a generic initial condition. Such a method would amount
to finding appropriate potentials, or sets of harmonic moments $\{ t_{k},\bar{t}_{k}\}$
such that the small droplets would have some characteristic length
scale. A different approach would be to start from a well developed
DLA or ST bubble and take the inverse time evolution, which is dispersively
regularized. A weak condition for the dispersively regularized evolution
to be in the same class as the DLA or ST evolution would be that the
aggregate retains its fractal dimension as its area is decreased.

A different route is to focus on the local properties of the ST singularities.
This viewpoint is motivated by studies of DLA and its generalization,
the dielectric breakdown model, which demonstrate that the fractal
dimension of the corresponding patterns is related to properties of
individual tip-splitting events (side-branching in these models is
negligible). Indeed, tip splitting appears, generically, after the
merging of two bubbles (see panel E. in Fig. \ref{cap:Evolution-of-the}),
and their form is believed to be universal. We studied the tip-splitting
which appears in the two-Miwa case and saw that its scaling properties
is congruent, through a naive argument, with a fractal of dimension
$5/3=1.66\ldots.$, while the measured fractal dimension of a DLA
clusters is $1.71$. One thus may hope that by a more rigorous treatment,
which nevertheless makes use of the ideas of the naive argument, would
produce results closer to the measured fractal dimension. 

\appendix

\chapter{Proof of 2DTL for $\tau_{N}$\label{cha:Proof-of-2DTL}}

In this Appendix we will derive Eq. (\ref{eq:1st2DTL}) showing that
$\tau$-function satisfy the 2DTL equations, our treatment follows
the one presented in Ref. \cite{35:Gerasimov:Matrix:Toda}, with the
necessary modifications appropriate for a two-dimensional Coulomb
gas. First, the normalization of the orthogonal polynomials is given
by:

\[
\int\overline{P_{n}(z)}e^{-|z|^{2}+\sum_{k}t_{k}z^{k}+c.c.}P_{m}(z)d^{2}z=e^{\phi_{n}}\delta_{nm}.\]
 We shall use the fact that orthogonal polynomials are also orthogonal
to any lower order polynomial:\[
k<N\Rightarrow\int\overline{P_{n}(z)}z^{k}e^{-|z|^{2}+\sum_{k}t_{k}z^{k}+c.c.}d^{2}z=0.\]
We wish now to find the equations that $\psi_{n}$ (given in equation
(\ref{eq:BA exact})) satisfies. We define the infinite vector $\vec{\psi}(z)$
with elements $\psi_{n}(z).$ The functions $\psi_{n}(z)$ satisfy:\[
\int\overline{\psi_{n}(z)}\psi_{m}(z)=e^{\phi_{n}}\delta_{nm}.\]
 We differentiate this relation with respect to $t_{1}$ and $\bar{t}_{1}$
and get:\begin{equation}
\int\left(\frac{\partial}{\partial t_{1}}\overline{\psi_{n}(z)}\right)\psi_{m}(z)+\int\overline{\psi_{n}(z)}\left(\frac{\partial}{\partial t_{1}}\psi_{m}(z)\right)=\frac{\partial\phi_{n}}{\partial t_{1}}e^{\phi_{n}}\delta_{nm}\label{t1-diff}\end{equation}
 \begin{equation}
\int\left(\frac{\partial}{\partial\bar{t}_{1}}\overline{\psi_{n}(z)}\right)\psi_{m}(z)+\int\overline{\psi_{n}(z)}\left(\frac{\partial}{\partial\bar{t}_{1}}\psi_{m}(z)\right)=\frac{\partial\phi_{n}}{\partial\bar{t}_{1}}e^{\phi_{n}}\delta_{nm}\label{bar-t1-diff}\end{equation}
 Note that differentiating $P_{n}(z)$ with respect to either $t_{1}$
or $\bar{t}_{1}$ we get a polynomial of degree $n-1$. Thus for example
in\[
\frac{\partial}{\partial t_{1}}\psi_{m}(z)=\frac{\partial}{\partial t_{1}}P_{m}(z)e^{-\frac{|z|^{2}}{2}+\sum_{k}t_{k}z^{k}}=\left(\frac{\partial}{\partial t_{1}}P_{m}(z)\right)e^{-\frac{|z|^{2}}{2}+\sum_{k}t_{k}z^{k}}+z\cdot P_{m}(z)e^{-\frac{|z|^{2}}{2}+\sum_{k}t_{k}z^{k}}\]
 only the second term on the RHS of the equation just above contributes
to the integral $\int\overline{\psi_{n}(z)}\left(\frac{\partial}{\partial t_{1}}\psi_{m}(z)\right)$
when $n>m$. Thus for $n>m$ we have:\[
\int\overline{\psi_{n}(z)}\left(\frac{\partial}{\partial t_{1}}\psi_{m}(z)\right)=\int\overline{\psi_{n}(z)}z\psi_{m}(z)\]
 Note that when $n>m,$ the expression on the RHS is non-zero only
for $n=m+1.$ If we calculate the quantity $\int\overline{\psi_{n}(z)}\left(\frac{\partial}{\partial t_{1}}\psi_{m}(z)\right)$
for all $n$ we will be able to calculate $\frac{\partial}{\partial t_{1}}\psi_{m}(z)$
using a completeness relation. In order to do so we look at $\int\left(\frac{\partial}{\partial t_{1}}\overline{\psi_{n}(z)}\right)\psi_{m}(z)$
in equation (\ref{t1-diff}). Using orthogonality we can again surmise
that $\int\left(\frac{\partial}{\partial t_{1}}\overline{\psi_{n}(z)}\right)\psi_{m}(z)=0$
for $n\leq m.$ Now since the RHS of (\ref{t1-diff}) is also zero
for $n<m$ we come to the conclusion that $\int\overline{\psi_{n}(z)}\left(\frac{\partial}{\partial t_{1}}\psi_{m}(z)\right)=0$
for $n<m$. For $n=m$ the RHS in (\ref{t1-diff}) gives $\frac{\partial\phi_{n}}{\partial t_{1}}e^{\phi_{n}}$
and so we have $\int\overline{\psi_{n}(z)}\left(\frac{\partial}{\partial t_{1}}\psi_{n}(z)\right)=\frac{\partial\phi_{n}}{\partial t_{1}}e^{\phi_{n}}.$
This is all the information we need in order to use the completeness
relation:\[
\frac{\partial}{\partial t_{1}}\psi_{m}(z)=\sum_{n}\frac{1}{\int\overline{\psi_{n}(z)}\psi_{n}(z)}\psi_{n}(z)\int\overline{\psi_{n}(z)}\left(\frac{\partial}{\partial t_{1}}\psi_{m}(z)\right)=\]
\[
=\sum_{n}e^{-\phi_{n}}\psi_{n}(z)\int\overline{\psi_{n}(z)}\left(\frac{\partial}{\partial t_{1}}\psi_{m}(z)\right)=\frac{\partial\phi_{m}}{\partial t_{1}}\psi_{m}(z)+e^{-\phi_{m+1}}\psi_{m+1}\int\overline{\psi_{m+1}(z)}z\psi_{m}(z)\]
 The term $c_{0}\equiv e^{-\phi_{m+1}}\psi_{m+1}\int\overline{\psi_{m+1}(z)}z\psi_{m}(z)$
is recognized as the leading term in the expansion of $z\psi_{m}$
in terms of orthogonal polynomials, $z\psi_{m}=c_{0}\psi_{m+1}+c_{1}\psi_{m}+c_{2}\psi_{m-1}+...$
and since we normalize the orthogonal polynomials to have a unit leading
coefficient we conclude that $c_{0}=1$. Thus \[
\frac{\partial}{\partial t_{1}}\psi_{m}(z)=\frac{\partial\phi_{m}}{\partial t_{1}}\psi_{m}(z)+\psi_{m+1}.\]
 This equation can be written in the form:\[
\frac{\partial}{\partial t_{1}}\vec{\psi}=\left(e^{\partial_{t}}+\frac{\partial\phi_{t}}{\partial t_{1}}\right)\vec{\psi},\]
where $e^{\partial_{t}}$ is the shift operator with respect to the
index of the vector $\psi$ (this index will be denoted by the letter
$t$) Similar arguments lead us, through the use of (\ref{bar-t1-diff})
to the equation:\[
\frac{\partial}{\partial\bar{t}_{1}}\vec{\psi}=-e^{\phi_{n}-\phi_{n-1}}e^{-\partial_{t}}\vec{\psi}\]
 Now consider the compatibility of these two equations:\[
\frac{\partial}{\partial\bar{t}_{1}}\frac{\partial}{\partial t_{1}}\vec{\psi}=\frac{\partial}{\partial t_{1}}\frac{\partial}{\partial\bar{t}_{1}}\vec{\psi}\]
 This leads to:\[
\left[-\frac{\partial}{\partial t_{1}}+e^{\partial_{t}}+\frac{\partial\phi_{t}}{\partial t_{1}},\frac{\partial}{\partial\bar{t}_{1}}+e^{\phi_{n}-\phi_{n-1}}e^{-\partial_{t}}\right]=0\]
 Which reduces to the following equations:\begin{equation}
\frac{\partial^{2}}{\partial t_{1}\partial\bar{t}_{1}}\phi_{t}=e^{\phi_{t}-\phi_{t-1}}-e^{\phi_{t+1}-\phi_{t}}\label{2DTL}\end{equation}
 Namely the 2-dimensional Toda lattice equations. Since $\tau_{N}=\Pi_{i=0}^{N}\tau_{i}$
this equation may be written as (\ref{eq:1st2DTL}).

\chapter{The KdV equation \label{cha:APNDXThe-KdV-equation}}

In this Appendix we outline the way in which the KdV equation can
be integrated using the algebro-geometrical approach. The starting
point is to represent the KdV equation (\ref{eq:KdV}) with $\beta=1$
in the Lax form \cite{Das:Solitons}. For this we define the Lax operator,
$L$:\[
L=-\partial_{x}^{2}-\frac{1}{6}u,\]
 which is just the Schr\"{o}edinger operator with a solution of the
KdV equation inserted as a potential. We also define the operator,
$A$:\[
A=4\partial_{x}^{3}+\frac{1}{2}\left(u\partial_{x}+\partial_{x}u\right).\]
 One can now check, by direct substitution, that \begin{equation}
\partial_{t}L=\left[A,L\right],\label{eq:KdVLax}\end{equation}
 is equivalent to the KdV equation:\[
\partial_{t}u=u\partial_{x}u+\partial_{x}^{3}u.\]

Let us look for solutions of the spectral equation \begin{equation}
L\psi=\lambda\psi,\label{eq:KdvSpectral}\end{equation}
 whose time evolution is given by:\[
A\psi=\partial_{t}\psi.\]
 These two equations are compatible by (\ref{eq:KdVLax}).

\begin{figure}
\includegraphics[%
  width=6cm]{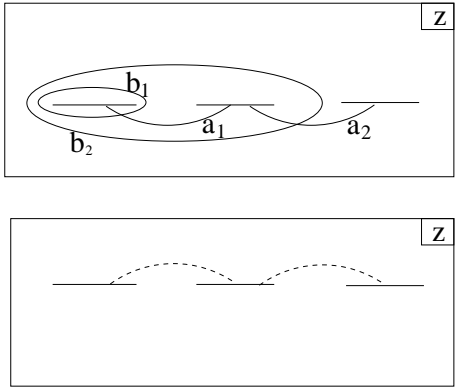}

\caption{Canonical cycles for a genus-2 surface (the generalization is easy
for higher genus). The canonical choice is different from the convention
used in the rest of the paper and shown in Fig (\ref{cap:abcycles}).
\label{cap:Canonical-cycles}}
\end{figure}

One may try to find different nonlinear integrable equations by choosing
appropriate Lax pairs. Namely we should search for a differential
anti-Hermitian operator $B_{n},$ $n\geq0,$ such that:\[
\partial_{t_{n}}L=\left[B_{n},L\right].\]
 The anti-Hermiticity of $B_{n}$ ensures that the spectrum of the
operator $L$ does not depend on time. If we manage to have the RHS
be a multiplicative operator (an operator which multiplies a function
by another function), then we would have an equation of the form:\[
\partial_{t_{n}}u=F_{n}(u,\partial_{x}u,\partial_{x}^{2}u,...).\]
 We have designated a different time, $t_{n},$ for evolutions generated
by different operators $B_{n}$. We may now treat $u$ as a function
of an infinite set of times, \[
u(t_{0},t_{1},t_{2},t_{3},...),\]
 where $x$ corresponds to $t_{0}$ and $t$ stands for the time,
$t_{1}.$ We already know $B_{0}$ and $B_{1}$:\[
B_{1}=A,\,\,\,\,\, B_{0}=\partial_{x}.\]
 Higher $B$'s may be found by formally expanding $L^{\frac{2n+1}{2}}$,
where $u$ is a small parameter, and retaining only the part which
has positive powers in the derivative (designated by the `+' subscript
below). Thus, for example:\[
B_{1}=\left(-L\right)_{+}^{\frac{3}{2}}=\left[\partial_{x}^{3}+\frac{1}{4}\left(\partial_{x}u+u\partial_{x}\right)+O\left(\partial_{x}^{-1}\right)\right]_{+}.\]
 See, e.g., \cite{Das:Solitons} for more details. We can now choose
the time evolution of the wave function to be consistent with the
generators associated with the infinite set of times:\begin{equation}
\partial_{t_{n}}\psi=B_{n}\psi.\label{eq:KdvFullLax}\end{equation}
 The wave function $\psi$ depends on the infinite set of times $\left\{ t_{k}\right\} _{k=0}^{\infty}$
and on the spectral parameter $\lambda$. Regarded as a function of
$\lambda$, $\psi$ has branch cuts in the spectrum of the operator
$L$. As explained in the body of the text, this motivates one to
define $\psi$ on an algebraic Riemann surface. To generate exact
solutions of the KdV equations, we may first construct a function
satisfying (\ref{eq:KdvFullLax}) and (\ref{eq:KdvSpectral}), defined
on some Riemann surface, and then find $u$ from the spectral equation.
To do this we must draw our attention to the analytic properties of
the wave function. The first analytic property we may derive for the
function $\psi$ is its $t_{0}$ (or $x$) dependence near infinity,
it is an essential singularity: \[
\psi\propto e^{\sqrt{\lambda}t_{0}}.\]
 This is easily seen from equation (\ref{eq:KdvSpectral}). The essential
singularities which are controlled by the higher times, can be similarly
shown to lead to\[
\psi\propto e^{\sqrt{\lambda}^{2n+1}t_{n}}.\]
 This follows from examination of the asymptotic behavior of equations
(\ref{eq:KdvFullLax}). We may now assume that these are the only
essential singularities of the wave function, namely that \begin{equation}
\psi e^{-\sum_{k}t_{k}\sqrt{\lambda}^{2n+1}}\label{eq:essentialSingKdv}\end{equation}
 is meromorphic around infinity. The other property that $\psi$ can
be shown to have (which we do not show here) is that the wave function
on a genus $g$ Riemann surface has $g$ poles. Given the location
of the poles and the form of the essential singularity (\ref{eq:essentialSingKdv})
the wave function is unique. The function with analytic properties
outlined above is known in the theory of Riemann surfaces as the Baker-Akhiezer
function. We now turn to a detailed exposition of the construction
of the BA. To do this we first define explicitly the Riemann surface
on which the wave function is defined. The Riemann surfaces occurring
in the KdV equations are hyper-elliptic surfaces, that is an algebraic
Riemann surface given by an equation of the form \[
y^{2}=\Pi_{i=1}^{2g+1}\left(\lambda-\lambda_{i}\right).\]
 One can think of this as $2g+2$ branch points on the complex plane
where the first $2g+1$ branch points are the $\lambda_{i}$'s and
the last branch point is at infinity. A Riemann surface of nonzero
genus has non-trivial cycles on it, that is close contours which cannot
be deformed continuously to a point. We choose the canonical set of
cycles, $a_{i}$ and $b_{i}$, $1\leq i\leq g$, shown in Fig.~\ref{cap:Canonical-cycles}.
Notice that this choice of $a$ and $b$ cycles does not conform with
the choice made in the body of this article.

To proceed, we need certain meromorphic differentials on the Riemann
surface. Differentials from the first group are holomorphic everywhere
on the Riemann surface (they are called differentials of the first
kind). On any smooth Riemann surface of genus $g$, the space of holomorphic
differentials is $g$-dimensional. A basis in this space, $\omega_{i}$,
can be fixed by imposing the normalization condition \begin{equation}
\oint_{a_{j}}\omega_{j}=\delta_{ij}.\label{eq:holoNorma}\end{equation}
 Explicitly, they are of the form\begin{equation}
\omega_{j}=\frac{\lambda^{g-1}a_{g-1}+\lambda^{g-2}a_{g-2}+...+a_{0}}{\sqrt{\Pi_{i=1}^{2g+1}\left(\lambda-\lambda_{i}\right)}}d\lambda\label{eq:HoloDiffGenus1}\end{equation}
 where $a_{i}$ are to be fixed by the normalization requirement.

The second group of differentials are those having poles of order
$m$, $m>1$ at a point $\zeta$ (known as differentials of the second
kind). As an example, we take differentials with the pole located
at infinity for a genus $1$ surface:\[
\omega_{\zeta}^{(2)}=\frac{\lambda d\lambda}{\sqrt{\Pi_{i=1}^{3}\left(\lambda-\lambda_{i}\right)}}+c\omega_{1},\]
 where the constant, $c$, should be chosen according to some normalization
of the differentials. The standard normalization we shall use here
is vanishing of all $a$-periods:\[
\oint_{a_{i}}\omega_{A}^{(m)}=0.\]

Differentials of the third kind are meromorphic differentials which
have simple poles at two points, $\zeta_{1}$ and $\zeta_{2}$, with
residues $1$ and $-1$ respectively. As before, the differential
is to be normalized.

Having introduced the meromorphic differentials on the Riemann surface,
we can try to find the wave function as:\[
\psi\sim e^{\int\hat{S}},\]
 Here $\hat{S}$ is a normalized differential with poles at infinity
of the form:\[
\hat{S}\sim\sum_{n}\frac{t_{n}d\lambda}{\sqrt{\lambda}^{1-2n}},\]
 as dictated by (\ref{eq:essentialSingKdv}). This differential can
be represented as a sum of normalized differentials of the second
kind. So the defined $\psi$-function has the correct essential singularity
but is not single-valued on the Riemann surface.

To amend this problem we multiply the wave function by a function
which picks up the opposite phase on each of the cycles, while not
contributing to the essential singularity at infinity. This can be
done with the help of the Riemann $\Theta$-functions. Their basic
properties are given below.

Given the normalized holomorphic differentials (\ref{eq:holoNorma}),
one defines a matrix $\tau_{ij}$ (known as the Riemann matrix) given
by\[
\tau_{ij}=\oint_{b_{i}}\omega_{j},\]
 where $b_{i}$ are the $b$-type cycles, as illustrated in Fig.~\ref{cap:Canonical-cycles}.
Associated with this Riemann matrix is the Riemann theta function
(see, e.g., Ref.~\cite{84:Dubrovin:Algebr:Geome}): \[
\theta_{B}(\vec{z})=\sum_{\vec{m}\in\mathbb{Z}^{g}}e^{2\pi i\left(\frac{1}{2}\left\langle \vec{m},\tau\vec{m}\right\rangle +\left\langle \vec{m},\vec{z}\right\rangle \right)}.\]
 We can use the Riemann theta function to construct functions defined
on the Riemann surface. Since $\theta$ has $\mathbb{C}^{g}$ as its
domain, we first must define a function from the Riemann surface to
$\mathbb{C}^{g}$. This function is given by the Abel map which is
actually a mapping from the $g$-torus to $\mathbb{C}^{g}/L$, where
$L$ is a discrete group%
\footnote{$L$ is given by $\left\{ \left(\oint_{d}\omega_{1},\oint_{d}\omega_{2},...,\oint_{d}\omega_{g}\right)\right\} $,
where $d$ runs over all cycles on the Riemann surface. It is easy
to calculate that $L=\mathbb{Z}^{g}+\tau\mathbb{Z}^{g}$. %
}, given by:\[
\zeta\rightarrow\left(\int_{\zeta_{0}}^{\zeta}\omega_{1},\int_{\zeta_{0}}^{\zeta}\omega_{2},...,\int_{\zeta_{0}}^{\zeta}\omega_{g}\right)\]
 Here $\zeta_{0}$ is some given point on the Riemann surface. We
shall designate this mapping by $\vec{F}(\zeta)$. The image of the
Riemann surface under $\vec{F}$ in $\mathbb{C}^{g}/L$ is known to
be isomorphic to the Riemann surface. We may use this isomorphism
to construct functions on the Riemann surface by considering functions
whose domain is $\mathbb{C}^{g}/L$. We first examine the quasi-periodicity
of the Riemann theta function: \[
\theta\left(\vec{z}+\vec{m}+B\vec{n}\right)=e^{2\pi i\left(-\left\langle \vec{n},\vec{z}\right\rangle a-\frac{1}{2}\left\langle \vec{n},\tau\vec{n}\right\rangle \right)}\theta(\vec{z}).\]
 Having this property, we can now construct a well-defined function
on the Riemann surface with essential singularities appropriate for
the BA function with poles at points $\zeta_{i}$, $1\leq i\leq g$
\cite{132:Krich:Intgr:AlgGeo,Baker}: \[
\psi=\frac{\theta\left(\vec{F}(\zeta)+\vec{U}-\vec{C}\right)}{\theta\left(\vec{F}(\zeta)-\vec{C}\right)}e^{\int_{\zeta_{0}}^{\zeta}\hat{S}}\]
 Here $U_{k}=\oint_{b_{k}}\hat{S}$ and $\vec{C}$ should be chosen
such that the zeros for the theta function are at the points $\zeta_{i}$.
It can be shown that this constant should be chosen as $\vec{C}=\vec{\Xi}+\sum_{i=1}^{g}\vec{F}(\zeta_{i})$,
where \[
\Xi_{i}=\frac{1-\sum_{i=1}^{g}\tau_{j,i}}{2}.\]
.

\chapter{The spectral operator of the 2DTL and the string equation \label{apndx:Sting2DTL}}

In this Appendix we show that the spectral operator ${\mathcal{L}}$
constructed in Sec.~\ref{sub:2DTL} is indeed the spectral operator
of the 2DTL hierarchy and prove the string equation (\ref{eq:string-equation}).
The proof is based upon the methods used e.g.~in Ref.~\cite{35:Gerasimov:Matrix:Toda}.

We begin by introducing equations of the 2DTL hierarchy \cite{58:Takasaki}.
Eq. (\ref{eq:1st2DTL}) is the first equation of the hierarchy. The
higher equations are generated by the Lax representation \begin{equation}
\frac{\partial{\mathcal{L}}^{k}}{\partial t_{q}}=\left[{\mathcal{L}}^{k},{\mathcal{L}}_{+}^{q}\right],\label{Lax}\end{equation}
\begin{equation}
\frac{\partial{\mathcal{L}}^{k}}{\partial\bar{t}_{q}}=\left[{\mathcal{L}}^{k},\bar{\mathcal{L}}_{+}^{q}\right]\label{eq:Lax2}\end{equation}
 where ${\mathcal{L}}^{k}$ is the operator ${\mathcal{L}}$ raised
to the power $k$, while the subscript $+$ means taking the lower
triangular part of the matrix represented in the basis of $\psi_{m}(z)$
(as we shall describe below). These equations manifest the fact that
the 2DTL is an integrable systems whose infinite number of constants
of motion are conjugate to the time variables $t_{q}$.

We turn now to prove that the operator ${\mathcal{L}}$ built in order
to satisfy the spectral equation, ${\mathcal{L}}\psi_{m}=z\psi_{m}$,
with \begin{equation}
\psi_{m}(z)=e^{-\frac{1}{\hbar}\left(\frac{|z|^{2}}{2}-\sum_{k}t_{k}z^{k}\right)}P_{m}(z),\end{equation}
 indeed solves Eqs. (\ref{Lax}) and (\ref{eq:Lax2}).

To shorten the notation, we designate the orthogonal polynomials $P_{m}(z)$
as $|m\rangle$, and write the inner product in the form \begin{equation}
\langle n|m\rangle=\!\int\! d^{2}z\overline{P_{n}(z)}P_{m}(z)e^{-\bar{z}z+\sum_{k}t_{k}z^{k}+C.c.}=e^{\phi_{n}}\delta_{mn}\label{eq:inner-product}\end{equation}
 where $e^{-\bar{z}z+\sum_{k}t_{k}z^{k}+C.c.}d^{2}z$ is the integration
measure. From the spectral equation have \begin{equation}
{\mathcal{L}}^{k}\frac{\left|m\right\rangle }{\sqrt{\left\langle m\right|\left.m\right\rangle }}=\frac{z^{k}\left|m\right\rangle }{\sqrt{\left\langle m\right|\left.m\right\rangle }}.\label{eq:Lax-Sato}\end{equation}
 where $z^{k}|m\rangle$ means the function $z^{k}P_{m}(z)$. In particular,
the matrix elements of ${\mathcal{L}}^{k}$, which we denote by ${\mathcal{L}}_{nm}^{k}$,
take the form \begin{eqnarray}
\frac{\langle n|{\mathcal{L}}^{k}|m\rangle}{\sqrt{\langle n|n\rangle\langle m|m\rangle}}=\frac{\langle n|z^{k}|m\rangle}{\sqrt{\langle n|n\rangle\langle m|m\rangle}}=e^{\frac{1}{2}\left(\phi_{m}-\phi_{n}\right)}\gamma_{nm}^{(k)},\label{eq:Lknm}\end{eqnarray}
 where we have used the inner product formula (\ref{eq:inner-product}),
and defined $\gamma_{nm}^{(k)}$ to be: \begin{equation}
\gamma_{nm}^{(k)}=\frac{\langle n|z^{k}|m\rangle}{\langle m|m\rangle}\end{equation}

To calculate the LHS of equation (\ref{Lax}), let us first expand
$\langle n|z^{k}$ using the matrix elements $\gamma_{nm}^{(k)}$:
\begin{equation}
\langle n|z^{k}=\sum_{m=0}^{n+k}\gamma_{nm}^{(k)}\langle m|,\label{basic}\end{equation}
 and then differentiate this equation with respect to $t_{q}$. We
get: \[
\frac{\partial}{\partial t_{q}}\left(\langle n|z^{k}\right)=\left(\frac{\partial}{\partial t_{q}}\langle n|\right)z^{k}=\]
\begin{equation}
=\sum_{m=0}^{\infty}\left(\frac{\partial}{\partial t_{q}}\gamma_{nm}^{(k)}\right)\langle m|+\sum_{m=0}^{\infty}\gamma_{nm}^{(k)}\left(\frac{\partial}{\partial t_{q}}\langle m|\right)\label{basic2}\end{equation}
 Now, in order to calculate the derivatives of the orthogonal polynomials
we differentiate the orthogonality condition (\ref{eq:inner-product}).
The result takes the form: \begin{eqnarray}
 & \frac{\partial}{\partial t_{q}}\langle n|m\rangle & =\frac{\partial\phi_{n}}{\partial t_{q}}e^{\phi_{n}}\delta_{nm}=\label{eq:temp}\\
 & \left(\frac{\partial}{\partial t_{q}}\langle n|\right) & |m\rangle+\langle n|\left(\frac{\partial}{\partial t_{q}}|m\rangle\right)+\langle n|z^{q}|m\rangle,\nonumber \end{eqnarray}
 where the last term in the RHS comes from the differentiation of
the measure $e^{-|z|^{2}+\sum t_{k}z^{k}+c.c.}$. Since the leading
monomial in $|l\rangle$ is $z^{l}$ (with unit coefficient), the
time differential $\frac{\partial|l\rangle}{\partial t_{q}}$ can
be written as combination of polynomials $|n\rangle$, with $n<l$.
This implies that the second term in the RHS of (\ref{eq:temp}) vanishes
for $m\geq n$ while the third one vanishes for $m\leq n$. Thus for
$n=m$: \begin{equation}
\frac{\partial\phi_{n}}{\partial t_{q}}=\frac{\langle n|z^{q}|n\rangle}{\langle n|n\rangle}=\gamma_{nn}^{(q)}\label{phideriva}\end{equation}
 and for $m<n$: \[
\frac{\partial\langle n|}{\partial t_{q}}|m\rangle=-\langle n|z^{q}|m\rangle=-\sum_{m=0}^{n-1}\gamma_{nm}^{(q)}\langle m|\]
 Substituting these results to (\ref{basic2}) we get:\[
-\sum_{m=0}^{n-1}\gamma_{nm}^{(q)}\langle m|z^{k}=\sum_{m=0}^{\infty}\frac{\partial\gamma_{nm}^{(k)}}{\partial t_{q}}\langle m|-\sum_{l=0}^{\infty}\sum_{m=0}^{l-1}\gamma_{nl}^{(k)}\gamma_{lm}^{(q)}\langle m|,\]
 Using (\ref{basic}) once again, and rearranging the terms, we conclude
that \[
\frac{\partial\gamma_{nm}^{(k)}}{\partial t_{q}}=\sum_{l=m+1}^{\infty}\gamma_{nl}^{(k)}\gamma_{lm}^{(q)}-\sum_{l=0}^{n-1}\gamma_{nl}^{(q)}\gamma_{lm}^{(k)}\]
 This equation is almost (\ref{Lax}). It still has to be corrected
by the normalization factors (\ref{eq:Lknm}). Taking these into account
we obtain,\[
\begin{array}{l}
\frac{\partial}{\partial t_{q}}\left(e^{\left(\phi_{m}-\phi_{n}\right)/2}\gamma_{nm}^{(k)}\right)=\\
e^{\left(\phi_{m}-\phi_{n}\right)/2}\left(-\sum_{l=0}^{n-1}\gamma_{nl}^{(q)}\gamma_{lm}^{(k)}+\sum_{l=m+1}^{\infty}\gamma_{nl}^{(k)}\gamma_{lm}^{(q)}+\frac{1}{2}\gamma_{nm}^{(k)}\frac{\partial\phi_{m}}{\partial t_{q}}-\frac{1}{2}\gamma_{nm}^{(k)}\frac{\partial\phi_{n}}{\partial t_{q}}\right).\end{array}\]
Substituting formula (\ref{phideriva}) for the derivatives $\frac{\partial\phi_{m}}{\partial t_{q}}$,
yields equation (\ref{Lax}), where \[
{\mathcal{L}}_{nm+}^{q}=\left\{ \begin{array}{ll}
e^{\frac{1}{2}(\phi_{m}-\phi_{n})}\gamma_{nm}^{(q)} & \mbox{if}~~~n>m\\
\\\frac{1}{2}e^{\frac{1}{2}(\phi_{m}-\phi_{n})}\gamma_{mm}^{(q)} & \mbox{if}~~~n=m\\
\\0 & \mbox{if}~~~n<m\end{array}\right.\]
 Thus we proved the first equation of the 2DTL hierarchy, (\ref{Lax}).
A similar proof holds also for the second equation (\ref{eq:Lax2}),
and therefore ${\mathcal{L}}$ is indeed the spectral operator of
the 2DTL and $\psi_{n}(z)$ is the $n$-th component of the BA function.

Solutions of the 2DTL corresponding to the matrix model are specified
by the additional constraint that the operator $\mathcal{L}$ satisfies
the {}``string equation\char`\"{} \begin{equation}
\left[{\mathcal{L}},{\mathcal{L}}^{\dagger}\right]=\hbar\label{eq:TheString}\end{equation}
 To see this, we write:\[
\left\langle m\right|\mathcal{L}^{\dagger}\left|n\right\rangle =\int\overline{\psi_{m}(z)}\bar{z}\psi_{n}(z)d^{2}z=\]
 \[
=-\int\left(\hbar\partial_{z}e^{-\frac{1}{\hbar}|z|^{2}}\right)P_{m}(\bar{z})P_{n}(z)e^{\frac{1}{\hbar}\left(\sum_{k}t_{k}z^{k}+\bar{t}_{k}\bar{z}^{k}\right)}d^{2}z=\]
 \[
=\!\hbar\!\!\int\!\! e^{-\frac{1}{\hbar}\left(|z|^{2}+\sum_{k}t_{k}z^{k}+\bar{t}_{k}\bar{z}^{k}\right)}P_{m}(\bar{z})\left(\partial_{z}\!-\! t_{k}kz^{k-1}\right)\! P_{n}(z)d^{2}z\]
 Thus $\mathcal{L}$ acts on polynomials by multiplication by $z$
while $\mathcal{L}^{\dagger}$ acts as $\hbar\left(\partial_{z}-\sum_{k}t_{k}z^{k-1}k\right)$.
This implies that $\mathcal{L}$ and $\mathcal{L}^{\dagger}$ satisfy
the canonical commutation relation (\ref{eq:TheString}).

\chapter{Pressure inside bubbles\label{sec:AppendixPressure}}

In this appendix we prove that the normalization condition (\ref{eq:normalization})
implies that all bubbles share the same pressure. For this purpose
we show that the derivative of $Q(z)$ defined in (\ref{eq:DefinitionQz}),
on the bubble contour, equals to the pressure inside the bubble. Differentiating
$Q(z)$ with respect to time we have that $\frac{dQ(z)}{dt}$ is equal
to the following expression: \[
\Re\left(-\frac{dz}{dt}\bar{z}+S(z)\frac{dz}{dt}+\int^{z}\frac{\partial S(z')}{\partial t}dz'\right)=\Re\int^{z}\frac{\partial S(z')}{\partial t}dz',\]
where the second equality follows from the fact that on the contour
$\bar{z}=S(z)$. To show that the expression on the right hand side
is the pressure, we first define $\Psi=P+i\Theta$ where $\Theta$
is the harmonic conjugate of the pressure, and show that \[
\int^{z}\frac{\partial S(z')}{\partial t}dz'\]
 is the pressure by showing that \begin{equation}
\frac{\partial S}{\partial t}dz=d\Psi.\label{eq:dSdtdPdz}\end{equation}
 To obtain this result we differentiate $S(z(t))=\bar{z}(t)$ with
respect to time, where $z$ is some point on the contour, then one
obtains:\begin{equation}
\frac{\partial S}{\partial t}dz=\frac{d\bar{z}}{dt}dz-\frac{dz}{dt}dS.\label{eq:DarcyDsDt2}\end{equation}
 One may obtain(\ref{eq:dSdtdPdz}) by choosing $dz$ to be in any
direction and performing an analytical continuation. We choose $dz$
to be a differential in the direction of the contour (and then $dS=d\bar{z})$
of length $dl.$ The RHS of (\ref{eq:DarcyDsDt2}) is clearly purely
imaginary on the contour and may be understood as $\frac{d\vec{r}}{dt}\times d\vec{l}$,
written in complex coordinates. Namely, it is proportional to the
normal velocity of the contour, thus \[
\frac{\partial S}{\partial t}dz=iv_{n}dl=i\frac{\partial P}{\partial n}dl=i\frac{\partial\theta}{\partial l}dl=id\theta=d\Psi,\]
 where $v_{n}$ is the normal velocity, $\frac{\partial P}{\partial n}$
is the normal derivative of the pressure which is equal by Cauchy-Riemann
to $\frac{\partial\Theta}{\partial l},$ the tangential derivative
of $\Theta$. Thus the equality $Q(\zeta_{1})=Q(\zeta_{2})$ where
$\zeta_{1}$ and $\zeta_{2}$ are points on the contour of two different
bubbles implies that the pressure difference between the bubbles is
zero.

\chapter{The 2-Miwa equations\label{sec:Appendix:-The-2-Miwa} }

In this Appendix we provide the details of the calculation of the
evolution of 2-Miwa genus-1 system. There are two main issues to consider:
The nonlinear equations (\ref{eq:qi}-\ref{eq:t}) and the normalization
condition (\ref{eq:normailzation}).

Let us first write down the conformal map from the torus $-a_{x}<\Re u\leq a_{x}$,
$-a_{y}<\Im u\leq a_{y}$ to the $z$-plane. The mapping, in this
case, would consist of three poles: \[
z(u)=\sum_{i=0}^{2}\alpha_{i}\zeta(u-u_{i})+\beta\,\,\,\,\,\,\,\mbox{where}\,\,\,\,\,\,\sum_{i=0}^{2}\alpha_{i}=0,\]
 $\zeta(u)$ is the Weierstrass zeta function, and $\beta$ is a constant.
Thus our problem is to determine seven unknowns: $\alpha_{0},\alpha_{1},u_{0},u_{1},u_{2},\beta$,
and the ratio $a_{y}/a_{x}$ (one of the cycles, say $a_{x}$ may
be taken to be any constant). Thus equations (\ref{eq:qi}-\ref{eq:t})
and the normalization condition provide the seven constrains needed
to determine these unknowns.

The location of the Miwa variables can be determined from Eq. (\ref{eq:qi}):
\[
q_{i}=z(-\bar{u}_{i})=-\sum_{j=0}^{2}\alpha_{j}\zeta(\bar{u}_{j}+u_{i})+\beta,\]
 where we have used the property $\zeta(-u)=-\zeta(u)$. The equation
for $t'_{1}$ (\ref{eq:t1}) is the same as the above equation with
the identification $q_{0}\equiv t'_{1}.$

The Miwa weights, $\mu_{i}$, can be computed from (\ref{eq:mui}):
\[
\mu_{i}=\bar{\alpha}_{i}z'(-\bar{u}_{i})=-\bar{\alpha}_{i}\sum_{j=0}^{2}\alpha_{j}\wp(\bar{u}_{i}+u_{j}),\]
 where $\wp(u)=-\zeta'(u)$, is the Weierstrass elliptic function.
Finally, equation (\ref{eq:t}) for $t$ is \[
t=-\bar{\alpha}_{0}\sum_{j=0}^{2}\alpha_{j}\wp(\bar{u}_{0}+u_{j})+\sum_{i}\mu_{i}.\]

Consider now the equation provided by the normalization condition
(\ref{eq:normailzation}). Using the representation of the Schwarz
function in term of the conformal mapping, $S(z)=\bar{z}(-u(z))$,
we may write the normalization condition as an integral in the $u$-plane:
\begin{equation}
\Re\left(\oint_{b}S(z)dz\right)=\Re\left(\oint\bar{z}(-u)z'(u)du\right),\label{eq:App:norm}\end{equation}
 The contour integral goes from $-a_{x}$ to $a_{x}$, as it follows
from the definition of the $b$-cycles of the Riemann surface and
the structure of the $u$-plane.

Now, by identifying the singular behavior of the elliptic function
$\bar{z}(-u)z'(u)$ on the torus, one may surmise that:\begin{equation}
\bar{z}(-u)z'(u)=\sum_{i=0}^{2}\left[\alpha_{i}\bar{z}'(-u_{i})\zeta(u-u_{i})-\bar{\alpha}_{i}z'(-\bar{u}_{i})\zeta(u+\bar{u}_{i})-\alpha_{i}\bar{z}(-u_{i})\wp(u-u_{i})\right]+C,\label{eq:surmize}\end{equation}
 where $C$ is a constant which can be determined by comparing the
right and left hand sides at some arbitrary point.

Thus to compute the normalization integral (\ref{eq:App:norm}) we
need to evaluate the integrals \[
\int_{-a_{x}}^{a_{x}}du\wp(u-u_{i}),~~~~\mbox{and}~~~\int_{-a_{x}}^{a_{x}}du\zeta(u-u_{i}).\]
 Using the quasi-periodicity of the Weierstrass $\zeta$-function
in the $x$ direction, \[
\zeta(u+2a_{x})=\zeta(u)+2\eta_{x}\]
 where $\eta_{x}$ is a constant which depends on $a_{x}$ and $a_{y}$,
we find that \begin{equation}
\int_{-a_{x}}^{a_{x}}du\wp(u-u_{i})=2\eta_{x}\label{eq:cycle-wp}\end{equation}
 The second integral can be deduced from the quasi-periodicity properties
of the $\sigma$-function: \[
\int du\zeta(u)=\log\sigma(u),\quad\sigma(u+2a_{x})=-\sigma(u)e^{\left(u+a_{x}\right)2\eta_{x}}.\]
 Therefore, \begin{equation}
\int_{-a_{x}}^{a_{x}}du\zeta(u\pm U)=i\pi+(a_{x}\pm U)2\eta_{x}.\label{eq:cycle-zeta}\end{equation}
 Substituting the integrals (\ref{eq:cycle-wp}) and (\ref{eq:cycle-zeta})
into the normalization integral (\ref{eq:App:norm}) with the integrand
(\ref{eq:surmize}), we arrive at the conclusion that: \begin{eqnarray*}
 & \Re & \left(\oint_{b}S(z)dz\right)=\\
 & \Re & \left(2a_{x}C-2\eta_{x}\sum_{i=0}^{2}\left(\alpha_{i}\bar{z}'(-u_{i})2u_{i}-\alpha_{i}\bar{z}(-u_{i})\right)\right).\end{eqnarray*}

\chapter{Merging bubbles.\label{cha:Merging-Droplets.}}

Let us take a closer look at the bubbles when they merge. We first
derive the conformal mapping at this merging moment. Since at this
point the torus degenerates, we expect a divergence of one of the
periods. It turns out that the period which diverges is $a_{y}$,
thus $z(u)$ which was periodic in $2a_{x}$ and $2ia_{y}$, is now
only periodic in $2a_{x}$. In the region $-a_{x}<\Re(u)<a_{x}$ the
singularities of $z(u)$ are poles at the points $u_{0}$, $u_{1}$
and $u_{2}$, with weights $\alpha_{0},$ $\alpha_{1}$ and $-(\alpha_{0}+\alpha_{1})$.
A mapping satisfying these conditions can be written as:\[
z(u)=\frac{i\pi}{a_{x}}e^{iu\frac{\pi}{a_{x}}}\left(\frac{\alpha_{0}}{e^{iu\frac{\pi}{a_{x}}}-e^{iu_{0}\frac{\pi}{a_{x}}}}+\frac{\alpha_{0}}{e^{iu\frac{\pi}{a_{x}}}-e^{iu_{1}\frac{\pi}{a_{x}}}}-\frac{\alpha_{0}+\alpha_{1}}{e^{iu\frac{\pi}{a_{x}}}-e^{iu_{2}\frac{\pi}{a_{x}}}}\right)+z_{*}.\]
 The two bubbles, which now merge, are the images of $\Re(u)=0$ and
of $\Re(u)=a_{x}$. Since these two bubbles now compose a single bubble,
$\Re(u)=0$ is mapped to a part of the contour which bounds the merged
bubble, while $\Re(u)=a_{x}$ is mapped to the other part. Both parts
merge at the point $z_{*}$. The fact that the pre-image of the contour
that bounds the bubbles is composed of two disconnected pieces, is
in contrast to the usual situation, where one maps a single contour
(usually the unit circle) to the boundary of the bubble. To bridge
this gap, we define the variable $p=e^{iu\frac{\pi}{a_{x}}}$. We
see that $\Re(u)=0$ corresponds to $p>0$, on the real axis, while
$\Re(u)=a_{x}$ corresponds to $p<0$ on the real axis. Thus in the
coordinate $p$ the pre-image of the contour is the real axis, while
the points $p=0$ and $p=\infty$ are mapped to $z_{*}$ . To obtain
a mapping from the unit circle to the boundary of the bubble, we have
to map the real line to the unit circle, using a M\"{o}bius transformation.
Adding the condition that $u_{1}$ is mapped to the point at infinity,
we obtain that the following M\"{o}bius transformation:\[
w(u)=\frac{1-e^{iu_{1}\frac{\pi}{a_{x}}}e^{iu\frac{\pi}{a_{x}}}}{e^{iu\frac{\pi}{a_{x}}}-e^{iu_{1}\frac{\pi}{a_{x}}}},\]
 which maps the real axis to the unit circle. To sum up, the mapping
$z\left(u\left(w\right)\right)$, which we shall denote for short
as $z(w)$, sends the exterior of the unit circle in $w$ plane to
the exterior of the merged bubbles in the $z$ plane. The point $z_{*}$
has two pre-images, $w_{1}\equiv-e^{iu_{1}\frac{\pi}{a_{x}}}$ (which
corresponds to $p\rightarrow\infty)$, and $w_{2}\equiv-e^{-u_{1}\frac{\pi}{a_{x}}}$
(which corresponds to $p\rightarrow0$). As can be straightforwardly
checked, the mapping $z(w)$ is regular in $w$ at these points. We
see that the point of merging of the two bubbles is the common point
of two smooth tangent lines (assuming that $\frac{\partial z(w)}{\partial w}|_{w_{i}}\neq0$
for $i=1,2,$ which holds in general). To describe the shape of the
aggregate near the merging point, we choose a Cartesian coordinate
system whose origin is at $z_{*}$, and whose $x$ axis is tangent
to the smooth contours of the bubbles. Since we have two smooth differentiable
contours meeting at the point $z_{*}$, we obtain that the generic
scaling is \[
y^{2}\sim x^{4},\]
 for small $x$ and $y$ on the contour.

\bibliographystyle{prsty}
\cleardoublepage\addcontentsline{toc}{chapter}{\bibname}\bibliography{mybib}

\end{document}